\documentclass[conference]{IEEEtran}

\usepackage{amsmath,amssymb,latexsym,amsthm}
\usepackage{graphicx}
\usepackage{psfrag}
\usepackage{caption}
\usepackage[labelformat=simple]{subcaption}
\usepackage[hyphens]{url}
\usepackage{ifthen}
\usepackage{draftwatermark}
\usepackage{color}
\usepackage[english]{babel}
\usepackage{amsfonts}
\usepackage{paralist}
\usepackage{algorithm}
\usepackage[]{algpseudocode}
\usepackage{colortbl}	
\usepackage[dvipsnames]{xcolor}
\usepackage{ulem} 
\usepackage{cancel}

\usepackage[inline]{enumitem}

\usepackage[pagewise]{lineno}

\newcommand{\ceil}[1]{\lceil #1 \rceil}

\newcommand{\Np}{\goodbreak\medskip\noindent}

\newcommand{\blockchain}{blockchain}

\newcommand{\showboth}{1}
\newcommand{\showbothR}{1}

\ifthenelse{\showboth = 0}
{\newcommand{\cremove}[2]{%
\ifthenelse{\equal{#1}{}}%
{\textcolor{red}{#2}}
{\textcolor{red}{\sout{#1}\ }\textcolor{red}{#2}}}
}
{\newcommand{\cremove}[2]{{}\textcolor{black}{#2}}}

\ifthenelse{\showbothR = 0}
{\newcommand{\cremoveR}[2]{%
\ifthenelse{\equal{#1}{}}%
{\textcolor{RoyalBlue}{#2}}
{\textcolor{RoyalBlue}{\sout{#1}\ }\textcolor{RoyalBlue}{#2}}}
}
{\newcommand{\cremoveR}[2]{{}\textcolor{black}{#2}}}

\newcommand{\rem}[1]{}
\newcommand{\var}{\mathrm{Var}}

 \SetWatermarkLightness{1.00}

\IEEEoverridecommandlockouts

\graphicspath{{./GRAPHICS/}}

\newcommand{\Prob}{\mathbb{P}}
\newcommand{\R}{\mathbb{R}}
\newcommand{\E}{\mathbb{E}} 

\title{\mbox{Block arrivals in the Bitcoin blockchain}}

\author{%
\IEEEauthorblockN{%
R. Bowden%
\IEEEauthorrefmark{1},
H.P. Keeler%
\IEEEauthorrefmark{1},
A.E. Krzesinski%
\IEEEauthorrefmark{2}
and P.G. Taylor%
\IEEEauthorrefmark{1}
}%
\IEEEauthorblockA{%
\IEEEauthorrefmark{1}%
School of Mathematics and Statistics,
University of Melbourne,
Vic 3010,  Australia \\
Email: \{rhys.bowden,hkeeler,taylorpg\}@unimelb.edu.au}%
\IEEEauthorblockA{%
\IEEEauthorrefmark{2}%
Department of Mathematical Sciences,
Stellenbosch University,
7600 Stellenbosch, South Africa \\
Email: aek1@cs.sun.ac.za}%
\thanks{%
The work of Anthony Krzesinski is supported by Telkom SA Limited.}
\thanks{%
The work of Peter Taylor is supported by the Australian Research Council
Laureate Fellowship FL130100039 and the ARC Centre of Excellence for
Mathematical and Statistical Frontiers (ACEMS).}
}

\begin{document}

\maketitle

\begin{abstract}

Bitcoin is a electronic payment system where payment transactions are
verified and stored in a data structure called the \blockchain. Bitcoin
miners work individually to solve a computationally intensive problem,
and with each solution a Bitcoin block is generated, resulting in a new
arrival to the blockchain.
The difficulty of the computational problem is updated every 2,016 blocks
in order to control the rate at which blocks are generated.
In the original Bitcoin paper, it was suggested that the blockchain
arrivals occur according to a homogeneous Poisson process. Based on
blockchain block arrival data and stochastic analysis of the
block arrival process, we demonstrate that this is not the case.
We present a refined mathematical model for block arrivals, focusing on
both the block arrivals during a period of constant difficulty and how
the difficulty level evolves over time.


\end{abstract}

%
%

\normalem 


\section{Introduction}

Bitcoin is an electronic currency and payment system which allows for
monetary transactions without a central authority. Bitcoin was first
proposed in a white paper \cite{nakamoto2008bitcoin} in 2008 by the
pseudonymous Satoshi Nakamoto, and as a functioning open source software
system in January  2009.  Since then it has grown rapidly and now has a
market capitalisation of over US\$160 billion.

The stated benefits of Bitcoin relative to payment over the Internet
with traditional currencies include immutable transactions, relative
anonymity, freedom against seizure by a central authority, faster and
cheaper international currency transfer, among others. The technology
used in Bitcoin also has potential to be applied to a wider range of
applications including identity management, distributed certification
and smart contracts.

Bitcoin employs a peer-to-peer network of \emph{nodes}, computers which
verify, propagate and store information about bitcoin transactions.
Bitcoin maintains a global ledger of all Bitcoin transactions ever
conducted -- the Bitcoin blockchain -- distributed over the network of
Bitcoin nodes. The problems of maintaining and updating a global ledger
without a trusted central authority, while still allowing any owner of
bitcoins to spend them as they wish are solved by a series of
cryptographic techniques. The technique of central interest for this
paper is \emph{proof-of-work}, whereby those who wish to contribute to
building a consensus about which transactions are included in the ledger
must perform some computational ``work''.  This work is referred to as
Bitcoin mining.


Bitcoin users conduct transactions by cryptographically signing
messages, which identify who is to be debited, who is to be credited,
and where the change (if any) is to be deposited. These messages are
then propagated through the Bitcoin network before being added to the
blockchain. Each node in the network keeps a list of valid outstanding
transactions called the transaction pool and passes it to neighbours in
the network. Before a transaction can be included in the blockchain, a
new \emph{block} containing that transaction must be mined. A block is a
list of transactions, together with metadata including the current time
and a reference to the most recent previous block in the blockchain
(hence the name blockchain). A block also includes a field called a
\emph{nonce}. The nonce contains no information, so it can be freely
varied in an attempt to find a valid block that satisfies the
requirements of the blockchain. To find such a block and publish it to
the bitcoin network is referred to as \emph{mining a block}.

\subsection{Bitcoin mining}\label{sec:bitcoin_mining}

Mining is a race between all the Bitcoin miners to find a valid block to
append to the blockchain. To have a good chance of winning this race
requires substantial computational resources, and the reward for doing
so is a bounty of (currently 12.5) new bitcoins. 

The requirements for a block to be valid are based on Bitcoin's hash
function. In general, a hash function $f$ maps a string $s$ (or
equivalently, an integer, if the binary representation of the string is
interpreted as an integer in base 2) to another string (integer) $f(s)$,
the \emph{hash} of $s$. The key property of the hash functions used in
Bitcoin is that given any $y$ in the codomain of $f$ it is
computationally infeasible to find any element of $f^{-1}(y)$.
Furthermore, the hash functions are designed such that any change in the
input string $s$ results in a totally different output $f(s)$.  That is,
$f(s_1)$ gives no information about $f(s_2)$ if $s_1 \neq s_2$, and so
to calculate hashes for a large number of input strings requires
calculating each hash separately. This property allows hashes to act as
short, tamper-proof summaries of large data files.  Calculating the
hashes of large numbers of arguments $s$ is the computational work in
Bitcoin's proof-of-work.


A Merkle tree is used to summarise and verify the integrity of the
transactions in a block.The Merkle tree is constructed by recursively
hashing pairs of transactions until there is only one hash, called the
Merkle root. The block header is formed from the Merkle root, metadata
(including the hash of the previous block) and the nonce.
For a block to be valid the hash of the block header must
be less than a certain value~$L$, referred to as the \emph{target}.
%
%
The hash of the candidate block is calculated. In the rare case that the
hash is less than the current target~$L$, the block is valid and is
appended to the miner's blockchain: the block is said to have been
\emph{mined}. In the far more likely case that the hash is not less than
the current target, the nonce is changed and the hash is recomputed.
This process is repeated (occasionally updating metadata or the list of
transactions) until some miner finds and publishes a block.  When a
block is mined it is broadcast to all the other miners who append the
new block to their blockchains and resume mining ``on top of'' the new
block. 

Bitcoin's hash function maps
$\mathbb{Z}^+ \cup \{0\}$ to $\{0,1,\ldots,2^{256}-1\}$.
The computational difficulty in mining arises because Bitcoin hashes are
effectively uniformly distributed in the interval $(0,2^{256}-1)$ while
the target~$L$ is much much less than $2^{256}$. Consequently, the
checking of a candidate block is a Bernoulli trial with a probability
of $L/2^{256}$ of being successful.  By design, this probability is at
most $2^{-32}$.  The trials are functionally independent, therefore the
number of hashes that need to be checked before a valid block is found
is a geometric random variable, with a very large mean, $2^{256}/L$
($\approx 6.2 \times 10^{21}$ at the time of writing). On the other
hand, computers around the world are currently performing hash function
calculations at a very large rate ($\approx 10^{19}$ hashes per second
at the time of writing).  The time taken to mine a block is therefore
very well-modelled by an exponential random variable.  Furthermore, over
any time interval where we can take the rate of trials to be constant,
the process whose events are given by the instants at which blocks are
mined should be well-approximated by a Poisson process, with a rate
given by the product of the number of hashes~$H$ calculated per second
globally (the global \emph{hash rate}), the value~$L$ of the target, and
a constant. 

The success of the blockchain method has resulted in Bitcoin becoming
increasingly popular and inspiring other electronic payment methods and
other systems, such as online education certification
\cite{dodd2017university}, that use blockchain mechanisms for
verification purposes. Indeed, although we will focus on the Bitcoin
blockchain, we point out that our analysis can apply to other systems
where similar blockchain policies are used. 

\subsection{Target (and difficulty) adjustment} \label{sec:difficultyAdjust}

The target~$L$ does not remain constant: it is adjusted every 2,016
blocks in an attempt to maintain an average mining rate of 6 blocks per
hour (2,016 blocks in 1,209,600 seconds) despite changes in the amount
of computational power being used for mining. Let $X_n$ be the time (in
seconds) at which the $n^\mathrm{th}$ block is mined and define $X_0 =
0$.  Then the target is adjusted at times $X_{2016i}$ for $i = 1,2,
\ldots $.  We call the time between adjustments of the target a
\emph{segment}. Let the target in the $i^{\mathrm{th}}$ segment
$(X_{2016(i-1)},X_{2016i})$ be $L_i$. Then
\begin{equation} \label{eq:adjustTarget}
L_{i+1} = L_{i} {(X_{2016i}-X_{2016(i-1)})} / {1209600}
\end{equation}
where $L_0 = 2^{224}$.
If $X_{2016i}-X_{2016(i-1)}$ is shorter than two weeks, the target~$L$
is reduced proportionally, and the average number of hashes checked to
mine a block is increased, slowing down mining. A common variable used
instead of the target is the \emph{difficulty}~$D$
defined as 
\begin{align} \label{eq:difficulty}
D_i &= \frac{2^{224}}{L_i} = D_{i-1} \frac{1209600}{(X_{2016i}-X_{2016(i-1)}}
\end{align}
so that $D_0 = 1$.
The difficulty~$D$ is $2^{32}$ times the average number of hashes
required to mine a block (see Equation~(\ref{eq:lambda_def})), and
denotes how much harder it is to mine a block than with the original
(maximum) target. This difficulty adjustment is part of what makes the
block arrival process for Bitcoin interesting: the adjustment is a
change in the block arrival rate based on the random block arrival
process itself, and the adjustment also occurs at a random time.



\subsection{Related work}

Nakamoto~\cite{nakamoto2008bitcoin} did not explicitly state that the
blocks arrive according to a homogeneous Poisson process. However, the
calculations in~\cite{nakamoto2008bitcoin} are based on the assumption
that the number of blocks that an attacker, mining at rate $\lambda_1$,
mines in the expected time $z/\lambda_2$ that it takes for the community
to mine $z$ blocks at rate $\lambda_2$, is Poisson with parameter
$\lambda_1 z/\lambda_2$, which is essentially equivalent to a Poisson
process assumption.  Rosenfeld~\cite{rosenfeld2014analysis} pointed out
that, under the assumption that blocks arrive in a Poisson process, the
number of blocks created by an attacker in the time that it takes the
community to create a fixed number of blocks is a random variable with a
negative binomial distribution; see
also~\cite[Section~1.4]{gobel2016bitcoin}


Rosenfeld also assumed that the block arrival process is a Poisson
process in~\cite{rosenfeld2011analysis} where he analysed reward
structures for pooled Bitcoin mining with a view to reducing the
variance in miner rewards while retaining fairness and preventing miners
from being incentivised to hop between pools. Lewenberg \textit{et
al.}~\cite{lewenberg2015bitcoin} also used a Poisson process model when
they addressed a similar problem, although unlike Rosenfeld they took a
game theory perspective and incorporated a constant, deterministic value
of the block propagation delay.


Eyal and Sirer~\cite{eyal2014majority} proposed an  attack strategy
called \emph{selfish-mine}, where certain dishonest miners keep
discovered blocks private in the hope that they can gain larger rewards
than honest miners. The majority of analysis on this topic has
explicitly or implicitly assumed a Poisson process model for the block
mining process, and typically used a Markov chain model for the number
of blocks mined by the selfish mining pool and those mined by the rest
of the miners. Sapirshtein \textit{et al.}~\cite{sapirshtein2016optimal}
found the optimal strategy for performing selfish mining. G\"obel
\textit{et al.}~\cite{gobel2016bitcoin} performed a stochastic and
simulation analysis of the selfish-mine strategy in the presence of
propagation delay. 

A Poisson process assumption for the block arrival process has been used
in other work. For example, Solat and
Potop-Butucaru~\cite{solat2016zeroblock} assumed a (homogeneous) Poisson
process for the block arrivals in their paper, where they developed a
timestamp-free method for combating attacks such as selfish-mine,
although this method does not crucially rely upon the Poisson
assumption. Decker and Wattenhofer assumed a (homogeneous) Poisson
process over periods where the difficulty is constant in their analysis
of the propagation of transactions and blocks through the Bitcoin
network~\cite{decker2013information}. Miller and
LaViola\cite{miller2014anonymous} also assumed a Poisson process when
modelling the Bitcoin system as a means of reaching fault-tolerant
consensus. 




\subsection{The goal and structure of the paper}

Despite the analyses mentioned in Section 1.3 above, there
has been little research to examine whether the block arrival process in
the Bitcoin blockchain actually does behave like a homogeneous Poisson
process.

The goal of this paper is to examine this assumption. It will turn out
that, at certain timescales, the block arrival process is not Poisson,
which will motivate us to propose refined point process models for the
block arrival process. We present a suite of point process models for
the block generation process, suitable for viewing the process over a
range of different timescales. Over long timescales, this comes down to
fitting time-varying models to the hash rate. Over shorter timescales we
need to take into account the dependencies introduced by the difficulty
update mechanism and the fact that the hash rate is varying within each
segment while the difficulty remains constant.

Our work here differs from previous modelling of the Bitcoin blockchain
because the difficulty adjustment is taken into account, and we also
examine the consequences of the interaction between the block
propagation delay and the hash rate upon the blockchain process. We also
infer miner-to-miner propagation delay from the available block
timestamp data.

The rest of the paper is structured as follows. In
Section~\ref{sec:data} we discuss the data available on the Bitcoin
blockchain and their limitations.  In Section~\ref{sec:hash rate} we
explain the drivers of the global hash rate, and how to estimate the
hash rate from the available data. In Section~\ref{sec:models} we
discuss several models for the block arrival process, including how to
approximate the hash rate in order to allow more tractable analysis of
the desired model.  We present results on the behaviour of the Bitcoin
system, and give a deterministic approximation.
Section~\ref{sec:summary} presents a summary of our various block
arrival models. Section~\ref{sec:simulation} presents the results of a
simulation of the Bitcoin blockchain. We summarise our results in
Section~\ref{sec:conc}.

\section{The block arrival process: the data}\label{sec:data}

Blocks arrive, that is, they are mined, published to the bitcoin network
and are added to the blockchain according to a random process. Examining
the block arrival process can serve as a means of diagnosing the health
of the blockchain, as well as detecting the actions of malicious
entities. 

There are several ways in which we can define when a block ``arrives''.
The two that are easiest to measure are: 
\begin{enumerate}
\item the time when the block arrives at our measurement node (or
nodes), and 
\item the timestamp encoded in the block header.
\end{enumerate}  
There are advantages and disadvantages associated with each of these
methods. We have forthcoming work on estimating the distribution of
errors in the arrival times from the combination of both sources of
data.

\subsection{Block arrival times}

The time when a block arrives at any given node in the Bitcoin
peer-to-peer network depends upon when the block was mined and how the
block propagates through the peer-to-peer network.  While propagation
adds a random delay component to the time when the block was mined, it
also represents when a block can truly be said to have arrived at the
node.  Once a block has propagated through the network (and to our
measurement nodes), then the information contained within the block is
visible to to the miners and the users of Bitcoin, and it is much less
likely that any miner will mine on top of a previous block and cause a
conflict. However, a substantial problem with using block arrival times
at measurement nodes is that it is impossible to get historical
information, or information from periods when the measurement nodes were
not functioning, or not connected to the network.

\subsection{Block timestamps} \label{sec:timestamps}

\begin{figure}[t!]

\centering

\begin{subfigure}[b]{0.48\textwidth} \centering
 \includegraphics[width=1.0\textwidth]{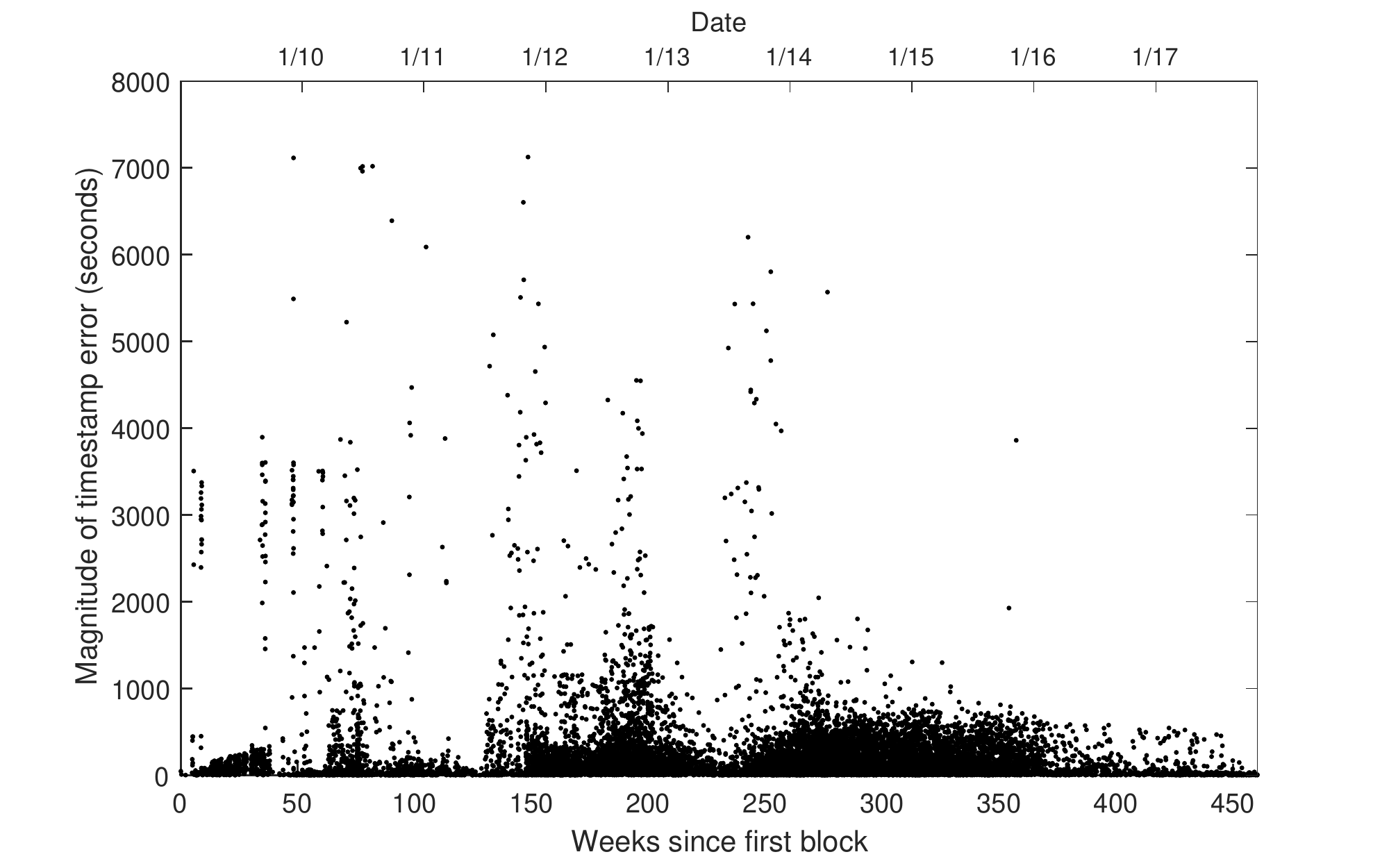}
\caption{Scatter plot of the sizes of the timestamp errors \emph{vs.}
when they occurred. Out-of-order timestamps are prevalent throughout the
history of Bitcoin.}
\label{fig_a1}
\end{subfigure}
~
\begin{subfigure}[b]{0.48\textwidth} \centering
 \includegraphics[width=1.0\textwidth]{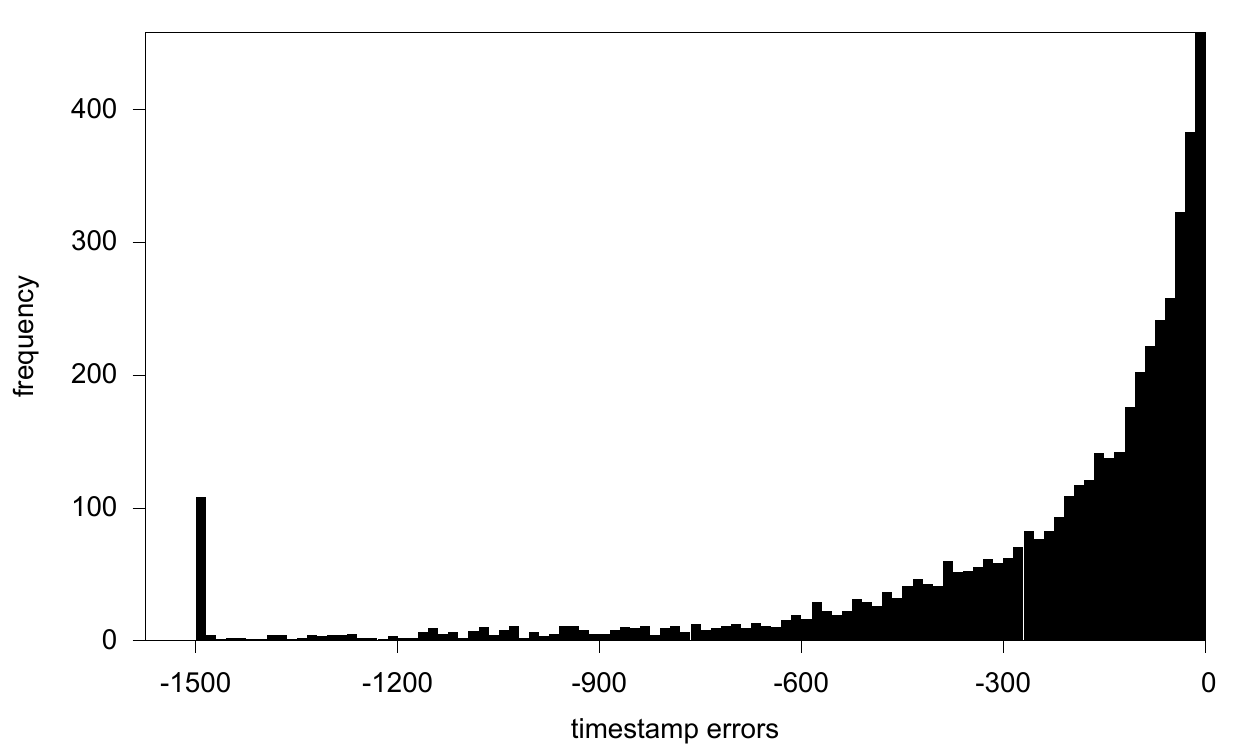}
\caption{Histogram of the sizes of the timestamp errors. The local
peak at -1500 includes all pairs of adjacent blocks $i,i+1$ whose
difference in timestamps $X_{i+1}-X_i < -1500$.}
\label{fig_a2}
\end{subfigure}
\caption{Known timestamp errors when timestamps for two consecutive
blocks in the blockchain are out of order.}
\label{fig_a}

\end{figure}
Using the timestamp in the block header as the authoritative time has
its own disadvantages.  This time is determined from the miner's clock
when the mining rig creates the block template.  If the miner's clock is
incorrect then this will introduce an error.  Timestamp errors are
limited in magnitude: blocks with timestamps prior to the median
timestamp of the previous 11 blocks, or more than 2 hours into the
future are rejected as being invalid and will not be propagated by the
network.

Nevertheless, block timestamp errors can be severe. For example, of
the~500,000 blocks mined up to November 2017, 13,618 of them have a
timestamp prior to that of the previous block and some 1,000 of them
have timestamps more than 10 minutes prior to the previous block
timestamp. If one considers that a miner has access to the previous
block (and hence the timestamp of the previous block) before mining the
next block, these errors seem particularly
anomalous\footnote{Out-of-order timestamps are often caused by a miner
using a timestamp from the future, and then miners for later blocks
using a correct timestamp that falls before the incorrect (future)
one.}. This suggests that some miners intentionally use inaccurate
timestamps. One potential reason for this is mining software that varies
the timestamp to use it as an additional nonce.\footnote{ Miners can
change both the nonce in the header and another nonce in the coinbase
transaction. These two nonces combined allow miners to cycle through
$2^{96}$ hashes before modifying the timestamp.}


Figure~\ref{fig_a1} shows a scatter plot of the magnitude of the
negative inter-arrival times versus the timestamp at which they
occurred. The horizontal axis covers the entire history of Bitcoin.
These errors are not localised to specific times in history. Incorrect
timestamps remain frequent even recently in the blockchain.

Figure~\ref{fig_a2} shows a truncated histogram of the sizes of the
negative inter-arrival times. Some inter-arrival times are more negative
than $-7,000$ seconds but most are between $-1,500$ and $0$ seconds.

We will later argue that while data for individual arrivals are
unreliable, our estimation of the hash rate is insensitive to errors in
the individual arrival times (especially if those errors are
independent) since it only relies upon the average rate of arrival of
large numbers of blocks.

\subsection{Cleaning the data}\label{sec:cleaning}

Blockchain timestamps are recorded in whole numbers of seconds since the
1st of January 1970, with the first block being mined on the 3rd of
January 2009.  The first step we took was to disregard all data prior to
the 30th of December 2009.  Owing to Bitcoin being newly adopted over
2009, the block arrival process in this period had a large number of
anomalies, including one day in which no blocks were mined.

\begin{figure}[t!]
\centering
 \includegraphics[width=0.5\textwidth]{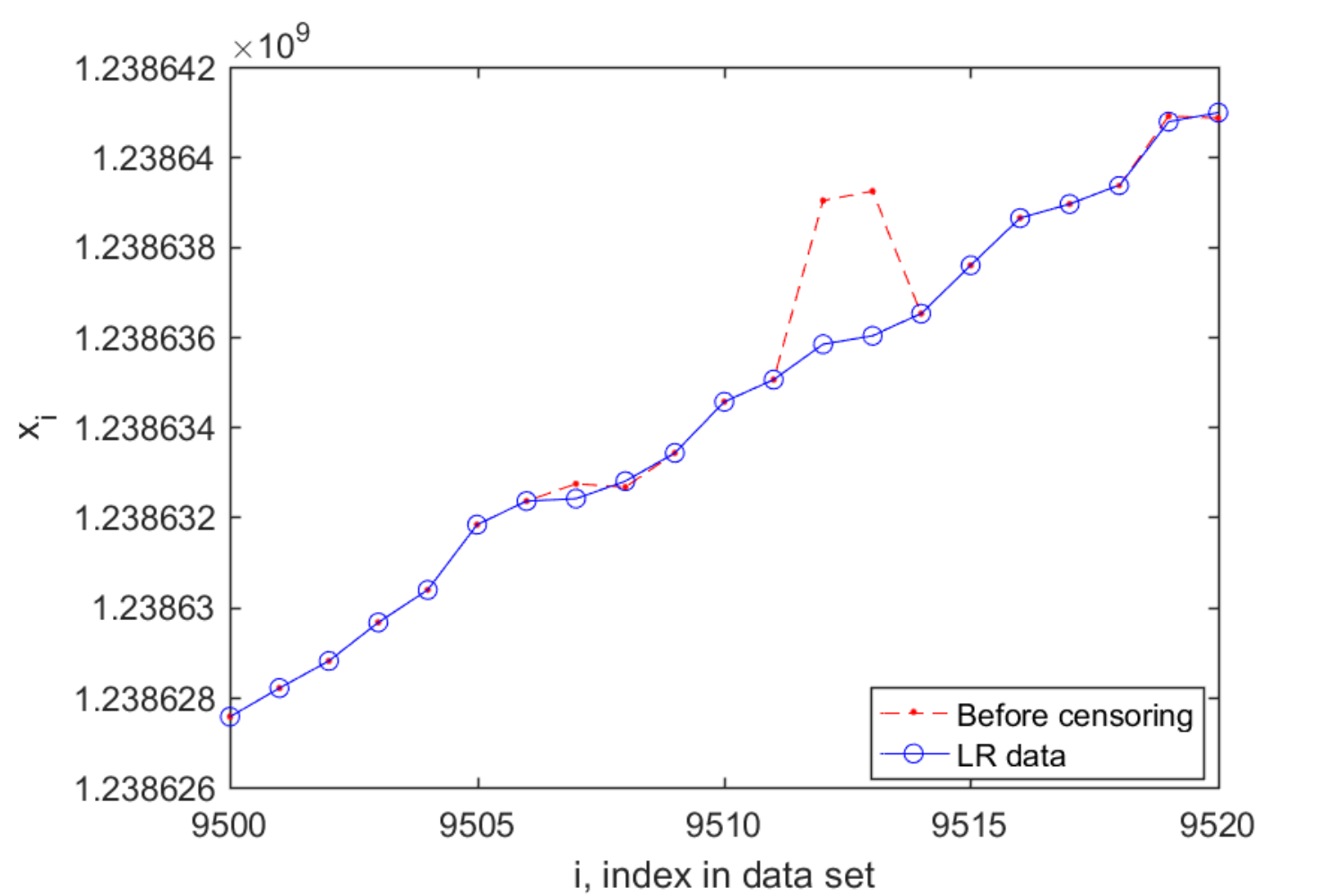}

\caption{An example of the effect of the resampling
scheme~(\ref{list:inc_sub}) from Section~\ref{sec:cleaning}. The index
$i$ is the number of blocks since 1 January 2009.
\label{fig:LRcase_study}}

\end{figure}

The next step was to deal with obvious errors. As discussed in
Section~\ref{sec:timestamps}, it is possible to have errors in the
timestamps. We know the order that the blocks \emph{must} have been
mined in, because each block header contains the hash of the previous
block header. Let the timestamp of the $i^\mathrm{th}$ block be $X_i$.
Due to errors, sometimes the timestamps are out of order: $X_i>X_j$ for
some $i<j$. We call $T_i := X_{i}-X_{i-1}$ the $i^{\mathrm{th}}$
\emph{inter-arrival} time, and if $T_i$ is negative then there must have
been an error. Some of these errors are both substantial and obvious:
for instance if $X_i-X_{i-1} = 7162$ and $X_{i+1}-X_i = -6603$ while
other nearby inter-arrival times are unremarkable, then it would
be sensible to assume that $X_i$ is an error. However, there are other
cases that are more complicated to deal with. There are several
approaches that we considered. 
\begin{enumerate}

\item \textbf{Ignore}: Ignore the fact that the data are out of order.
If we then look at the block inter-arrival times, some of these will be
negative. Also, some will be very large. If we are studying the variance
of the inter-arrival times then this will be inflated relative to any of
the methods discussed below.  Furthermore, the empirical distribution
function for inter-arrival times will not make physical sense, and will
certainly not be exponentially distributed.

\item \textbf{Reorder}: Sort the timestamps into ascending order. This
results in non-negative inter-arrival times, but we are essentially
removing random points and inserting them elsewhere in the process. This
has a far from obvious effect, and the effect is dependent on the
distribution of these already unreliable timestamps.

\item \textbf{Resample}: Remove those timestamps which we deem
unreliable (by some criterion), then resample them uniformly in the
interval between the adjacent timestamps that we consider reliable. For
instance, say $X_{i+1},X_{i+2},X_{i+3}$ are deemed unreliable but $X_i$
and $X_{i+4}$ are not. Then we would independently uniformly sample
three timestamps in the interval $(X_i,X_{i+4})$, sort those new
timestamps, and replace the previous values of $X_{i+1},X_{i+2},X_{i+3}$
by the three new timestamps.  This is essentially replacing the
unreliable timestamps with timestamps sampled from a Poisson process on
$(X_i,X_{i+4})$, conditional on there being three timestamps in this
interval.  There are a few ways we can determine which timestamps are
unreliable:

\begin{enumerate}

\item Try to guess which timestamps are unreliable based on the
surrounding timestamps. It is not clear how to do this, especially if we
wish to do it empirically.

\item Declare any timestamp that is adjacent to a negative inter-arrival
time to be unreliable. That is, if $X_{i+1}-X_{i}<0$, both $X_{i}$ and
$X_{i+1}$ are unreliable.  \label{list:neg_adj}

\item Sort the timestamps and consider any timestamp that changes
position when sorted to be unreliable. This addresses the case of a
single wrong timestamp the same way as in~(\ref{list:neg_adj}) and
provides a consistent approach for more complicated
scenarios.\label{list:sort}

\item Find the longest increasing subsequence of the data (if $X_1,X_2,
\ldots$ is treated as a sequence). Such a subsequence might not be
unique. If there is more than one longest increasing subsequence, then
take the intersection of all longest increasing subsequences. Any
timestamps not in this intersection are considered unreliable.
\label{list:inc_sub}

\end{enumerate}
Once the unreliable timestamps are determined, resample them between the
adjacent reliable timestamps.
\end{enumerate}
To see why~(\ref{list:inc_sub}) is reasonable, consider the case of
having two adjacent timestamps a long way out of order, see for instance
Figure~\ref{fig:LRcase_study}: $X_1<X_2<X_5<X_6<X_7<X_8<X_3<X_4$. This
is a scenario that actually occurs in the blockchain. Here it is clear
that $X_3$ and $X_4$ should be the timestamps under suspicion,
but~(\ref{list:neg_adj}) will replace $X_4$ and $X_5$, and
(\ref{list:sort}) will replace all of $X_3,\ldots,X_8$, whereas
(\ref{list:inc_sub}) replaces exactly $X_3$ and $X_4$. The algorithm for
computing this is available at~\cite{bowden17lis}.

These are the steps we followed to clean the block arrival time data:
\begin{itemize}

\item Disregarded all data prior to the 30th of December 2009.

\item Performed~(\ref{list:inc_sub}) for the rest of the data. 

\end{itemize}
We refer to the data after these two steps as LR (later, resampled).  Of
the 464,372 blocks after the 30th of December 2009, 441,225 or 95\% of
them are in every longest increasing subsequence. The remaining 5\% were
resampled. While the longest increasing subsequences span the whole
length of the blockchain, the errors that are corrected by taking the
intersection of all longest increasing subsequences are all of a local
nature. Most of those timestamps that were resampled were in consecutive
groups of 1, 2 or 3.  Two consecutive timestamps being resampled was the
most common: this occurs when one timestamp is incorrect but with a
small enough error to be only one place out of order. The other
timestamp that it is transposed with will also be considered unreliable
and resampled. An isolated timestamp being resampled is caused by that
timestamp having a particularly large error.  If a timestamp has a large
error it can have two or more timestamps between its observed value and
its correct value.  In this case, if none of the other nearby timestamps
are out of order the incorrect timestamp will be the only one resampled. 

We provide this data (along with annotation as to which points have been
resampled) at \cite{bowden17lis}. We suggest that this sequence could
serve as a reference version of cleaned blockchain timestamp data for
other research on this topic.

\section{Estimating the global hash rate}\label{sec:hash rate}

The global hash rate $H(t)$ is a key factor driving the block arrival
process. It is the total processing power (hashes calculated per second)
that all the Bitcoin miners in the world are dedicating to mining at
time~$t$. In the long term, $H(t)$ is largely increasing exponentially
in a Moore's Law fashion.  The main drivers of change in the hash rate
are:
\begin{enumerate}

\item Miners switching new machines on. This typically involves ordering
mining machines from one of a small number of ASIC manufacturers, often
while those ASICs are still in development, and waiting for the
manufacturer (who is prone to delays) to manufacture and ship the ASICs
in batches. Then the ASICs arrive at the miners and they install them
and switch them on and start mining. Each generation of new mining
machines is faster than the last, although this increase is not steady,
even in an exponential sense. It is widely believed that the rate of
increase in the hash rate is petering out now.

\item Mining machines becoming uneconomical because of increasing
electricity prices, increasing global hash rates, reduced bitcoin
prices, or halving\footnote{The Bitcoin mining reward is halved every
210,000 blocks mined. The coin reward will decrease from 12.5 to 6.25
bitcoins around June 2020.} of the mining reward, and being switched
off.

\item Mining machines becoming more economical due to both to
technological advancements and, at times, the increasing value of
bitcoins, and when electricity prices are reduced.

\end{enumerate}


\subsection{Empirical estimates for the global hash rate}

\begin{figure}[t!]

\centering

 \includegraphics[width=0.5\textwidth]{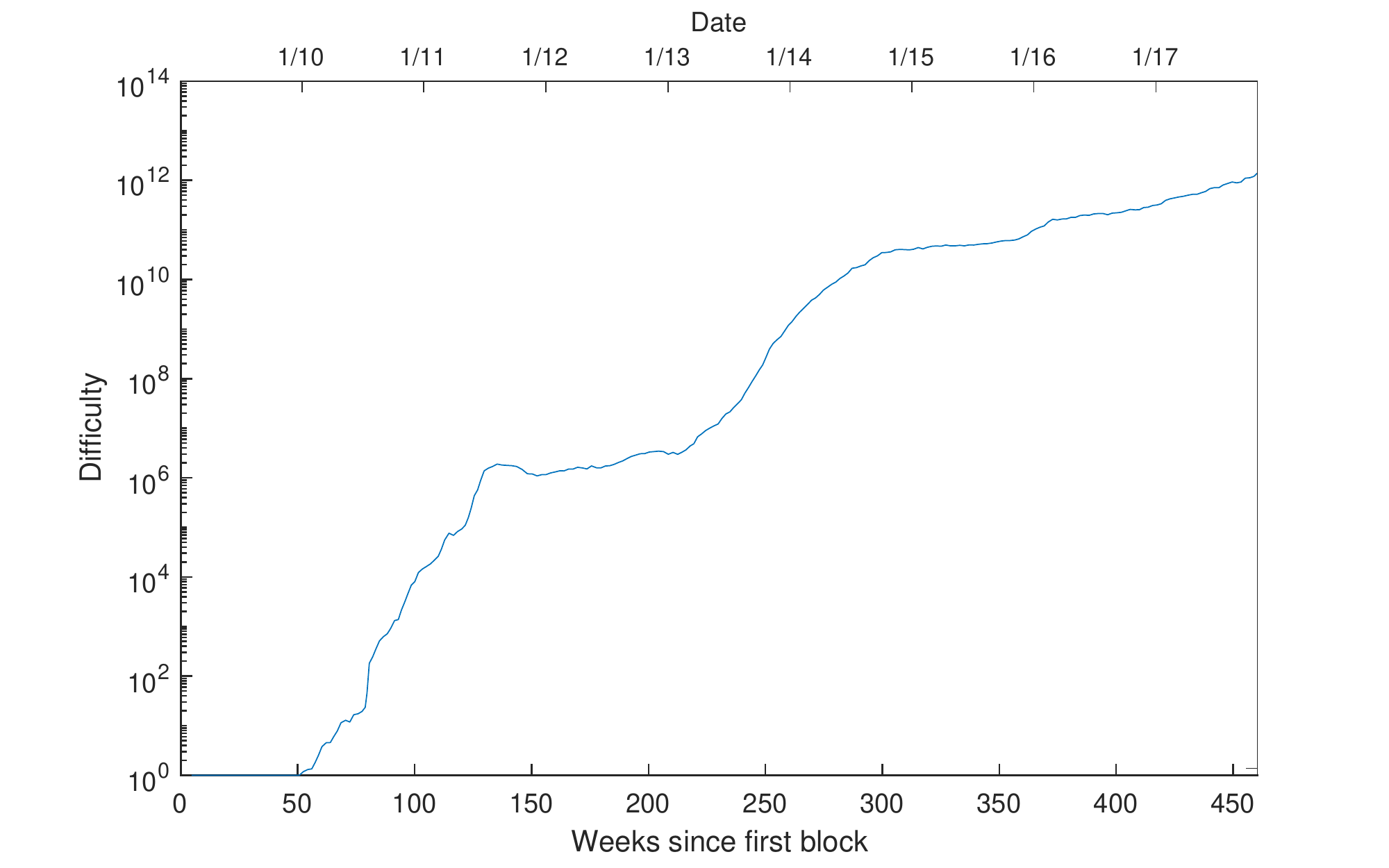}

\caption{Plot of the difficulty over the life of the blockchain. This
serves as an estimate of a constant multiple of the global hash rate.}

\label{fig:difficulties}

\end{figure}

The global hash rate $H(t)$ cannot be measured directly.  However, the
difficulty $D_i$ is recorded in the header of every block.
Equations~(\ref{eq:adjustTarget}) and~(\ref{eq:difficulty}) can be
combined to give
\begin{align}\label{eq:difficulty2}
D_{i+1} = \frac{1209600 D_i}{X_{2016i}-X_{2016(i-1)}}.
\end{align}
The ratio $2016/(X_{2016i}-X_{2016(i-1)})$ is the average rate of block
discovery in the period $[X_{2016i},X_{2016(i-1)})$, and $2^{32}D_i$ is
the expected number of hashes that must be checked to find a valid
solution when the difficulty is $D_i$. A reasonable estimate
$\widehat{H}_i$ for the average of the global hash rate $H(t)$ over the
segment $(X_{2016i},X_{2016(i-1)})$ is therefore given by the product of
those two quantities, namely
\begin{align}
\widehat{H}_i =
\frac{2016 \times 2^{32}D_i}{X_{2016i}-X_{2016(i-1)}} =
\frac{2^{32}}{600} D_{i+1}.
\end{align}
Thus we can get a sense of how the hash rate has changed over time by
examining the graph of the difficulty over time presented in Figure
\ref{fig:difficulties}.


We can refine this approach to obtain a sliding window estimate
$H^W_{k,i}$ of the hash rate, with the window being fixed at~$k$ blocks
long, centred around the $i^{\mathrm{th}}$ block. 
\begin{align}\label{eq_2}
H^W_{k,i} =
\frac{2^{32}}{X_{\ceil{i+k/2}}-X_{\ceil{i-k/2}}} \sum_{j =
\ceil{i+k/2}}^{\ceil{i-k/2}} D_{\ceil{j/2016}}.
\end{align}
If there was no difficulty change between $X_{\ceil{i+k/2}}$ and
$X_{\ceil{i-k/2}}$, then
\begin{align}\label{eq_3}
H^W_{k,i} =\frac{2^{32} D_{\ceil{i/2016}}k}{X_{\ceil{i+k/2}}-X_{\ceil{i-k/2}}}.
\end{align}
We can consider $H^W_{k,i}$ to be an estimate of $H(t)$ at times $t_i =
(X_{\ceil{i+k/2}}+X_{\ceil{i-k/2}})/2$. To get a continuous estimator we
can linearly interpolate between these points to get $H^W_k(t)$ for
$t\geq X_{k/2}$.

The estimation of $H(t)$ from Equation~(\ref{eq_2}) is not without its
problems due to the stochastic nature of the block arrival process, and
inaccuracies in the values of the block timestamps $X_i$.
%
%
If the window $k$ is longer, there is less stochastic variation in the
estimate, but the time resolution of the resulting estimate of $H(t)$ is
lower so that it is more difficult to pick up short-term fluctuations in
$H(t)$, yielding the standard bias/variance trade-off when performing
smoothing. Note that regardless of what window we use, the time
resolution of the estimate will be limited by the frequency of block
arrivals, and the noise caused by the stochastic variation inherent in
the process. 

Another method for estimating $H(t)$ via smoothing the block arrival
data is by using a method essentially equivalent to weighted kernel
density estimation (KDE), but without the requirement that the output is
a density. For each block $i=1,2,\ldots$ we centre a kernel $K(x)$ at
the measured time  $X_i$ of the arrival of the block, and weight it with
$w_i = 2^{32}D_{\ceil{i/2016}}$ where $w_i$ is the number of hashes that
must be checked on average to mine a block with the same difficulty as
block $i$. The kernel estimator is given by 
\begin{align}
H^K_h(t)
&= \frac{1}{h}\sum_{i} w_i K({(t-X_i)}/{h}) \nonumber \\
&= \frac{2^{32}}{h}\sum_{i} D_{\ceil{i/2016}} K({(t-X_i)}/{h}).
\end{align}
Some examples of appropriate kernel functions are the rectangular
function $K(x) = \frac{1}{2} 1_{|x|<1}$, the normal density function
$\frac{1}{\sqrt{2\pi}}e^{-x^2/2}$, and the Epanechnikov
kernel~\cite{epanechnikov1969non} $\frac{3}{4}(1-x^2) 1_{|x|<1}$. Here
$h$ is referred to as the bandwidth of the kernel estimator.  The
bandwidth~$h$ performs a role similar to that of the window size $k$
above, so the value of~$h$ is important. If~$h$ is too small then the
data has spurious variability from measurement error (and more
substantially, from the stochastic variability of block discoveries). If
$h$ is too large then the data will be oversmoothed and fail to capture
true short-term variability in $H(t)$.  However, we are primarily
interested in the long term behaviour of $H(t)$, so our conclusions
later in the paper are relatively insensitive to which estimate for
$H(t)$ we use, the kernel function $K(x)$ and the window size $k$ or the
bandwidth $h$.

\begin{figure}[t!]

\centering

 \includegraphics[width=0.5\textwidth]{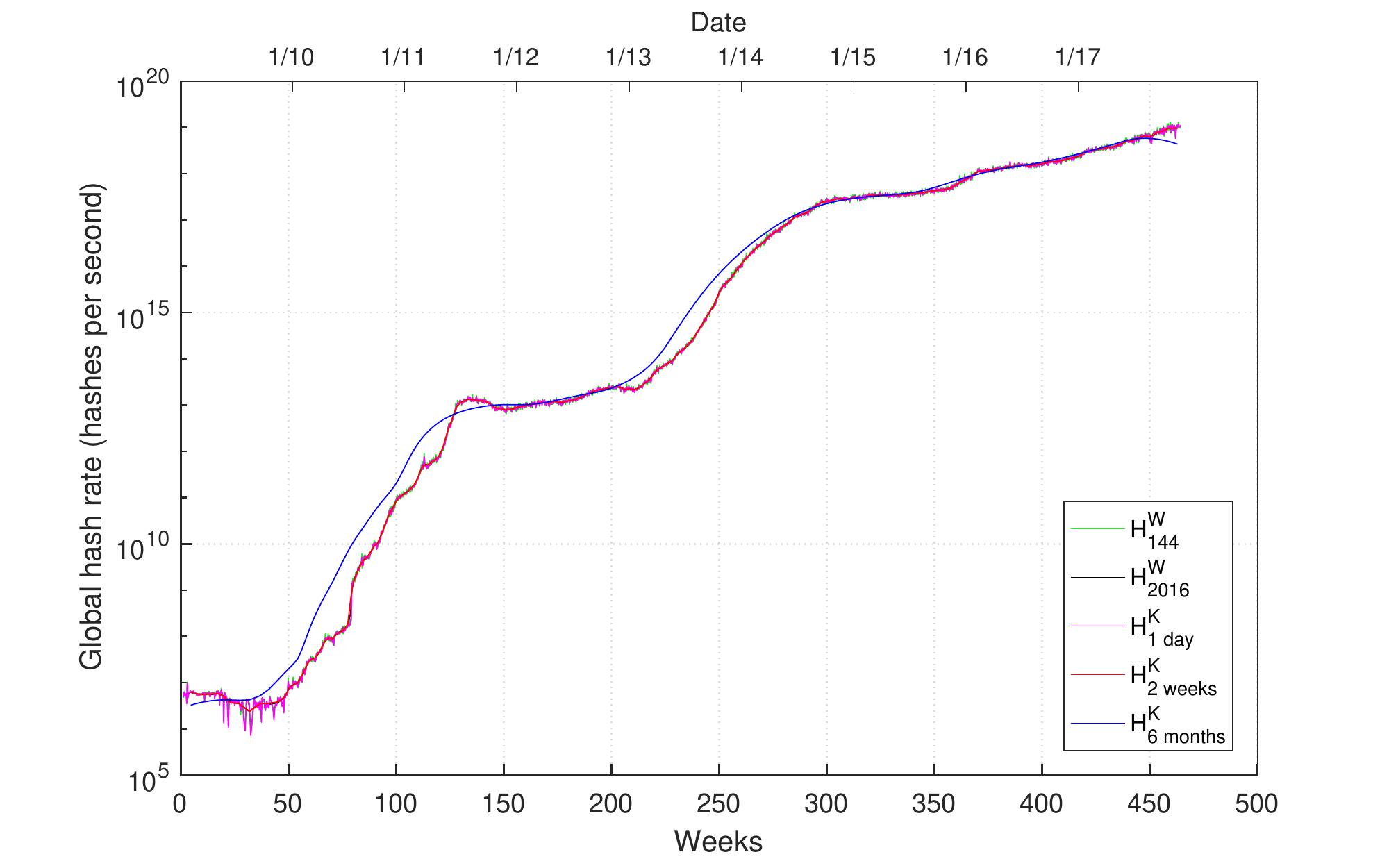}


\caption{Estimates of the hash rate $H^W_{2016}(t)$, $H^W_{144}(t)$,
$H^K_{\texttt{1 day}}(t)$, $H^K_{\texttt{2 weeks}}(t)$ and
$H^K_{6\texttt{ months}}(t)$ respectively from January 2009 to April
2017. The sliding window estimate with $k = 2016$ is not identical to
the kernel-based estimate with $h=\texttt{2 weeks}$, but it is almost
indistinguishable at this scale, so the two curves overlay and only one
of them is visible on the graph.}

\label{fig_1}

\end{figure}

Figure~\ref{fig_1} shows a comparison of $H^W_{144}(t)$,
$H^W_{2016}(t)$, $H^K_{\texttt{1 day}}(t)$, $H^K_{\texttt{2 weeks}}(t)$
and $ H^K_{6\texttt{ months}}(t)$ over the life of the Bitcoin
blockchain where~$K$ denotes the Epanechnikov kernel.  At this scale it
is difficult to distinguish between $H^W_{2016}(t)$ and $H^K_{\texttt{2
weeks}}(t)$. The more smoothed kernel-based estimate ($h = 6\texttt{
months}$) appears to lead the other estimates because the rapid increase
in the hash rate is dispersed by the width of the kernel into the time
before it has truly happened. This makes the smoother (higher bandwidth)
estimate misleading and we will use a lower bandwidth estimate for the
remainder of the paper. The behaviour of the blockchain in the first 50
weeks is erratic.

\subsection{Intervals of exponential growth in the global hash rate}
\label{sec:exponential_consequences}

\begin{figure*}[tbph]

\centering

\begin{subfigure}[b]{0.48\textwidth} \centering
 \includegraphics[width=\textwidth]{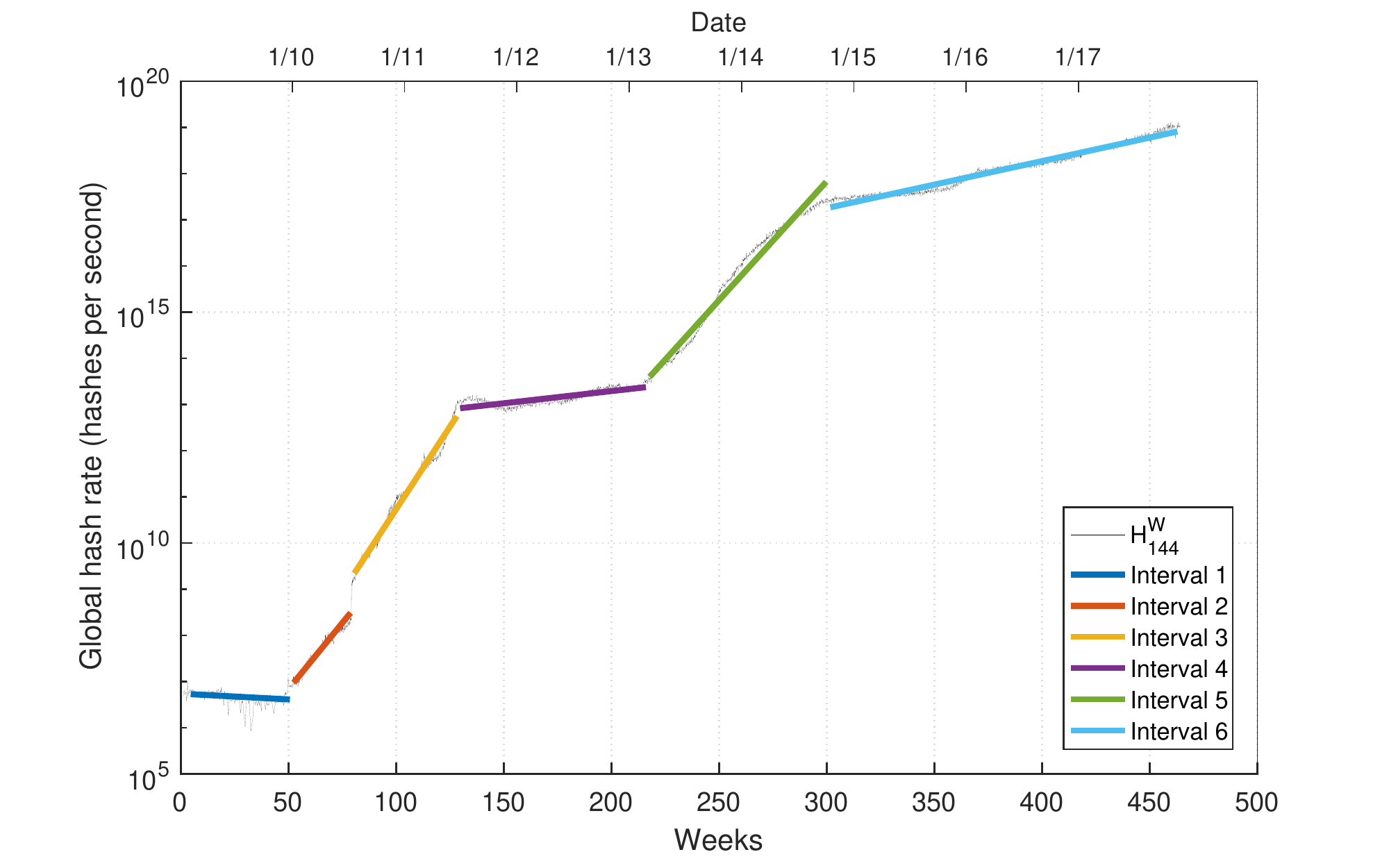}
\caption{All intervals}
\label{fig_2a}
\end{subfigure}
~
\begin{subfigure}[b]{0.48\textwidth} \centering
 \includegraphics[width=\textwidth]{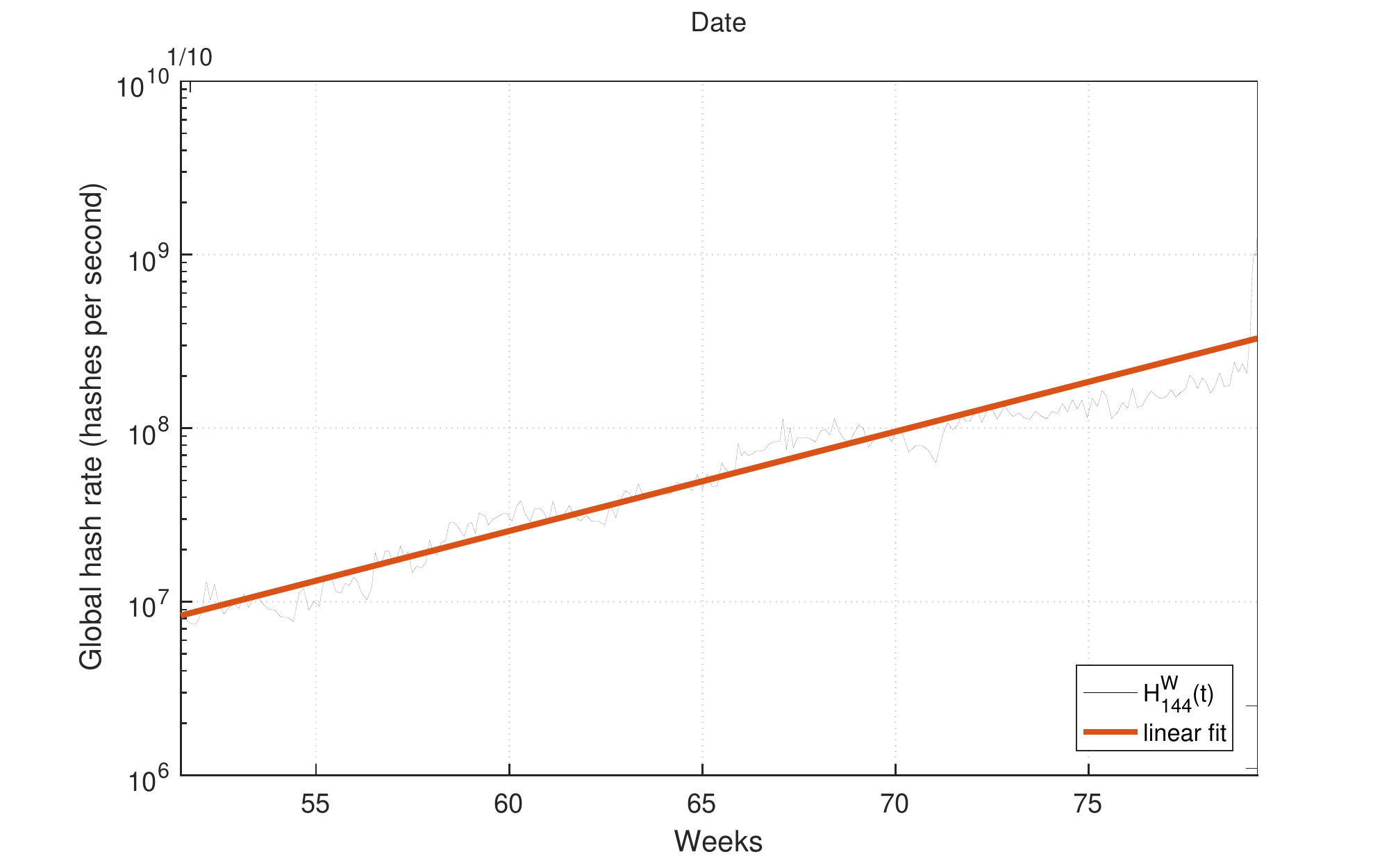}
\caption{Interval 2}
\label{fig_2a2}
\end{subfigure}

\begin{subfigure}[b]{0.48\textwidth} \centering
 \includegraphics[width=\textwidth]{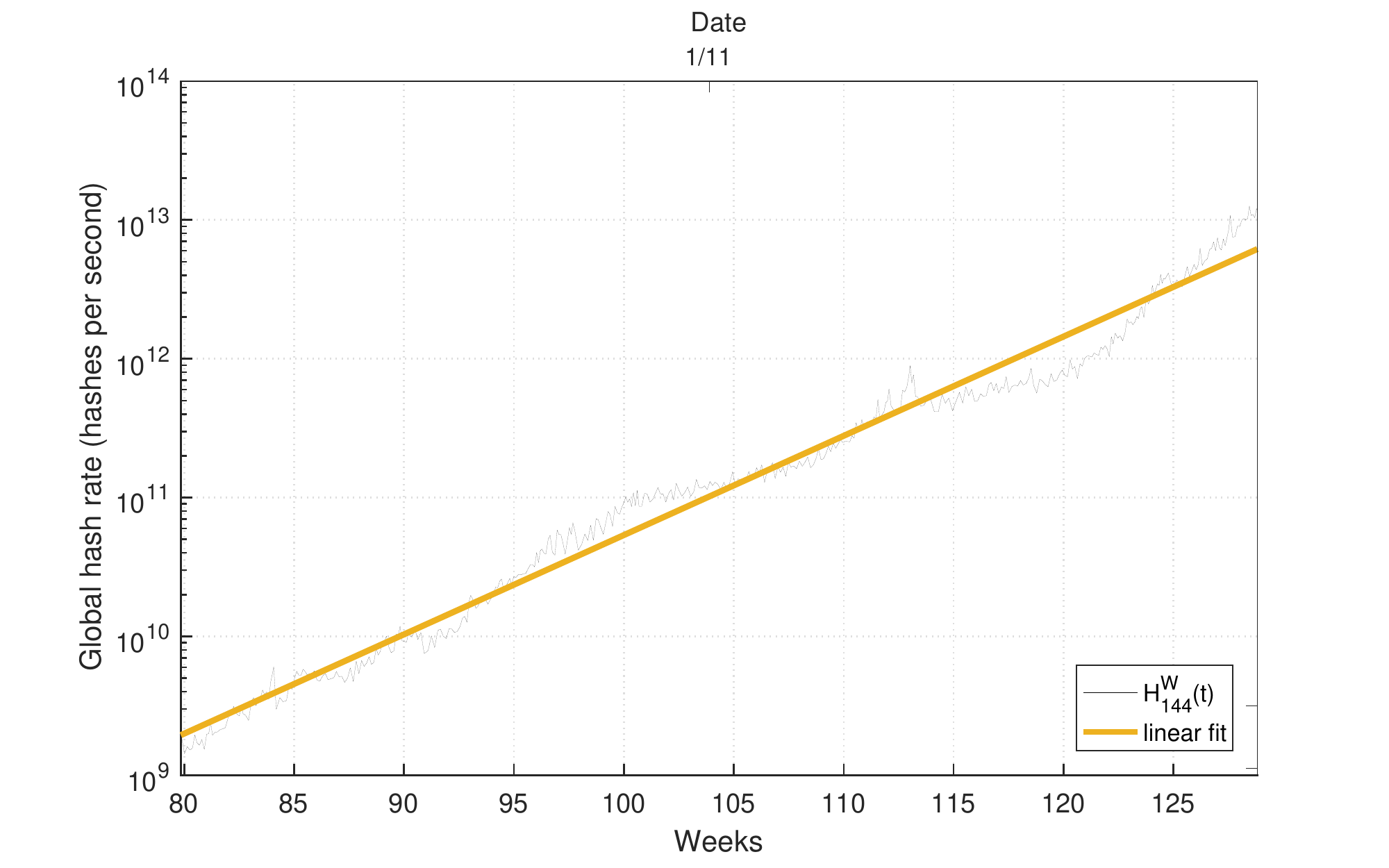}
\caption{Interval 3}
\label{fig_2b}
\end{subfigure}
~
\begin{subfigure}[b]{0.48\textwidth} \centering
 \includegraphics[width=\textwidth]{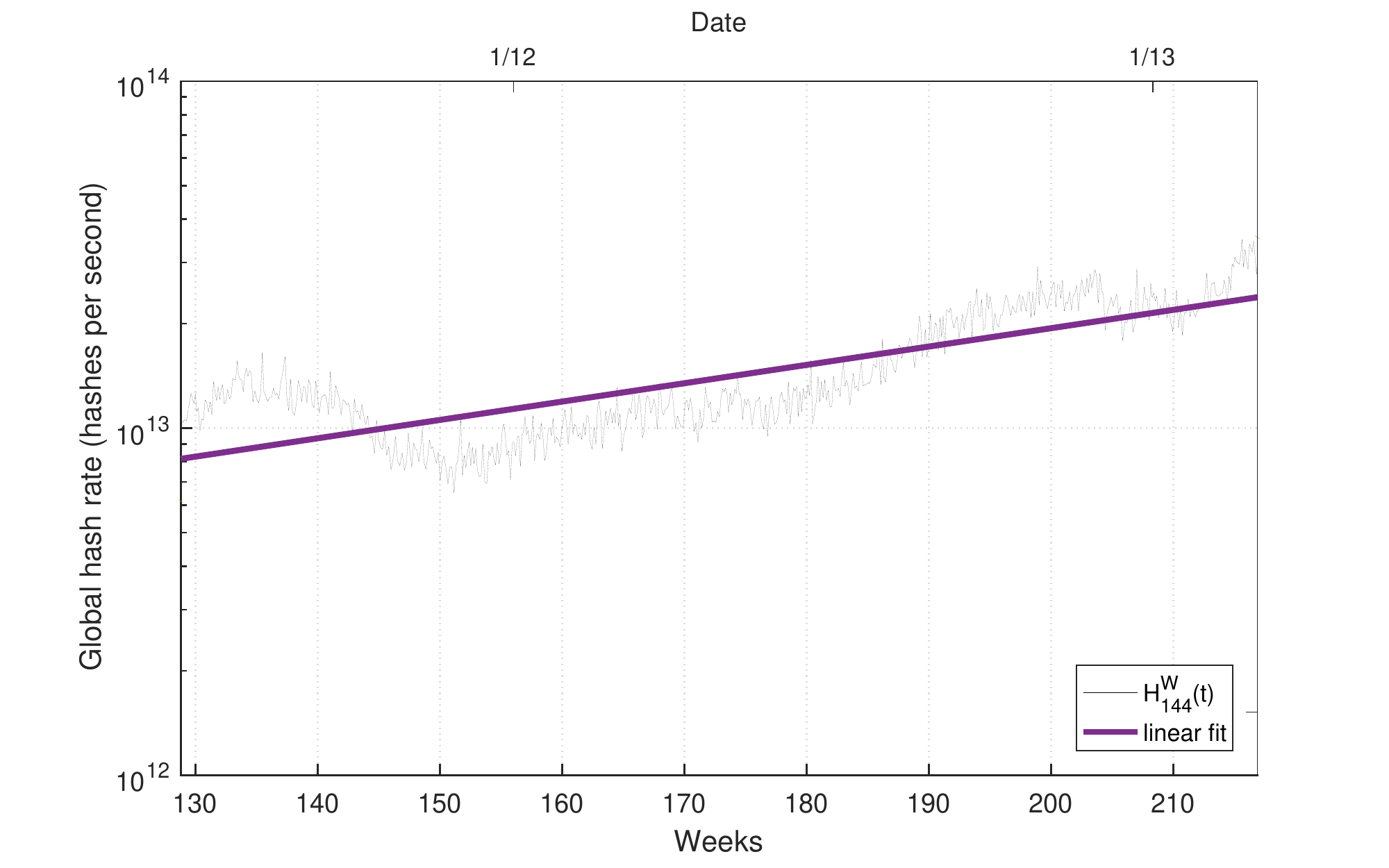}
\caption{Interval 4}
\label{fig_2c}
\end{subfigure}

\begin{subfigure}[b]{0.48\textwidth} \centering
 \includegraphics[width=\textwidth]{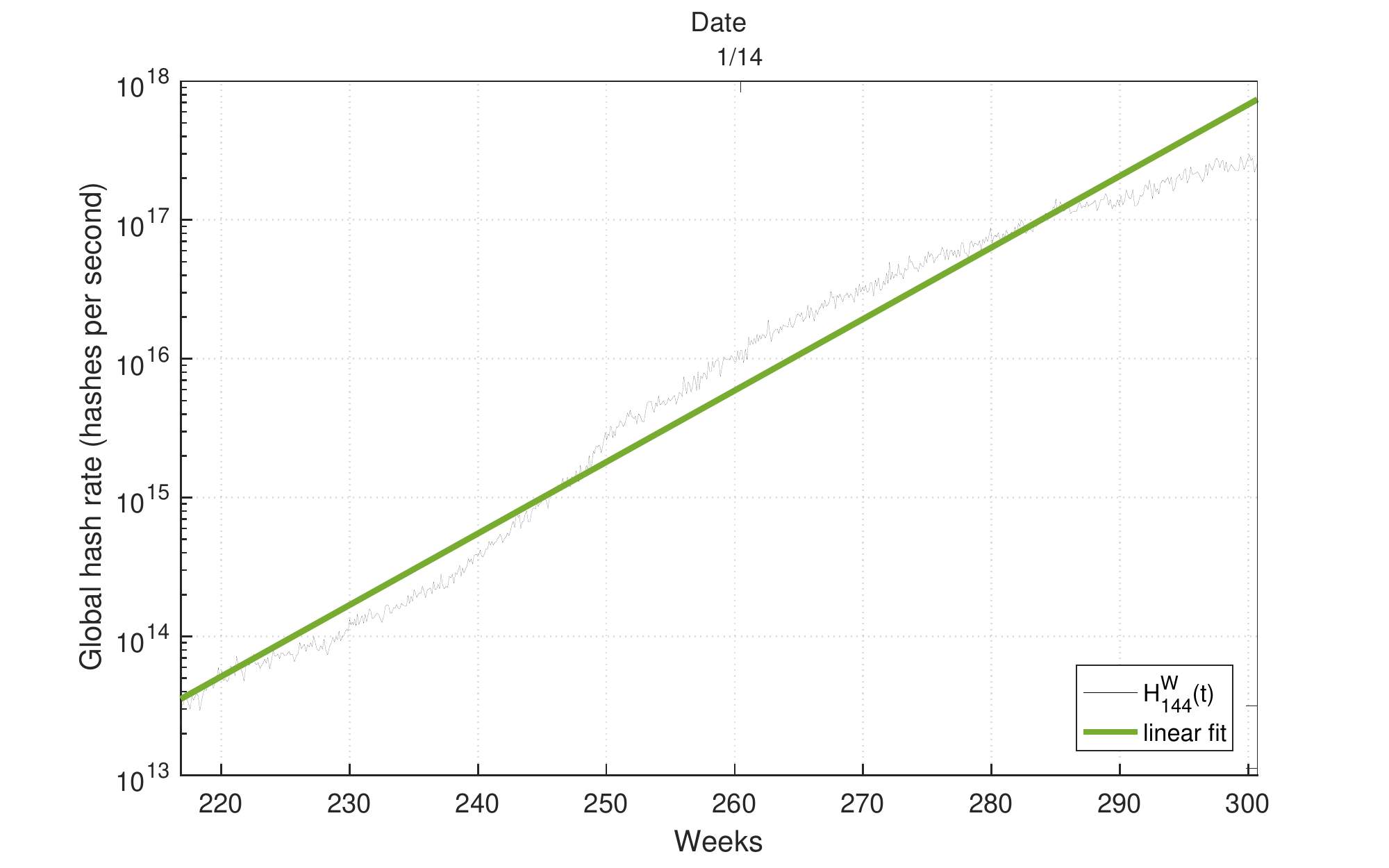}
\caption{Interval 5}
\label{fig_2d}
\end{subfigure}
~
\begin{subfigure}[b]{0.48\textwidth} \centering
 \includegraphics[width=\textwidth]{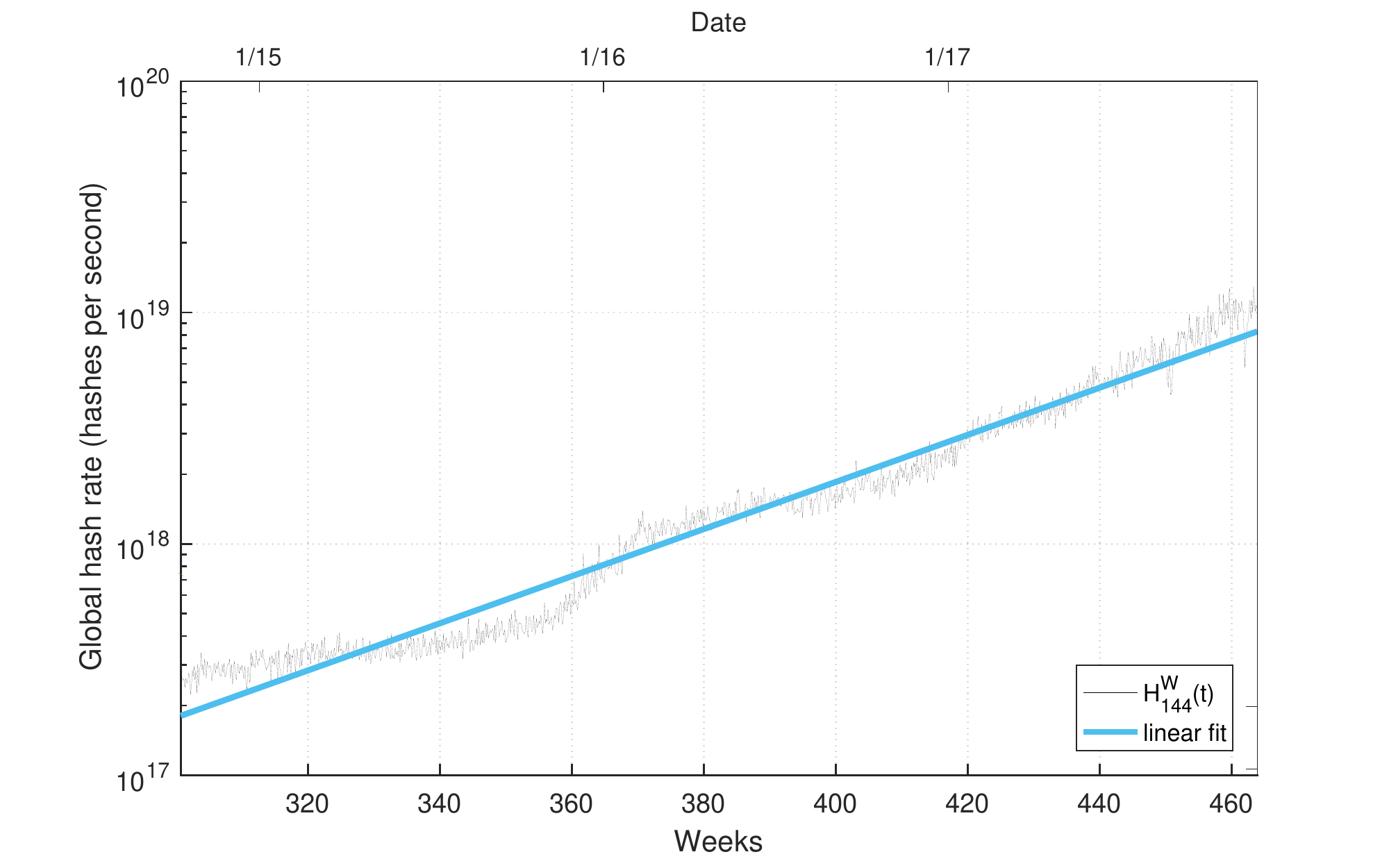}
\caption{Interval 6}
\label{fig_2e}
\end{subfigure}

\caption{Linear fits to $H^W_{144}(t)$ over a sequence of six intervals,
with the endpoints of the intervals given in Table~\ref{tab:piecewise}.
A closer view of the fitted hash rate $\widehat{H}(t) = e^{at+b}$ and a
more fine-grained estimate of the hash rate $H^W_{144}(t)$ over the
intervals is given in subfigures (b) through (f).}

\label{fig_2}

\end{figure*}

By inspecting the plot of an estimate of the hash rate from 2009 to 2017
(Figure~\ref{fig_1}) we see that the logarithm of the global hash rate
is approximately linear for long periods, that is, the hash rate is
increasing exponentially. We approximate the global hash rate $H(t)$ by
$e^{at+b}$ on intervals where it can be estimated that $a$ and $b$
remain constant.  In Figure~\ref{fig_2} we partition the
history of the blockchain into such periods, and for each of these
periods we fit a linear function to $\log(H^W_{144}(t))$ using standard
linear regression. The values for this fit are given in
Table~\ref{tab:piecewise}.


\begin{table}[b]

\centering

{\small
\begin{tabular}{|c|r|r|r|r|}  \hline
\multicolumn{1}{|c|}{interval} &
\multicolumn{1}{|c|}{start} &
\multicolumn{1}{|c|}{end} &
\multicolumn{1}{|c|}{$a$}  &
\multicolumn{1}{|c|}{$b$} \\ \hline
 1 & 3 Jan 2009 & 30 Dec 2009 & $-9.44\times 10^{-9}$ & 27.1\\
2 & 30 Dec 2009 & 13 Jul 2010 & $2.18\times 10^{-7}$ & -259\\
3 & 13 Jul 2010 & 24 Jun 2011 & $2.72\times 10^{-7}$ & -326\\
4 & 24 Jun 2011 &  1 Mar 2013 & $2.01\times 10^{-8}$ & 3.38\\
5 & 1 Mar 2013 &  9 Oct 2014 & $1.96\times 10^{-7}$ & -236\\
6 & 9 Oct 2014 &  24 Nov 2017 & $3.88\times 10^{-8}$ & -15.1\\\hline
\end{tabular}
}

\caption{The parameters of the piecewise exponential hash rate
$\widehat{H}(t) =e^{at+b}$ in each interval, with $t$ in seconds.  Each
interval starts and ends on a segment boundary.}

\label{tab:piecewise}

\end{table}

%
%
%

The assumption of exponential growth in the hash rate is natural for two
reasons. First, the increase in Bitcoin mining power is partially driven
by technological developments: the processors used in mining have become
increasingly cost- and energy-efficient over time in a manner
reminiscent of Moore's Law.
Second, we observe in the actual blockchain that the average time taken
for a segment is consistently under 2 weeks. If we want the average time
taken for a segment to be stationary with a mean not equal to 2 weeks
then $H(t)$ is required to grow exponentially. Intuitively, if the time
taken for each segment was roughly constant at a value of $T_{2016}$,
then the difficulty $D(t)$ would increase by a factor of $({\texttt{2
weeks}})/T_{2016}$ each segment, and since the difficulty is increasing
exponentially, the hash rate would also have to increase exponentially
to maintain constant segment time.
Approximating $\log(H(t))$ with a linear function also has the benefit
of tractability for modelling and analysis. 

We can calculate the limiting average for segment times by recognizing
that the average change in the difficulty of successive segments is
cancelled out by the growth in the hash rate over the duration of each
segment.  The faster the increase in the hash rate, the lower the
limiting average inter-arrival time.

\begin{figure}[t!]
\centering
 \includegraphics[width=0.5\textwidth]{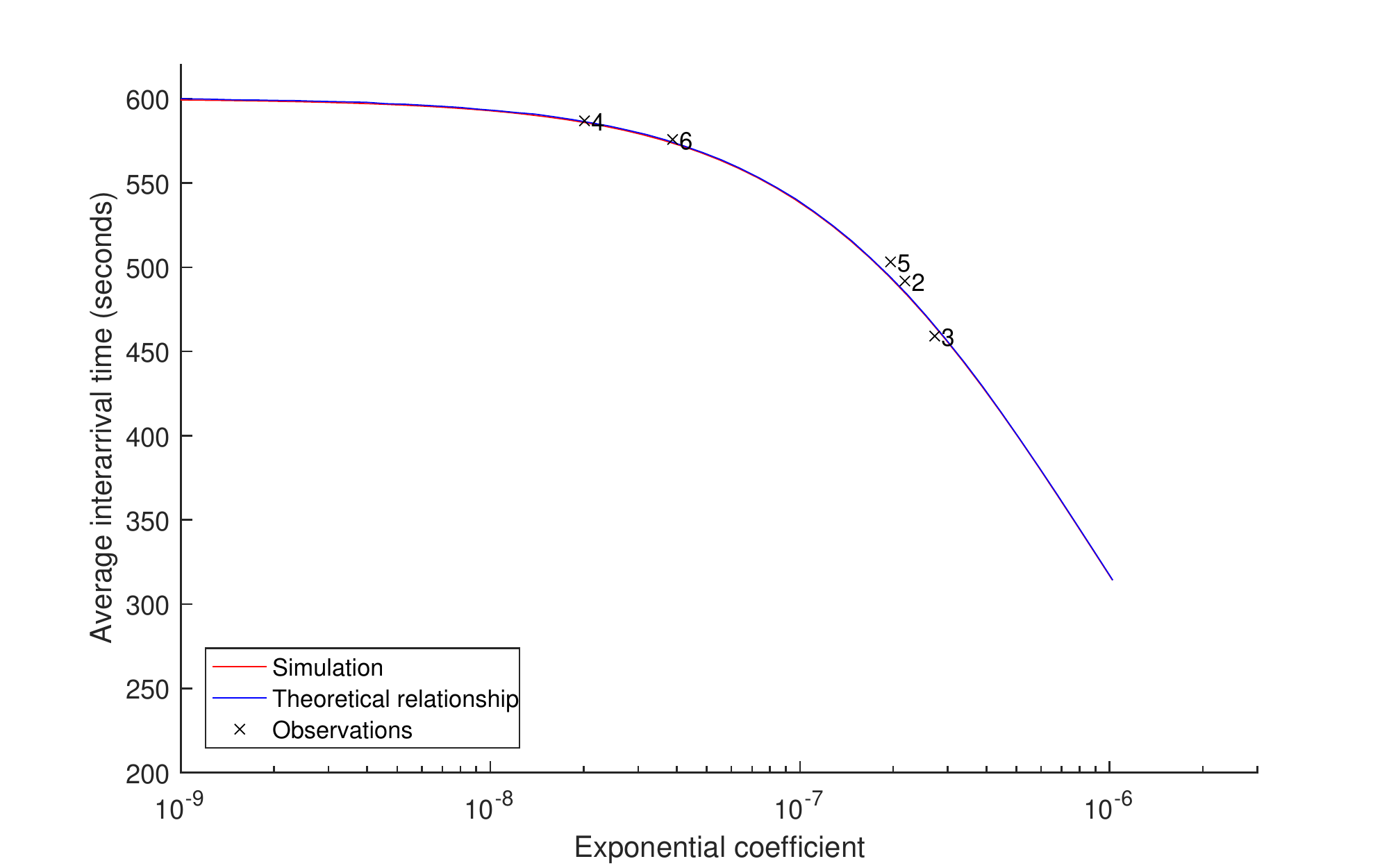}

\caption{The relationship between the inter-arrival time and $a$, the
slope of the logarithm of the hash rate over time, in theory,
simulations and observations.  The global hash rate at time $t$ is
modelled by $\widehat{H}(t) = e^{at+b}$. The system reaches a steady
state where the average time $T_{\texttt{ave}}$ for a block to arrive is
related to $a$ by Equation~(\ref{eq:time_vs_a}). The simulated values
are given by a simulation of 40,000 blocks with exponential hash rate,
random difficulty changes and no propagation delay. The observed values
of the average inter-arrival time are calculated from the observed
timestamps for each interval, and the values of $a$ are the fitted
values given in Table \ref{tab:piecewise}. The segment time is 2016
times the average block inter-arrival time for the relevant interval.}

\label{fig:time_vs_a1} \end{figure}

Let $\widehat{H}(t) = e^{at+b}$.  Then the multiplicative growth in hash
rate over a segment of duration $T_{2016}$ is $e^{aT_{2016}}$. The ratio
of the new difficulty to the previous difficulty, given by
Equation~(\ref{eq:difficulty2}), is $(\texttt{2 weeks})/T_{2016}$. In
seconds, this gives the relationship between the average time taken to
mine a segment and the slope of the logarithm of the hash rate (see
Equation~(\ref{eq:lambert}))
\begin{equation}\label{eq:seg_time_vs_hr_growth}
e^{aT_{2016}} = 1209600/T_{2016}
\end{equation} 
or equivalently,
\begin{equation}\label{eq:seg_time_vs_hr_growth2}
a = \frac{1}{T_{2016}}\log\left({1209600} / {T_{2016}}\right).
\end{equation} 
Note that the long-term average time to mine a block is $T_{2016}/2016$,
although the average inter-arrival time decreases over the course of
each segment as shown in Figure~\ref{fig_2016} below. This means that
the long-term average time to mine a block $T_{\texttt{ave}}$ when the
hash rate is $H(t) = e^{at+b}$ is given by solving the relationship
\begin{align}\label{eq:time_vs_a}
e^{2016aT_{\texttt{ave}}} = 600/T_{\texttt{ave}}.
\end{align}
We simulated the arrival of 40,000 blocks with random difficulty changes
(exponential hash rate function, no block propagation delay) using a
range of values of~$a$.  Figure~\ref{fig:time_vs_a1} shows that the long
term average block inter-arrival time agrees with
Equations~(\ref{eq:seg_time_vs_hr_growth}) through (\ref{eq:time_vs_a}).
Further, we also plot the average inter-arrival time of the observed
blockchain timestamp data for each of the intervals in
Table~\ref{tab:piecewise} versus the fitted value of $a$ for each
interval. Figure~\ref{fig:time_vs_a1} shows that the observed data also
fits the predicted relationship in Equation
(\ref{eq:seg_time_vs_hr_growth}).
We provide a theoretical basis for
Equation~(\ref{eq:seg_time_vs_hr_growth}) in
Section~\ref{ss:determ_approx}.


\section{Is the block arrival process a Poisson process?} \label{sec:models}

\subsection{Poisson processes} \label{sec:poisson}

A point process $N$ is a Poisson point process on $\R$  if it has the
following two properties~\cite[Chapter~2.4]{daley2003introduction}.
\begin{enumerate}

\item The random number of points $N([a,b))$ of the point process $N$
located in a bounded interval $[a,b) \subset \R$ is a Poisson random
variable with mean $\Lambda([a,b))$, where $\Lambda$ is a non-negative
Radon measure.

\item The numbers of points of the point process $N$ located in $k$
intervals $[a_1,b_1),\dots, [a_k,b_k)$ form $k$ independent Poisson
random variables with means $\Lambda([a_1,b_1)),\dots,
\Lambda([a_k,b_k))$.

\end{enumerate}
From now on we will write $N([a,b))$ as $N(a,b)$ and $\Lambda([a,b)) =
\Lambda(a,b)$ for convenience. The first property implies that 
\begin{equation}
\Prob(N(a,b)=n)=\frac{\Lambda(a,b)^n \textrm{e}^{-\Lambda(a,b)}}{n!},
\end{equation}
and $\E[N(a,b)] = \Lambda(a,b)$, while the second property is the
principal reason for the tractability of the Poisson point process and
it is usually the basis of statistical tests that measure the
suitability of Poisson models.  The Poisson distribution of $N(a,b)$
implies its variance is $\var[N(a,b)]=\Lambda(a,b)$, a fact that is also
used as a statistical test.

The measure $\Lambda$ is known as the \emph{intensity measure} or
\emph{mean measure} of the Poisson point process.  We will assume that a
function $\lambda(t)$ exists such that 

\begin{align} 
\Lambda(a,b)=\int_a^b \lambda(t)\textrm{d}t.
\end{align}
Then $\lambda(t)$ is known as a \emph{rate function}.  If $\lambda(t)$
is a constant $\lambda >0$ then the point process is called a
homogeneous Poisson point process. Otherwise, the point process is
called an \emph{nonhomogeneous} or \emph{inhomogeneous } Poisson point
process.  Restricting our attention to the interval of non-negative
numbers $[0,\infty)$, the intensity measure is given by
\begin{equation}
\Lambda(t):=\Lambda([0,t))=\int_0^t \lambda (t)\textrm{d}t,
\end{equation}
For a Poisson process $N$ with intensity measure $\Lambda$, the
probability of $n$ points existing in the interval $[a,b)$ is
\begin{equation}
\Prob(N(a,b)=n)=\frac{[\Lambda(b)-\Lambda(a)]^n
\textrm{e}^{-[\Lambda(b)-\Lambda(a)]}}{n!}.
\end{equation}

\subsubsection{Arrival times and inter-arrival times}

Consider a point process $\{X_{(i)}\}_{i\geq1}$ defined on the
non-negative reals with an almost certainly finite number of points in
any bounded interval. Then we can interpret the points of the process as
arrival times and put them in increasing order, $X_1 \leq X_2\leq
\ldots$. Then the distances between adjacent points are $T_i :=
X_{i}-X_{i-1}$ for $i=2,3,\ldots$ and $T_1 = X_1$. The random variables
$T_i$ are known as \emph{waiting} or \emph{inter-arrival} times.  For a
homogeneous Poisson process with rate $\lambda$, the corresponding
inter-arrival times are independent and identically-distributed
\cremove{(i.i.d.)}{} exponential random variables with mean $1/\lambda$
\cremoveR{The (cumulative probability) distribution (function) of the
$k^\mathrm{th}$ waiting time $T_k$, where $T_1=X_{(1)}$ and
$T_k=X_{(1)}-X_{(k-1)}$ for $n\geq 2$, is}{}
\begin{equation}\label{eq:expiat}
\Prob (T_k\leq t)=1-\textrm{e}^{-\lambda t},
\end{equation}
where the memoryless property of the exponential distribution has been
used. This is not the case for an inhomogeneous Poisson point process
with intensity $\lambda(t)$, where the first inter-arrival time
$T_1=X_1$ has the distribution
\begin{equation}
\Prob (T_1\leq t_1)=1-\textrm{e}^{-\int_0^{t_1} \lambda (s)\textrm{d}s
}.
\end{equation}
%
%
Given the first waiting time $T_1=t_1$, the conditional distribution of
the second waiting time $T_2$ is
\begin{equation}
\Prob (T_{2} \leq t_2|T_1\leq t_1)=1-\textrm{e}^{-\int_{t_1}^{t_2} \lambda (s)\textrm{d}s }\,, 
\end{equation}
and so on for $k\geq 2$,
\begin{equation}
\Prob (T_{k}\leq t_k|T_{k-1}\leq t_{k-1})=1-\textrm{e}^{-\int_{t_{k-1}}^{t_k} \lambda (s)\textrm{d}s }\,.
\end{equation}
It can be shown that the $k^{\mathrm th}$ arrival time $X_{k} $
has the distribution
\begin{equation}\label{eq:nhpp_distr}
\Prob(X_{k} \leq t)=\textrm{e}^{-\Lambda(t)}\sum_{n=k}^{\infty}\frac{\Lambda(t)^k }{k!}\, ,
\end{equation}
with density
\begin{equation}\label{eq:nhpp_dens}
f_{X_{k}}(t)=\frac{\lambda(x)\Lambda(t)^{k-1} }{(k-1)!} \textrm{e}^{-\Lambda(t)}\, .
\end{equation}
Condition on $n$ points $\{U_i\}_{i=1}^n$ of a Poisson process existing
in some bounded interval $[0,t]$. We call these points \emph{conditional
arrival times}. If the Poisson process is homogeneous, then the
conditional arrival times are uniformly and independently distributed,
forming $n$ uniform random variables on $[0,t]$.  This difference
between waiting times $T_i$ and conditional arrival times $U_i$ plays a
role in a Poisson test, originally proposed by Brown \emph{et
al.}~\cite{brown2005statistical} and later examined in detail by Kim and
Whitt~\cite{kim2014choosing}.

For a nonhomogeneous Poisson process, each point $U_i$ is independently
distributed on the interval $[0,t]$ with the distribution
\begin{equation}
\Prob (U_i\leq u)=\frac{\Lambda (u)}{\Lambda (t) }, \qquad u\in[0,t].
\end{equation}
If the distribution of each $U_i$ is known and invertible, then each
$U_i$ can be transformed into a uniform random variable on $[0,1]$,
resulting in $n$ independent uniform random variables.
In other words, $\Lambda(t)$ maps a Poisson process to a homogeneous
Poisson process with density one on the real line. Consequently,
statistical methods for nonhomogeneous Poisson processes often involve
transforming the data, before performing the analysis.

The monograph by Kingman~\cite{kingman1992poisson} is  recommended for
background reading on the Poisson point process.
Interpreted as a stochastic process, standard books on stochastic
processes  cover the Poisson process. The papers by Brown \emph{et
al.}~\cite{brown2005statistical}  and Kim and
Whitt~\cite{kim2014choosing} are good starting points to get up-to-date
on methods for fitting nonhomogeneous Poisson processes and tests for
Poissonness in the one-dimensional setting. 

\subsection{Block arrivals modelled as a homogeneous Poisson
process with rate six blocks per hour}

At first glance, a Poisson process appears to be a reasonable model
for block arrivals, see Section~\ref{sec:bitcoin_mining}.

The simplest model is that the blocks arrive at the instants of a
homogeneous Poisson process with an arrival rate equal to the designed
rate of six blocks per hour. However, the empirical average since
Bitcoin began has been 6.4 blocks per hour over hundreds of thousands of
blocks, almost certainly indicating that the hypothesised parameter of
six blocks per hour is incorrect.

\begin{figure*}[t!]

\centering

\begin{subfigure}[b]{0.48\textwidth} \centering
 \includegraphics[width=1.0\textwidth]{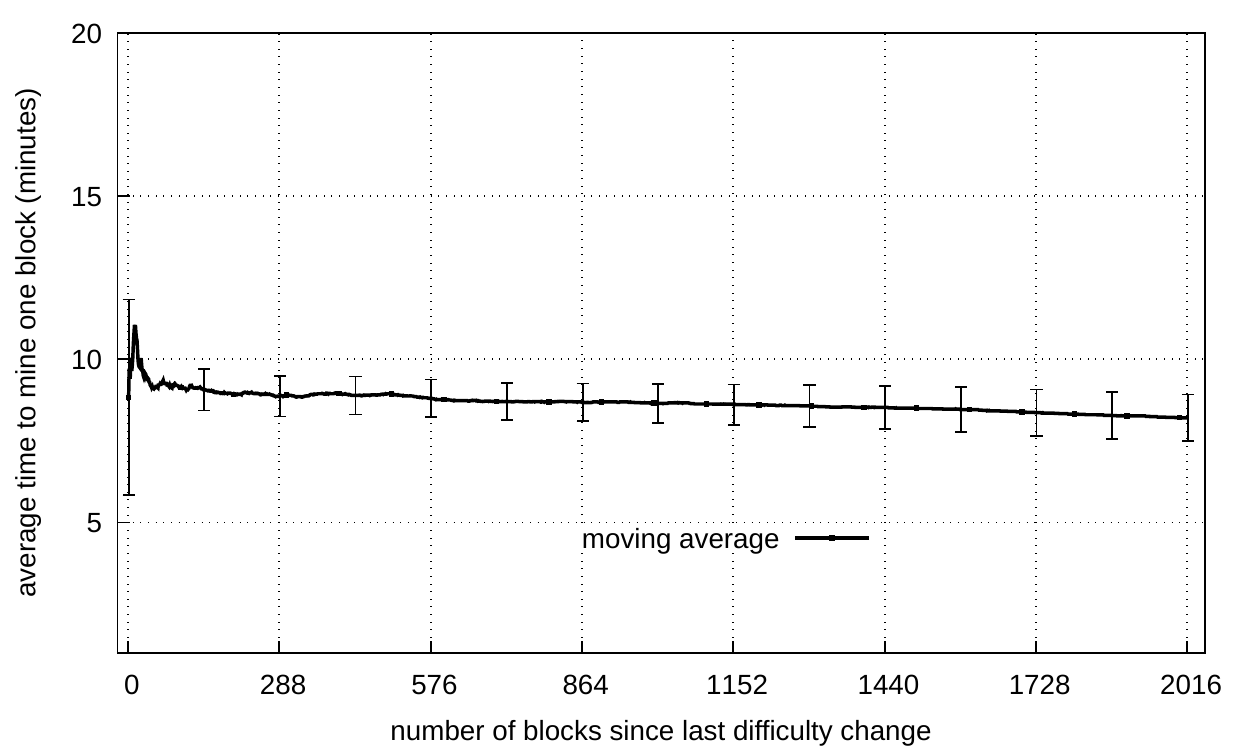}
\caption{Interval 2}
\label{fig_2016a}
\end{subfigure}
~
\begin{subfigure}[b]{0.48\textwidth} \centering
 \includegraphics[width=1.0\textwidth]{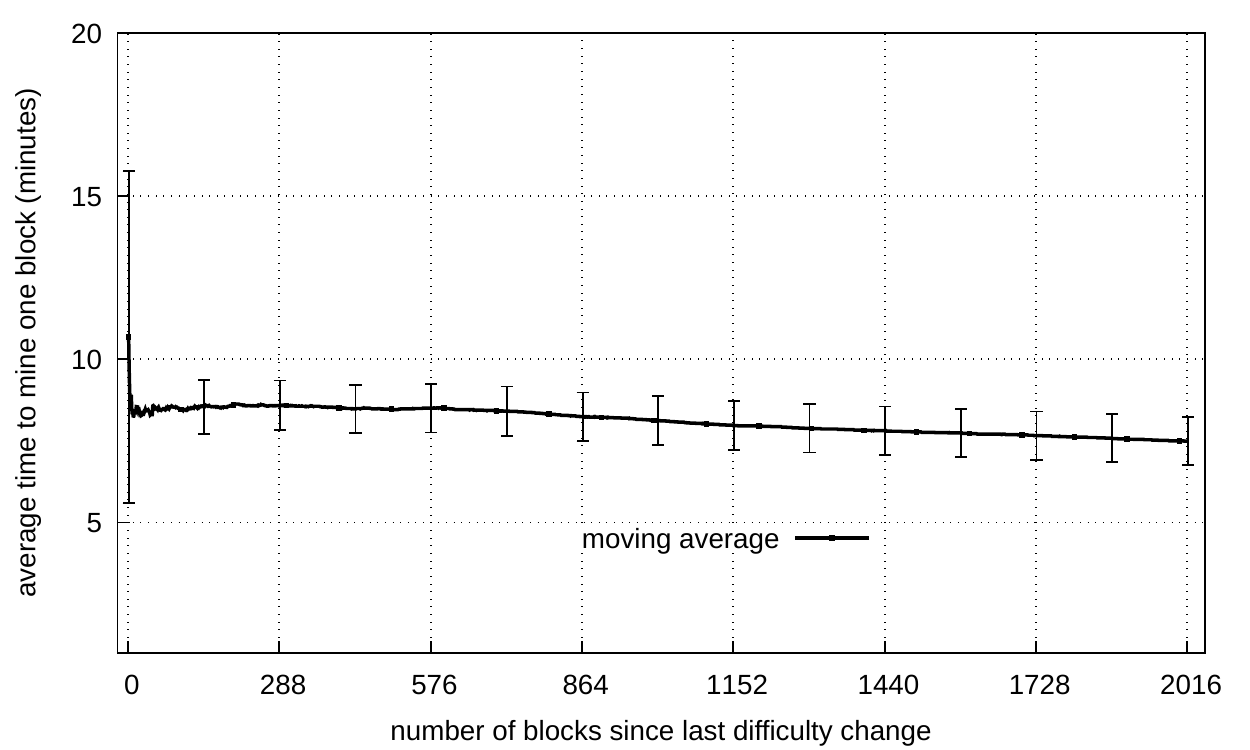}
\caption{Interval 3}
\label{fig_2016b}
\end{subfigure}

\begin{subfigure}[b]{0.48\textwidth} \centering
 \includegraphics[width=1.0\textwidth]{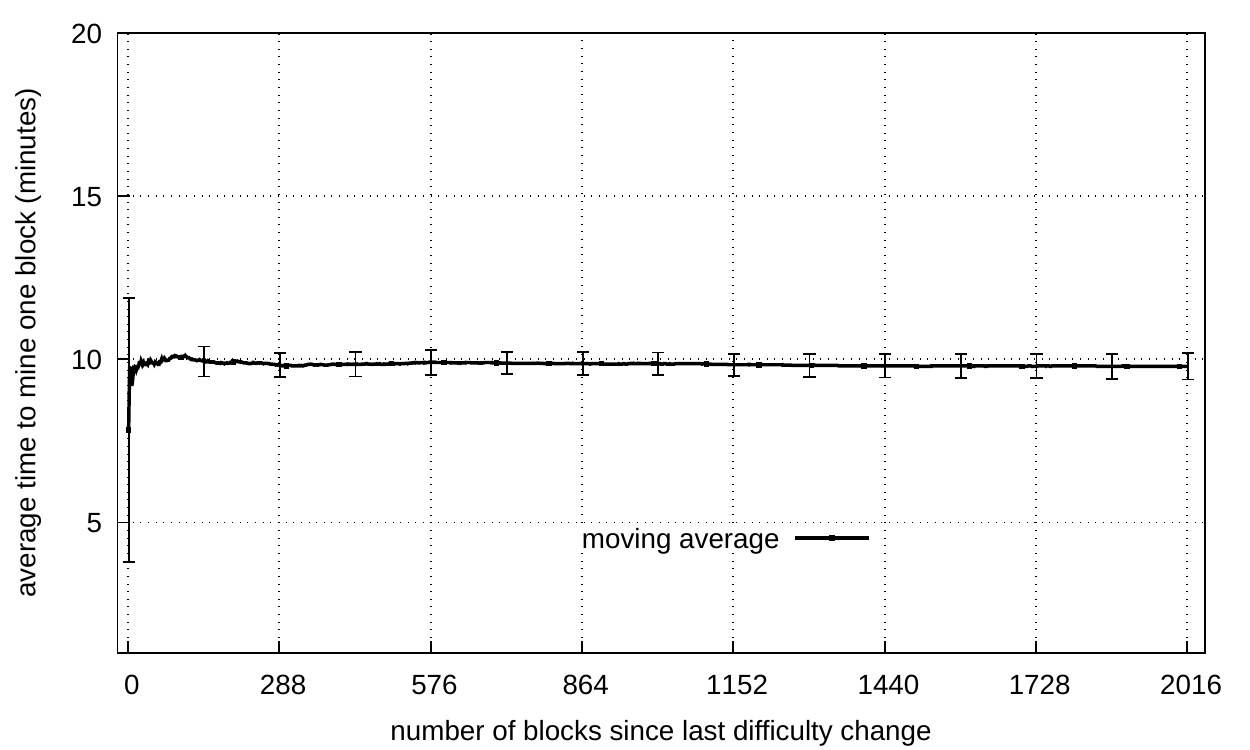}
\caption{Interval 4}
\label{fig_2016c}
\end{subfigure}
~
\begin{subfigure}[b]{0.48\textwidth} \centering
 \includegraphics[width=1.0\textwidth]{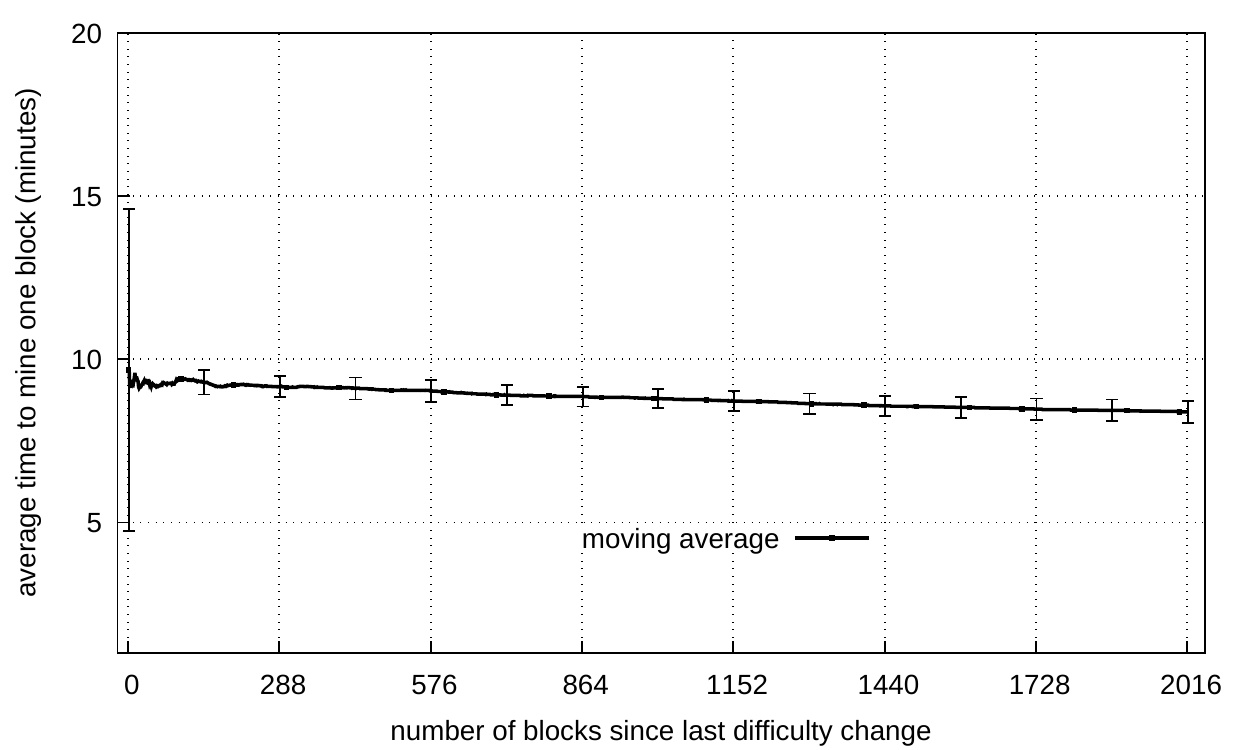}
\caption{Interval 5}
\label{fig_2016d}
\end{subfigure}
~
\begin{subfigure}[b]{0.48\textwidth} \centering
 \includegraphics[width=1.0\textwidth]{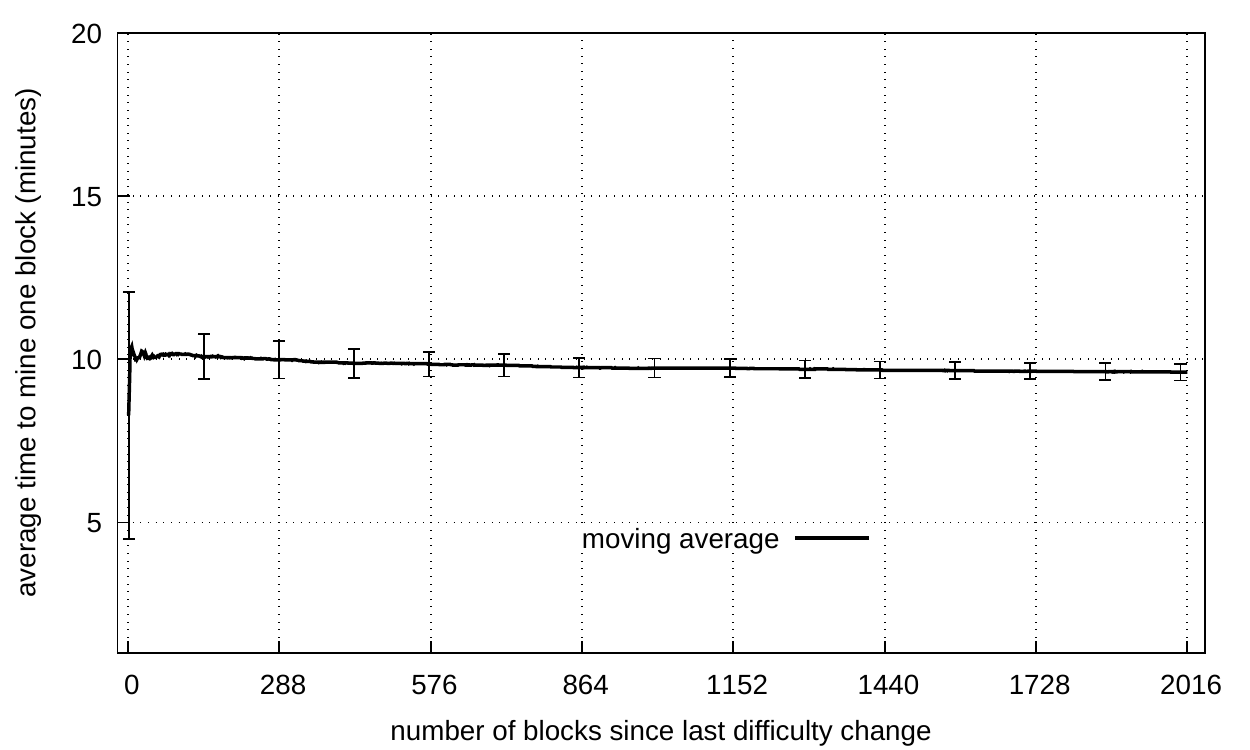}
\caption{Interval 6}
\label{fig_2016e}
\end{subfigure}

\caption{The average block inter-arrival times \emph{vs.}
the number of blocks since the last difficulty change.}

\label{fig_2016}

\end{figure*}

A possible reason for this is that the hash rate is generally increasing
over time, and the method by which the difficulty is updated means that
the difficulty lags behind the hash rate by about two weeks.
Consequently, the average rate at which blocks are mined is greater than
six blocks per hour due to the delay in adjusting the difficulty.
Furthermore, the hash rate and hence the block arrival rate usually
increases over the course of a segment until the end of the segment when
the difficulty is increased to compensate, and the arrival rate is
reduced.
Figure~\ref{fig_2016} shows the average block inter-arrival time versus
the number of blocks since the last difficulty change for the second to
sixth intervals depicted in Figure~\ref{fig_2}, see also
Table~\ref{tab:piecewise}.
Figures~\ref{fig_2016a}, \ref{fig_2016b} and \ref{fig_2016d} show that
when the hash rate is increasing rapidly the average inter-arrival time
decreases as the segment of 2,016 blocks progresses.
Figures~\ref{fig_2016c} and \ref{fig_2016e} show that when the hash rate
is increasing slowly the average inter-arrival time decreases slightly
as the segment of 2,016 blocks progresses.

\subsection{Block arrivals modelled as a homogeneous Poisson
process with any rate}

The next simplest model is a homogeneous Poisson process with an arrival
rate equal to the empirical average block arrival rate.

A key characteristic of a homogeneous Poisson process is that the times
between arrivals $T_i$ are independently exponentially distributed with
identical mean, see Equation (\ref{eq:expiat}). We can test this
hypothesis using the Kolmogorov-Smirnoff test (K-S test). This test uses
$\sup|F_n(x)-F(x)|$ as the test statistic, where $F_n(x)$ is the
empirical cumulative distribution function of the $n$ data points, and
$F(x)$ is the true distribution. The Lilliefors test
\cite{lilliefors1969kolmogorov} allows us to apply the K-S test if we do
not know the mean of the true distribution.

Performing the Lilliefors test on the LR data rejects the null
hypothesis that block mining intervals are exponentially distributed, at
a significance level of $\alpha = 0.05$. In fact, the $p$-value for the
test is less than $0.001$, the smallest reportable $p$-value for the
test when performed in MATLAB. Thus we can be very confident that the LR
data are not generated by a homogeneous Poisson process\footnote{Note
that the uncleaned timestamp data are obviously not generated by a
homogeneous Poisson process since the mean and standard deviation of the
inter-arrival data are significantly different.}.  

Furthermore, we know the process is not Poisson due to the dependence
between disjoint intervals inherent in the design of the difficulty
adjustments: the time taken for a segment is random, which determines
the difficulty for the next segment, and that difficulty influences the
rate of block arrivals over that segment (Equation
(\ref{eq:difficulty})).

\begin{figure}[t!]
\centering
 \includegraphics[width=0.45\textwidth]{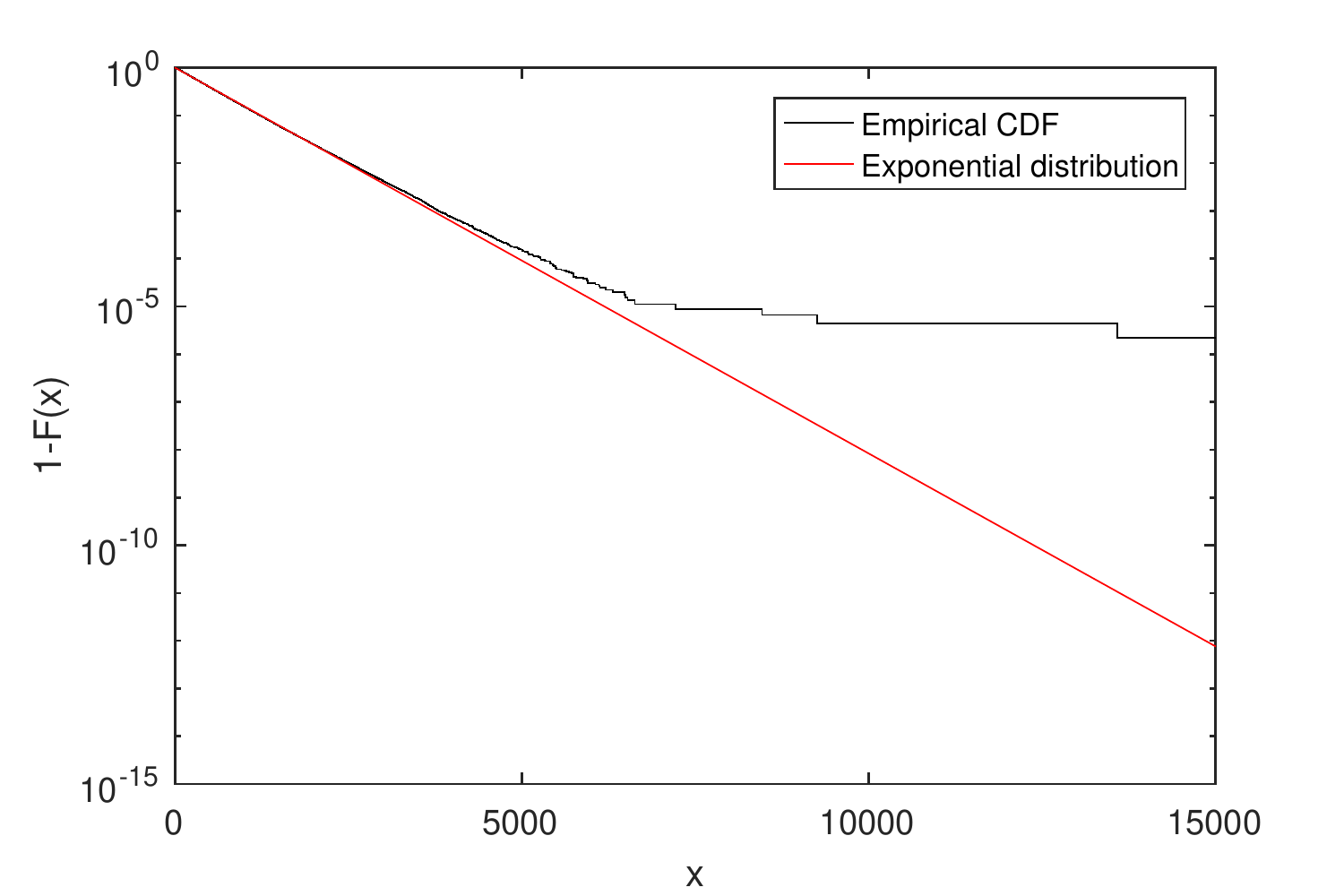}
\caption{Comparison of the survivor function ($1-F(x)$) for the
empirical LR data and an exponential distribution on a log axis.
\label{fig:LRtails}}
\end{figure}

We can examine the distribution function of the inter-arrival times to
try to see what is happening. Figure~\ref{fig:LRtails} shows a
comparison between the empirical survivor function (the difference
between the empirical distribution function and one) for our
inter-arrival time data and the survivor function for an exponential
random variable with the same mean on a logarithmically scaled plot. We
can see that a few abnormally large inter-arrival times $X_{i+1}-X_i$
are hindering the agreement between the empirical distribution function
and the exponential distribution.  If we resample all inter-arrival
times that are greater than 6500 as exponential random variables
conditioned to be greater than 6500, then the agreement is closer but
the Lilliefors test still has $p$-value less than $0.001$.

%

\subsection{Block arrivals modelled as a piecewise combination
of nonhomogeneous Poisson processes}

A natural way to account for the global hash rate changing over time is
to use a nonhomogeneous Poisson process to model the block discovery
rate. The instantaneous rate at which blocks are discovered is 
\begin{equation}\label{eq:lambda_def}
\lambda(t) = \frac{H(t)}{2^{32} D(t)}.
\end{equation}
To simulate such a process, we start with a form for the function
$H(t)$, for example the model defined in Section
\ref{sec:exponential_consequences} or the empirical process. Then for
each segment of 2,016 blocks: 
\begin{enumerate}

\item Calculate $D(t)$ for the current segment from
the previous difficulty and the time taken for the previous segment.

\item Calculate $\lambda(t)$ from $H(t),D(t)$ and
Equation~(\ref{eq:lambda_def}).

\item Simulate the 2,016 block discovery times for the segment according
to a nonhomogeneous Poisson process with rate $\lambda(t)$.

\end{enumerate}
%
This is not a  Poisson process over a timescale spanning successive
segments because the block arrival rate depends on the time of the first
and last block arrival in the previous segment. 


\subsubsection{Distributions for the time of arrival of the
$n^\mathrm{th}$ block in a segment}\label{sec:distributions}

To model a piecewise combination of nonhomogeneous Poisson processes we
will start by examining the behaviour of the block arrival process
within a segment, where the difficulty remains constant.

Let the random variable $X_n$ be the time of the $n^{\mathrm{th}}$ point
of an nonhomogeneous Poisson process with rate $\lambda(t)$ and
cumulative intensity function $\Lambda(t)$. The distribution
function of each random variable $X_n$ is given by Equation
(\ref{eq:nhpp_distr}) and its density is given by Equation
(\ref{eq:nhpp_dens}).
%
%
%
For some choices of $\lambda(t)$ we can simplify the density function,
and also calculate $\E\lbrack X_n\rbrack$. For example, if $\lambda(t) =
at$ then 
\begin{equation}
f_{X_n|X_0 = 0}(x)
=
ax e^{-ax^2/2} (ax^2/2)^{n-1} \frac{1}{(n-1)!} \, ,
\end{equation}
and 
\begin{equation}
\E\lbrack X_n\rbrack = \sqrt{\frac{2 \pi}{a}} \binom{2n-2}{n-1} \frac{1}{2n-2} \, .
\end{equation}
If instead $\lambda(t) = e^{at}$ then for $1\leq n\leq 2016$
\begin{equation}
f_{X_n|X_0 = 0}(x)
=
e^{ax} e^{-(e^{ax}-1)/a} ((e^{ax}-1)/a)^{n-1} \frac{1}{(n-1)!}\notag \,,
\end{equation}
and for $a\neq 0$
\begin{eqnarray}\label{eq:expon_expect}
\lefteqn{ \E\lbrack X_n|X_0 = 0 \rbrack = \sum_{i=0}^{n-1} (-a)^{-i} } \\
& =
\left(\frac{-e^{1/a}}{i!a}\mathrm{Ei}(-{1}/{a}) +\sum_{j=1}^i \frac{(-a)^{j}}{j(i-j)!}\sum_{k=0}^{j-1} \frac{a^{-1-k}}{k!} \right),
\end{eqnarray} 
where $\mathrm{Ei}(x)$ is the \emph{Exponential Integral}
$\mathrm{Ei}(x)=-\int_{-x}^{\infty} e^{-t}/t\,dt$.  See
Appendix~\ref{app:a} for a proof of Equation~(\ref{eq:expon_expect}).
Note that while Equation~(\ref{eq:expon_expect}) is not defined for
$a=0$, the limit as $a\rightarrow 0^+$ of
Equation~(\ref{eq:expon_expect}) is $n$, the value for a rate one
homogeneous Poisson process as expected.

Equation~(\ref{eq:expon_expect}) is easy to compute for small values of
$n$, but for larger values of $n$ (and small values of $a$) direct
evaluation becomes unmanageable. Instead of considering the expected
time of the $n^{th}$ arrival, another approach would be to find the time
at which the expected number of arrivals was equal to $n$ by inverting
the relationship $\Lambda(t) = n$, which we use in the following
subsection. 

\subsubsection{A deterministic approximation for the segment times}
\label{ss:determ_approx}

\begin{figure}[t!]
\centering
 \includegraphics[width=0.5\textwidth]{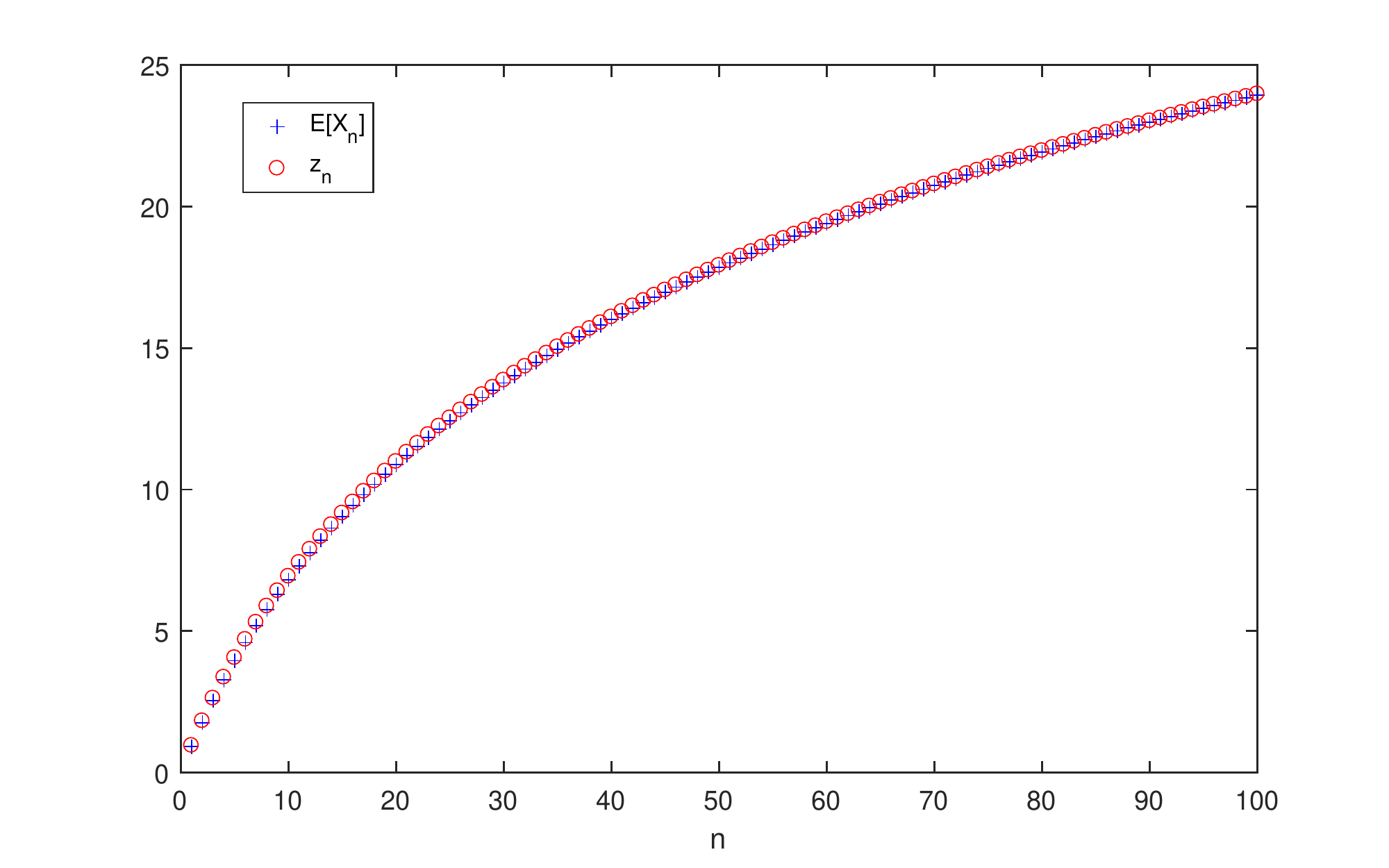}

\caption{Comparison of $\E[X_n]$, the expected time that the
$n^\mathrm{th}$ block arrives, with $z_n$, the time at which the
expected number of blocks arrived is $n$, for $\lambda(t) = e^{at}$ with
$a=0.1$.}

\label{fig:mean_compare}
\end{figure}

We now address the question of what happens over successive segments
when difficulty adjustments are taken into account. We approximate
$\E[X_n]$ with the time~$z_n$ at which the expected number of blocks
that have arrived is $n$, which is defined by the relation
\begin{align}
n = \int_{X_0}^{z_n} \lambda(t)\, dt \,,
\end{align} 
where $\lambda(t) = H(t)/(2^{32} D(t))$. If $X_0 = 0$ we have $z_n =
\Lambda^{-1}(n)$. We have reason to believe that this is a good
approximation over one segment, especially for the $\lambda(t)$ that we
consider. For example, Figure \ref{fig:mean_compare} shows a comparison
between $\E[X_n]$ and $z_n$ on a single segment where $\lambda(t) =
e^{at}$ with $a=0.1$. In the case of Bitcoin, $a$ is typically much
smaller (see Table~\ref{tab:piecewise}), which increases the accuracy of
the approximation. 

The results above for $\E[X_n]$ and $z_n$ are for single segments only,
we have not included the dependency caused by changes in the difficulty.
Now we investigate the behaviour of the times at which difficulty
changes occur. While these are random variables, we approximate them
with a deterministic version $y_n$: the times such that the average
number of arrivals since the last difficulty change is 2,016, that is,
for $n \geq 1$
\begin{align} 2016 &= \int_{y_{n-1}}^{y_n} \lambda(t)\, dt =
\int_{y_{n-1}}^{y_n} \frac{H(t)}{2^{32}D(t)} \,dt =
\int_{y_{n-1}}^{y_n} \frac{H(t)}{2^{32}D_{n}} \,dt \label{eq:delta_int}.
\end{align}
We fix $H(t)$ and the starting difficulty $D_1 = d_1 \geq 1$, and
determine each subsequent difficulty change as in the random version,
but instead of using the random arrival times to adjust the difficulty,
we use the deterministic approximation. The difficulty on
$(y_{n},y_{n+1})$ is then given by 
\begin{align}
d_{n+1} = d_{n} \frac{\texttt{2 weeks}}{y_{n}-y_{n-1}} \,,
\end{align} for $n \geq 1$. 

Let $\delta_n = y_n-y_{n-1}$, and thus $y_n =
y_0+\sum_{i=1}^{n}\delta_i$ and use fortnights (2 weeks) as our unit of
time. We have for $n\geq 2$
\begin{equation}
d_n = d_{n-1}/\delta_{n-1} = d_{1}/\prod_{i=1}^{n-1}\delta_{i} \,.
\end{equation}
If we assume an exponential
hash rate $\widehat{H}(t) = e^{at+b}$ with $a>0$ we can use
Equation~(\ref{eq:delta_int}) to find a recursion for the
sequence $(\delta_n)_{n=2}^{\infty}$
\begin{align}
2016 &=  \int_{y_{n-1}}^{y_n} \frac{e^{at+b}}{d_{n}} \,dt
      = \frac{e^b}{ad_{n}}(e^{ay_{n}}-e^{ay_{n-1}}) \nonumber \\
	   &= \frac{e^{ay_{n-1}+b}}{ad_{n}}(e^{a\delta_n}-1)
\end{align}
so that
\begin{align}
e^{a\delta_n}-1 &=
\frac{ad_{1}/\prod_{i=1}^{n-1}\delta_{i}}{\exp(ay_0+b+a\sum_{i=1}^{n-1}\delta_i)} \nonumber \\
	&= \frac{ad_{1}/\prod_{i=1}^{n-2}\delta_{i}}{b\exp(ay_0+a\sum_{i=1}^{n-2}\delta_i)} \cdot \frac{1}{e^{a\delta_{n-1}}\delta_{n-1}} \nonumber \\
	&= \frac{e^{a\delta_{n-1}}-1}{e^{a\delta_{n-1}}\delta_{n-1}}\label{eq:exp_recur}.
\end{align}
Thus
\begin{equation}
\delta_{n} = \frac{1}{a} \log \left(\frac{e^{a\delta_{n-1}}-1}{e^{a\delta_{n-1}}\delta_{n-1}} +1 \right).
\end{equation}
Consider the sequence $A_n = (e^{a\delta_n}
-1)_{n=1}^{\infty}$. Equation~(\ref{eq:exp_recur})  shows
that
\[
A_n = A_{n-1}/ {(e^{a\delta_{n-1}}\delta_{n-1})}.
\]
Thus if $\delta^* = \lim_{n \rightarrow \infty}\delta_n$ exists, then
$\lim_{n \rightarrow \infty} A_n=0$ or $\delta^*$ satisfies 
\begin{align}
e^{a\delta^*}\delta^*=1 \label{eq:limit} 
\end{align}
which is consistent with Equation~(\ref{eq:seg_time_vs_hr_growth}).
However, $\lim_{n \rightarrow \infty} A_n=0$ implies that $\lim_{n
\rightarrow \infty} \delta_n=0$, but $\lim_{\delta_{n-1} \rightarrow 0}
A_{n-1}/(e^{a\delta_{n-1}}\delta_{n-1})=a$, a contradiction. Thus we
have shown that if $\delta^*$ exists then it must satisfy
Equation~(\ref{eq:limit}).

An alternative means of expressing Equation~(\ref{eq:limit}) is
\begin{align}
\delta^* = \frac{1}{a}W(a) \label{eq:lambert}
\end{align}
where $W(\cdot)$ is the Lambert-W function
\cite{corless1996lambertw,lambert1758observationes}, defined so that $x
= W(xe^x)$. We now show that $\delta^* = \lim_{n \rightarrow
\infty}\delta_n$ exists.

Consider the function $f(\delta) = \frac{1}{a} \log
\left(\frac{e^{a\delta_{}}-1}{e^{a\delta_{}}\delta_{}} +1 \right)$,
defined so that $\delta_{n+1} = f(\delta_n)$. We will show that
$f^{(n)}(\delta_1)$ tends to a limit regardless of $\delta_1
> 0$.  Taking the derivative,
\begin{align}
f'(\delta) = \frac{a\delta - e^{a\delta}+1}{a\delta((\delta+1)e^{a\delta}-1)}.
\end{align}
Then $f'(\delta)$ is increasing from $\frac{-a}{2(1+a)}$ to $0$ as
$\delta$ goes from $0$ to $\infty$, thus
$-\frac{1}{2}<\frac{-a}{2(1+a)}<f'(\delta)<0$ and consequently
$f(\delta)$ is a contraction for $\delta \in (0,\infty)$. Therefore, by
the Banach fixed point theorem, we know that $\delta^* =
\lim_{n\rightarrow \infty}f^{(n)}(\delta_1) = \lim_{n\rightarrow \infty}
\delta_n$ exists and satisfies $f(\delta) = \delta$, and thus it is
given by the unique solution of Equation~($\ref{eq:limit}$). 

Equation~($\ref{eq:limit}$) has the following interpretation: if a
steady state is to occur, the ratio $e^{a\delta}$ by which the hash rate
increases over the period of a segment should be equal to the
corresponding increase in the difficulty. This increase in the
difficulty is given by the ratio $1/\delta$ of a fortnight to the actual
time taken for the segment. In the case of this deterministic
approximation, the system settles down to a steady state regardless of
initial conditions.


We can use this deterministic approximation for the times and magnitudes
of the difficulty changes as the basis of a nonhomogeneous Poisson
process model over a timescale spanning successive segments. In such a
model the difficulty changes are now deterministic, and so there is no
dependency on disjoint intervals to prevent the block arrival process
from being Poisson.

\subsection{Including the block propagation delay in the
block arrival model} \label{sec:delay}

Our model does not yet account for the time taken for a newly discovered
block to propagate through the network and have its transactions
validated at each node.  While this affects the times that blocks arrive
at a measurement node, it does not directly change the timestamps of the
blocks.  However, it does affect the block inter-arrival process as a
whole. When a new block is mined, it has to propagate through the
Bitcoin network to each other miner and be validated before that miner
can start mining on it\footnote{Sometimes miners will speed up the
process by mining blocks while waiting for transaction validation to be
completed.}. Block propagation delay thus has an effect on the rate at
which mining is performed.
We will use a simple method for modelling the delay: we treat it as an
effective reduction in the block arrival rate for a period after a block
is mined. We provide two options:
\begin{enumerate}

\item The simplest option is to model it as a constant delay to every
miner, and ignore the hash power contribution from the miner of the
block (who sees no propagation delay). Thus we would take any of our
previous models and add a period after each block arrival when no blocks
could be mined.

\item The second option is to assume that the block does not arrive at
all nodes at once, but gradually reaches a greater and greater
proportion of the miners (or more importantly, a greater and greater
proportion of the mining power). If we assume that the rate at which
miners receive the block is proportional to the number of miners who
have not yet received it, we obtain an exponential model for the delay.
In this case, if the most recent block arrived at time $s$, then the
effective global hash rate at time $t$  is $(1-e^{c(t-s)})H(t)$. 

This approach is consistent with the way in which
\cite{decker2013information} and \cite{croman2016scaling} refer to
propagation delay: they quantify it in each case by saying ``on average
a block will arrive at 90\% of nodes within $x$ seconds'', where $x$ is
26 and 79 respectively\footnote{The values given for median and 90\%
block propagation times in \cite{decker2013information} are
approximately consistent with exponential delay, but those given in
\cite{croman2016scaling} do not exactly fit an exponential distribution
for delay. This might be due to some nodes having much higher bandwidth
and connectivity than others in the network. Also, these figures are
only for node-to-node communication rather than miner-to-miner.}.


\end{enumerate}
We examine the consequences of this approach in more detail in the
following section. In the simulation experiments presented in
Section~\ref{sec:simulation} we will only consider option~2.

\section{Summary of the block arrival models} \label{sec:summary}

There are several aspects of the block arrival models proposed in this
paper. We provide a summary of these differing options here before using
a simulation to evaluate them in the following section. 
\begin{enumerate}

\item The hash rate function can be modelled
\begin{enumerate*}[label=\theenumi({\alph*})]

\item parametrically, or \label{label:1a}

\item empirically. \label{label:1b}

\end{enumerate*}

We will use $\widehat{H}(t) = e^{at+b}$ as the parametric hash rate
model and
$H^W_{144}(t)$ as the empirical hash rate model.

\item The difficulty adjustments are
\begin{enumerate*}[label=\theenumi({\alph*})]

\item deterministic, or \label{label:2a}

\item random. \label{label:2b}

\end{enumerate*}

The real blockchain has random difficulty adjustments, in that they
occur every 2,016 blocks of a random process, and their magnitude
depends on the time those blocks took to arrive. However, we saw in
Section~\ref{ss:determ_approx} that given the hash rate function, we can
find the expected times and magnitudes of these difficulty adjustments,
and use those instead to approximate the block arrival process.

\item The
\begin{enumerate*}[label=\theenumi({\alph*})]

\item absence, or \label{label:3a}

\item presence \label{label:3b}

\end{enumerate*}
of block propagation delay and how the delay is modelled
(Section~\ref{sec:delay}).

\end{enumerate}
Regardless of whether the global hash rate $H(t)$ is modelled
empirically or parametrically, for each of the block arrival models in
this paper the hash rate is determined before the random process
$X_i(t)$ is sampled.
\begin{itemize}

\item For the models~\ref{label:2a} with deterministic difficulty
adjustments,
the difficulty is adjusted at deterministic time instants $y_n$ which do
not correspond to the random instants of block arrivals,
and the block arrival rate $\lambda(t)$ is independent of the block
arrivals in the preceeding segments.
If there is no delay~\ref{label:3a}, then on each interval the model is
a nonhomogeneous Poisson process. 

\item For the models~\ref{label:2b} with random difficulty adjustment,
the difficulty is adjusted at random time instants corresponding to the
end of each 2,016 block segment using Equation~(\ref{eq:adjustTarget}).
If there is no delay~\ref{label:3a}, then on each segment the process
is a nonhomogeneous Poisson process with rate given by $\lambda(t) =
{H(t)} / {D_i}$.
Since the block arrival rate $\lambda(t)$ depends upon the first and
last arrivals in the previous segment, the process is not a Poisson
process over a timescale spanning successive segments.

\item If propagation delay is present~\ref{label:3b}, then
the block arrival process is not a nonhomogeneous Poisson process even
on a single segment.

\end{itemize}
In the following section we will compare simulations of these
models to the timestamp data from the Bitcoin blockchain.

\section{Simulations and comparison with blockchain
data} \label{sec:simulation}
\subsection{Exponential hash rate, random difficulty
changes, no block propagation delay}

We approximate the hash rate $H(t)$ with an exponential function
$\widehat{H}(t)=e^{at+b}$.  The block arrival rate is $\lambda(t) =
{\widehat{H}(t)}/{(2^{32}D(t))}$ as in Equation~(\ref{eq:lambda_def}),
where the difficulty $D(t)$ changes every 2,016 blocks as in
Equation~(\ref{eq:adjustTarget}).
We developed two independent implementations of the simulation model.
The outputs of the two simulators agreed closely. The average results
for each interval are given in Table~\ref{tab:blockArrival}.
While the estimate $H^W_{144}(t)$ of the real hash rate fluctuates
around the exponential approximation $\widehat{H}(t)$ (see
Figure~\ref{fig_2}), the average block inter-arrival time in each
interval closely matches the observed mean with the exception of
interval~5 from March 2013 to October 2014.

We can also see (Figure~\ref{fig:time_vs_a1}) that the mean
inter-arrival time from the simulations closely matches the mean
predicted by the steady state approximation given in
Equation~(\ref{eq:lambert}) regardless of the value of the exponential
coefficient $a$.

\subsection{Modifications for simulating the block propagation delay}
\label{sec:simulating_with_delay}

When simulating our models with propagation delay, we must take into
account the effect of that delay on the apparent hash rate, and
re-estimate the hash rate taking the (hypothesised) value of the delay
into account. Let the estimate of the underlying hash rate (the number
of hashes being checked per second, valid or otherwise) be
$\widehat{H}(t)$ and the effective hash rate be $\widetilde{H}(t) =
(1-e^{c(t-s)})\widehat{H}(t)$.

If we assume an exponential hash rate function $\widehat{H}(t) =
e^{at+b}$ with random difficulty changes, then for the period between
$X_i$ and $X_{i+1}$, the number of effective hashes checked is
\begin{eqnarray}
\lefteqn{\int_{X_i}^{X_{i+1}} \widetilde{H}(t)\, dt } \nonumber \\
&=& \int_{X_i}^{X_{i+1}}  (1-e^{c(t-X_i)})\widehat{H}(t) \,dt \nonumber \\
&=& \int_{X_i}^{X_{i+1}}  (1-e^{c(t-X_i)})e^{at+b} \,dt \nonumber \\
&=& e^{aX_i+b}\int_{0}^{X_{i+1}-X_i}   (e^{at}-e^{(a+c)t}) \,dt\nonumber \\
&=& \widehat{H}(X_i)\bigg[\frac{e^{at}}{a}-\frac{e^{(c+a)t}}{c+a}\bigg]_{0}^{X_{i+1}-X_i}\nonumber \\
&=& \widehat{H}(X_i)\bigg(\frac{e^{a(X_{i+1}-X_i)}-1}{a}-\frac{e^{(a+c)(X_{i+1}-X_i)}-1}{a+c}\bigg).
\end{eqnarray}
If we approximate the $X_i$ by their measured values we can invert this
relationship to obtain a corrected estimate for $H(t)$ in the
exponential case. The process is similar if we use an empirical hash
rate function.

If we want to simulate a model with deterministic difficulty changes, we
must employ a slightly more complicated version of this process to
estimate the times at which difficulty changes occur.  This is because
the arrival rate is not constant over the interval. The adjusted arrival
rate $\widetilde{\lambda}(t)$ is given by 
\begin{align} \widetilde{\lambda}(t) =
\frac{\lambda(t)}{1+{\lambda(t)}/{c}} =
\frac{1}{{1}/{\lambda(t)} + {1}/{c}} \end{align}
where $\lambda(t) = {H(t)}/{(2^{32}D(t))}$. Thus we must solve for
$y_{i+1} = X_{2016(i+1)}$ in 
\begin{align} \int_{y_i}^{y_{i+1}} \widetilde{\lambda}(t) dt \end{align}
iteratively for each $i$ to get the expected difficulty change times
$y_{i+1}$.

\begin{table}

\centering

{\small
\begin{tabular}{| l | r | r | r | r | r |} \hline
interval &
 \multicolumn{1}{|c|}{2} &
 \multicolumn{1}{|c|}{3} &
 \multicolumn{1}{|c|}{4} &
 \multicolumn{1}{|c|}{5} &
 \multicolumn{1}{|c|}{6} \\ \hline

Observed mean & 491.8 & 459.1 & 586.9 & 503.2 & 575.7\\
Simulated mean & 491.6 & 462.6 & 589.2 & 493.8 & 576.9\\
Observed s.d. & 503.3 & 477.6 & 578.1 & 494.7 & 567.3\\
Simulated s.d. & 493.7 & 465.6 & 589.7 & 494.7 & 578.0\\
\hline

\end{tabular}
}

\caption{Simulation experiments using an exponential hash rate function,
random difficulty changes and zero block propagation delay: the mean and
standard deviation of the block inter-arrival times in seconds for each
interval.}

\label{tab:blockArrival}

\end{table}

\subsection{Discrepancies between the observed and simulated values of
the standard deviation of the inter-arrival times}

Table~\ref{tab:blockArrival} shows that the average block inter-arrival
time of the observed data and the simulation using the random difficulty
change model match reasonably well in all the intervals except
interval~5 from March 2013 to October 2014, but the standard deviation
of the simulated data often does not match well with the observed data.
There are two possible reasons for this.   

\begin{itemize}

\item The first reason is the error in the block arrival times. Let
$X_i$ be the $i^{\mathrm{th}}$ block timestamp, let $W_i$ be the time
that block $i$ was actually created, and let $\epsilon_i$ be the error
in timestamp $i$, so that 
\begin{align}
X_i = W_i + \epsilon_i.
\end{align}
Then if the $\epsilon_i$ are independent and identically distributed 
with variance $\var(\epsilon)$, we would have the variance of the
timestamp inter-arrival times 
\begin{align}
\var(X_i-X_{i-1}) = \var(W_{i+1}-W_i) + 2\, \var(\epsilon).
\end{align} 
Thus the timestamp errors would increase the variance. This conclusion
is true even if the variance of the errors is not constant over time.
The data cleaning (Section~\ref{sec:cleaning}) reduces this effect but
does not eliminate or reverse it. If we examine Table
\ref{tab:blockArrival} we see that in some intervals the standard
deviation of the timestamp inter-arrival data is lower than that
predicted by simulation, so it cannot be entirely a result of errors in
the block timestamps. 

\item The second explanation as to why the timestamp inter-arrival time
data has a different variance to that predicted by the models is the
effect of the block propagation delay.  Propagation delay reduces the
effective global hash rate because some time is used mining blocks which
will not be included in the long-term blockchain. However, the
difficulty adjusts to compensate for the loss of hash rate. Small values
of the delay (relative to the mean inter-arrival time) therefore have
little effect on the overall mean, but do change the shape of the
inter-arrival time distribution. This reduces the probability that
blocks have very low inter-arrival times, so it might be expected that
it reduces the variance of the inter-arrival times, and in fact this is
what happens. 

\end{itemize}
To test the effect of block propagation delay on the distribution of
inter-arrival times we simulate delay as in option 2 in
Section~\ref{sec:delay}, where the effective global hash rate is
$(1-e^{c(t-s)})H(t)$ where the most recent block was mined at time $s$.
We plot the effects on the mean and standard deviation of the simulated
inter-arrival times $X_i-X_{i-1}$ for a range of values of~$c$ in
Figures~\ref{fig:int2} through \ref{fig:int6}.  Rather than displaying
the actual values of $c$ on the horizontal axis we instead show the
median delay because it is more intuitive.

Figures~\ref{fig:int2} through \ref{fig:int6} show the mean of 100
simulations for each set of parameters. The 95\% confidence intervals
are too small to be seen on the graph. The solid lines in each plot are
the simulation results. The horizontal dashed lines are the values
calculated from the observed LR resampled timestamp data for each
interval. For the simulated model to fit the real data, the solid lines
must intersect the dashed lines of the same colour.  Ideally the
simulated mean and standard deviation will intersect the observed mean
and standard deviation for the same value of the delay, that is, the
intersections will line up vertically.  The dotted line is the mean
block arrival interval $\delta^* = W(a)/a$ predicted by the steady state
deterministic approximation given in
Equation~(\ref{eq:limit})\footnote{We plot $\delta^*/2016$, the
predicted limiting average \emph{block} inter-arrival time rather than
the limiting average \emph{segment} inter-arrival time in order to be
comparable to the other values on the plot.}. The average block arrival
interval $\delta^\ast$ varies with $c$ because $a$ is the slope of the
line fitted to the logarithm of the estimated hash rate, and the
estimate of the hash rate depends on~$c$ (see
Section~\ref{sec:simulating_with_delay}).

\subsection{Exponential hash rate, random difficulty changes with delay}

Figures~\ref{fig:int2} through \ref{fig:int6}, subfigures~(a) shows what
happens when we use an exponential function and random difficulty
changes to approximate the hash rate.  In all intervals except interval
5 the observed mean is well approximated by using a reasonable value of
the block propagation delay. This model is also consistent with the
observed values of the standard deviation in the more recent intervals
4, 5 and 6, typically for a median block propagation delay of around 7
seconds. However, the standard deviation of the observed inter-arrival
data in intervals 2 and 3 is much higher than that the standard
deviations of the simulated model for all values of the block
propagation delay. This could be due to the highly variable hash rate in
intervals 2 and 3 being poorly modelled by the exponential
approximation. To address this we can use a more highly fitted,
empirical hash rate function.



\subsection{Empirical hash rate with delay}

Figures~\ref{fig:int2} through \ref{fig:int6}, subfigure (c) shows what
happens when we use the empirical function $H^W_{144}(t)$ to approximate
the hash rate, with random difficulty changes. We consider this to be
the most detailed and accurate model of those presented here.  In each
interval it is clear that this model is generally successful at matching
the observed block inter-arrival time statistics by our criterion
defined above: the median delay value at which the simulated lines
intersect the measured (``real'') values of the parameters is usually
around 10 seconds (it is lower for interval~3). Note that in most cases
the true delay is likely to be higher, due to the random errors in the
timestamp data artificially increasing the measured standard deviation
(the true dashed line would be lower, making the solid line intersect it
at a higher difficulty value).  

When comparing the values of median delay inferred here to those
generated in the measurement studies~\cite{decker2013information} and
\cite{croman2016scaling}, the delay values mentioned in this paper are
for miner-to-miner propagation delay, rather than general node-to-node
propagation delays.  Miners have a strong economic incentive to be
better connected to the Bitcoin network than a general Bitcoin node,
therefore in the present day they would be expected to have a lower
median delay.

\subsection{Deterministic difficulty changes}

\begin{figure*}
\centering
\begin{subfigure}[b]{0.475\textwidth}
\centering
 \includegraphics[width=\textwidth]{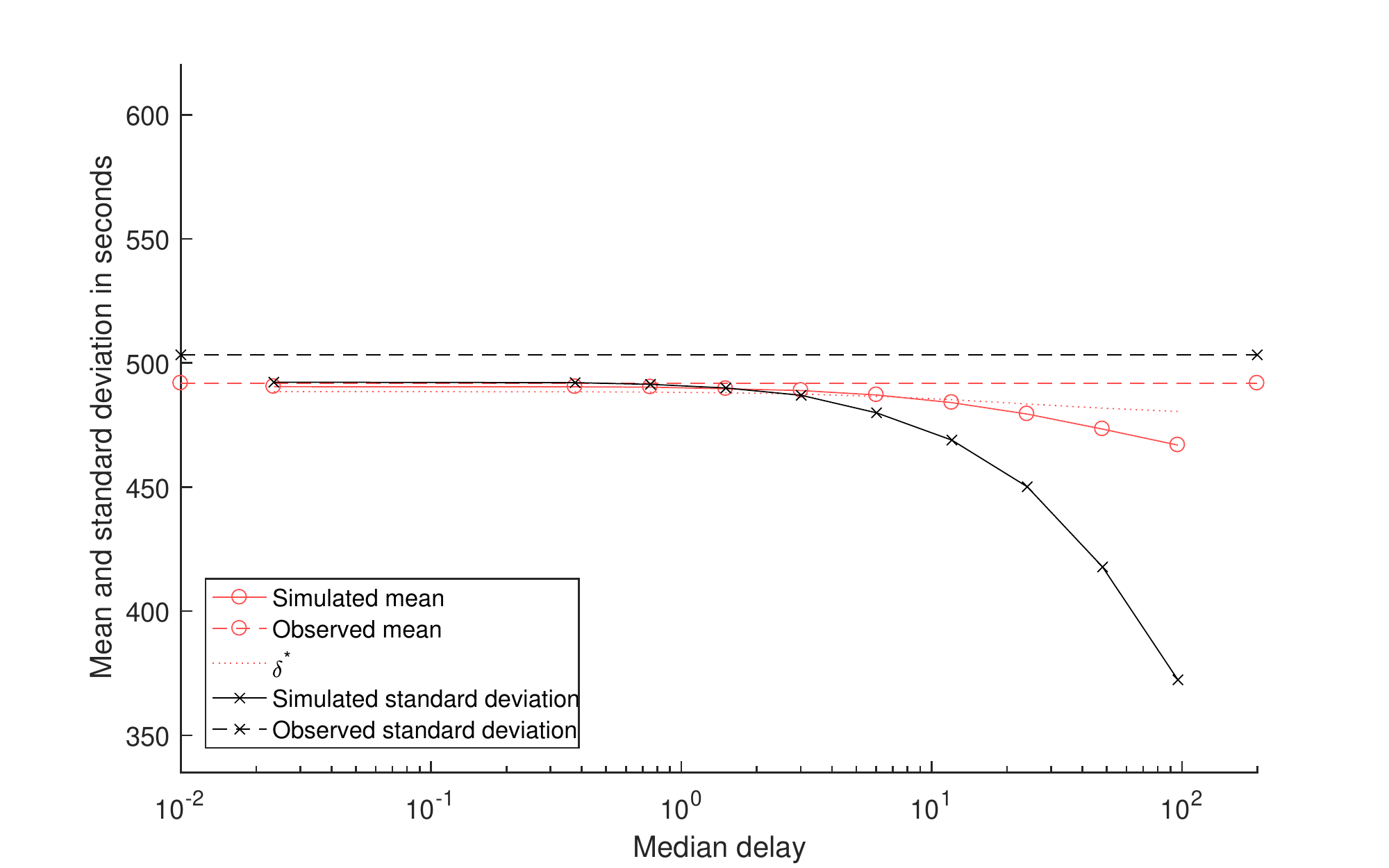}
\caption[]%
{{\small Exponential hash rate function, random difficulty changes}} 
\label{fig:exp2}
\end{subfigure}
\hfill
\begin{subfigure}[b]{0.475\textwidth} 
\centering 
 \includegraphics[width=\textwidth]{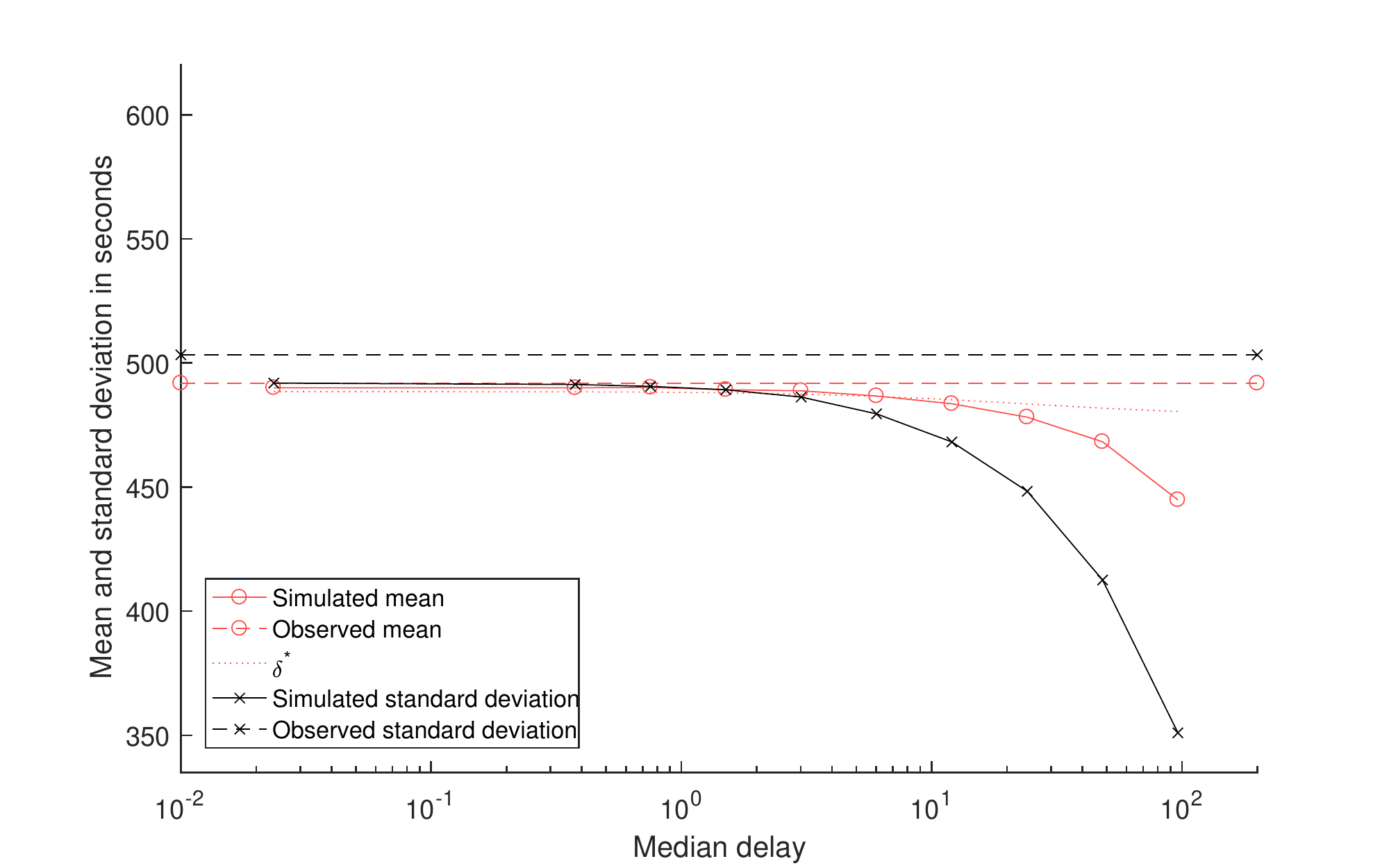}
\caption[]%
{{\small Exponential hash rate function, deterministic difficulty changes}} 
\label{fig:dexp2}
\end{subfigure}
\vskip\baselineskip
\begin{subfigure}[b]{0.475\textwidth} 
\centering 
 \includegraphics[width=\textwidth]{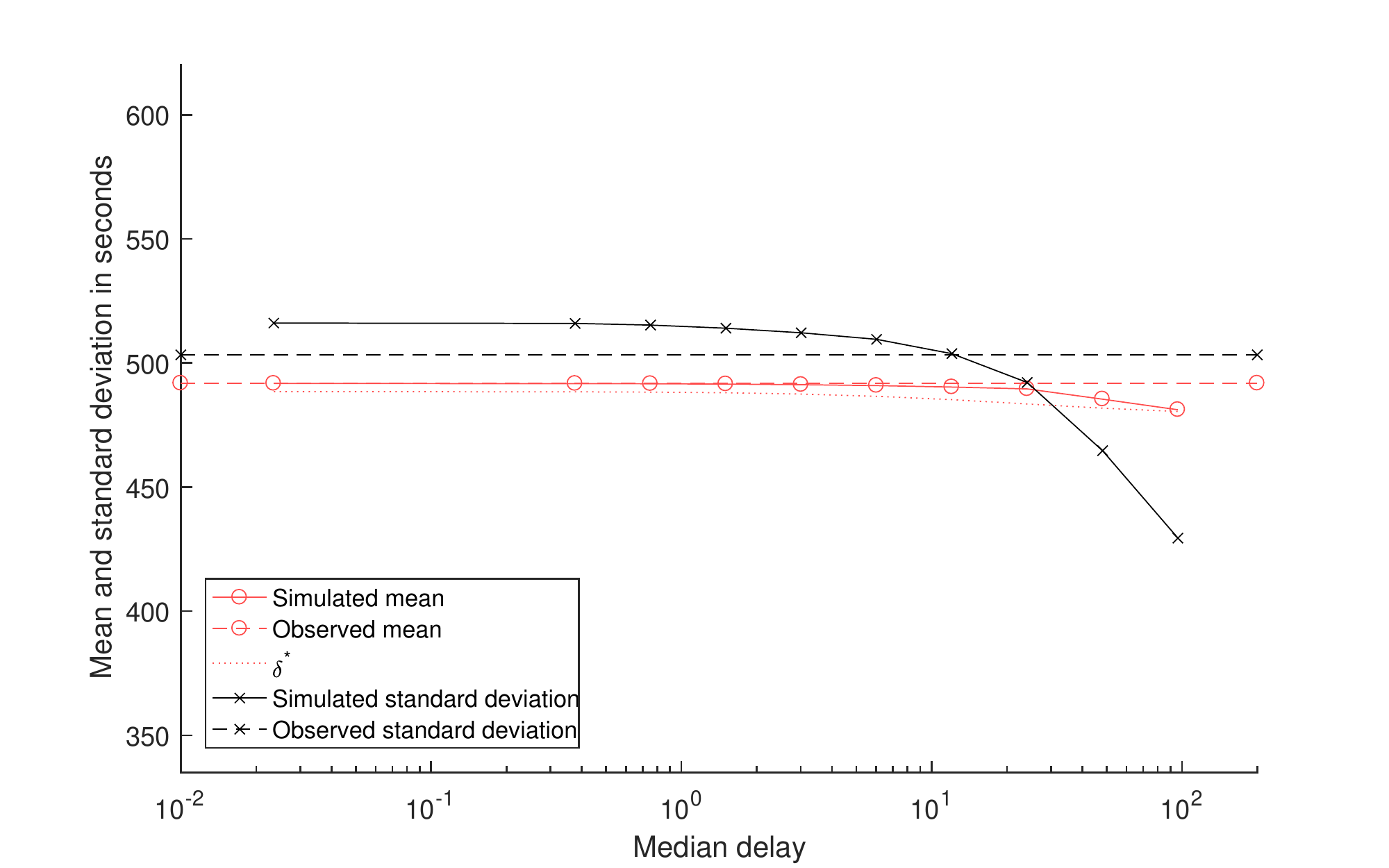}
\caption[]%
{{\small Empirical hash rate function, random difficulty changes}} 
\label{fig:np2}
\end{subfigure}
\quad
\begin{subfigure}[b]{0.475\textwidth} 
\centering 
 \includegraphics[width=\textwidth]{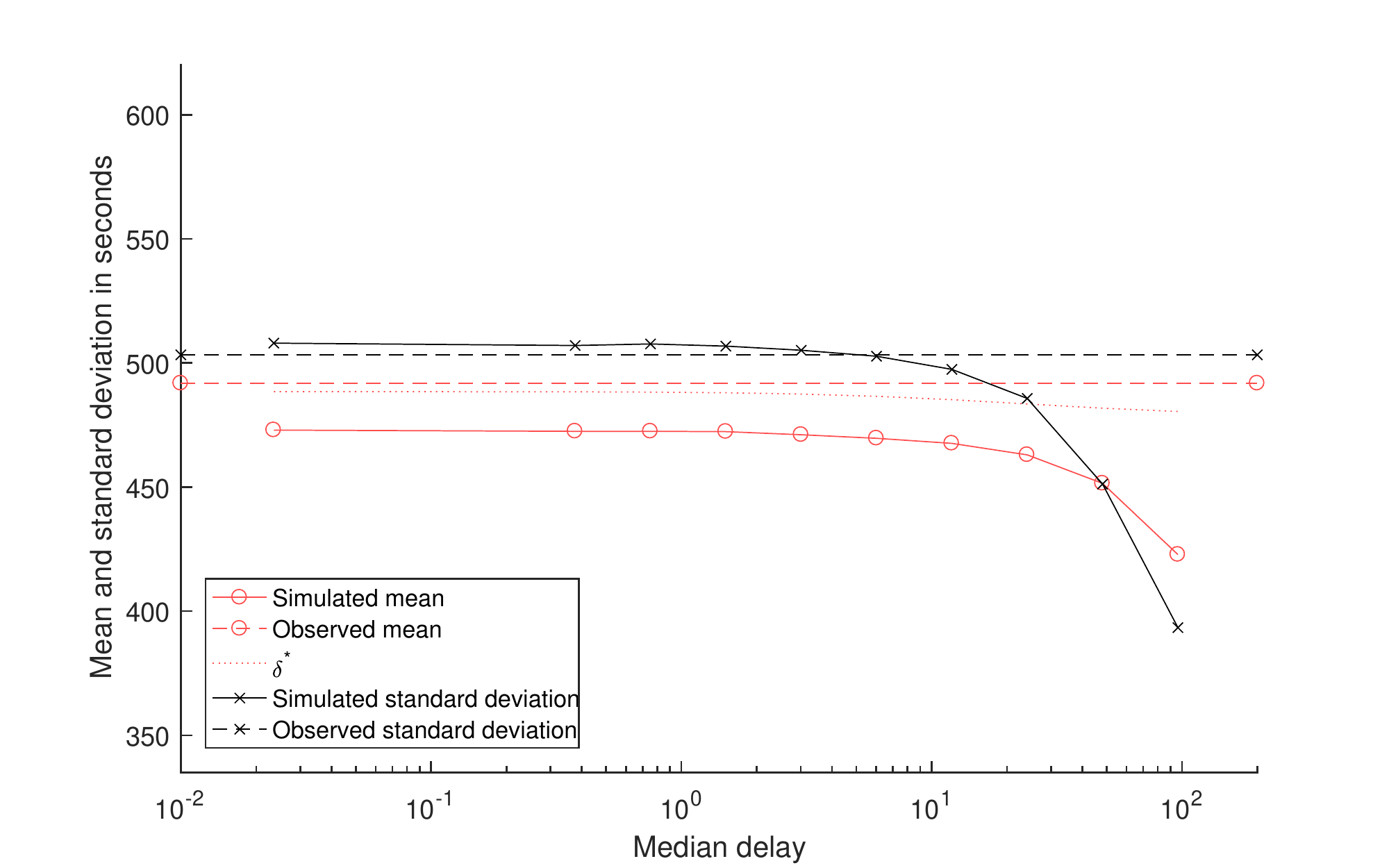}
\caption[]%
{{\small Empirical hash rate function, deterministic difficulty changes}} 
\label{fig:dnp2}
\end{subfigure}
\caption[]%
{\small Interval 2.  Simulating the effect of block propagation delay
(seconds) on the statistics of the inter-arrival time distributions. The
dashed lines indicate the values measured in the blockchain. The dotted
line is the value of the mean inter-arrival time predicted in
Equation~(\ref{eq:lambert}) by substituting in the value of $a$
estimated from the blockchain timestamp data and converting from mean
segment time in fortnights to mean inter-arrival time in seconds.
} 

\label{fig:int2}
\end{figure*}

\begin{figure*}
\centering
\begin{subfigure}[b]{0.475\textwidth}
\centering
 \includegraphics[width=\textwidth]{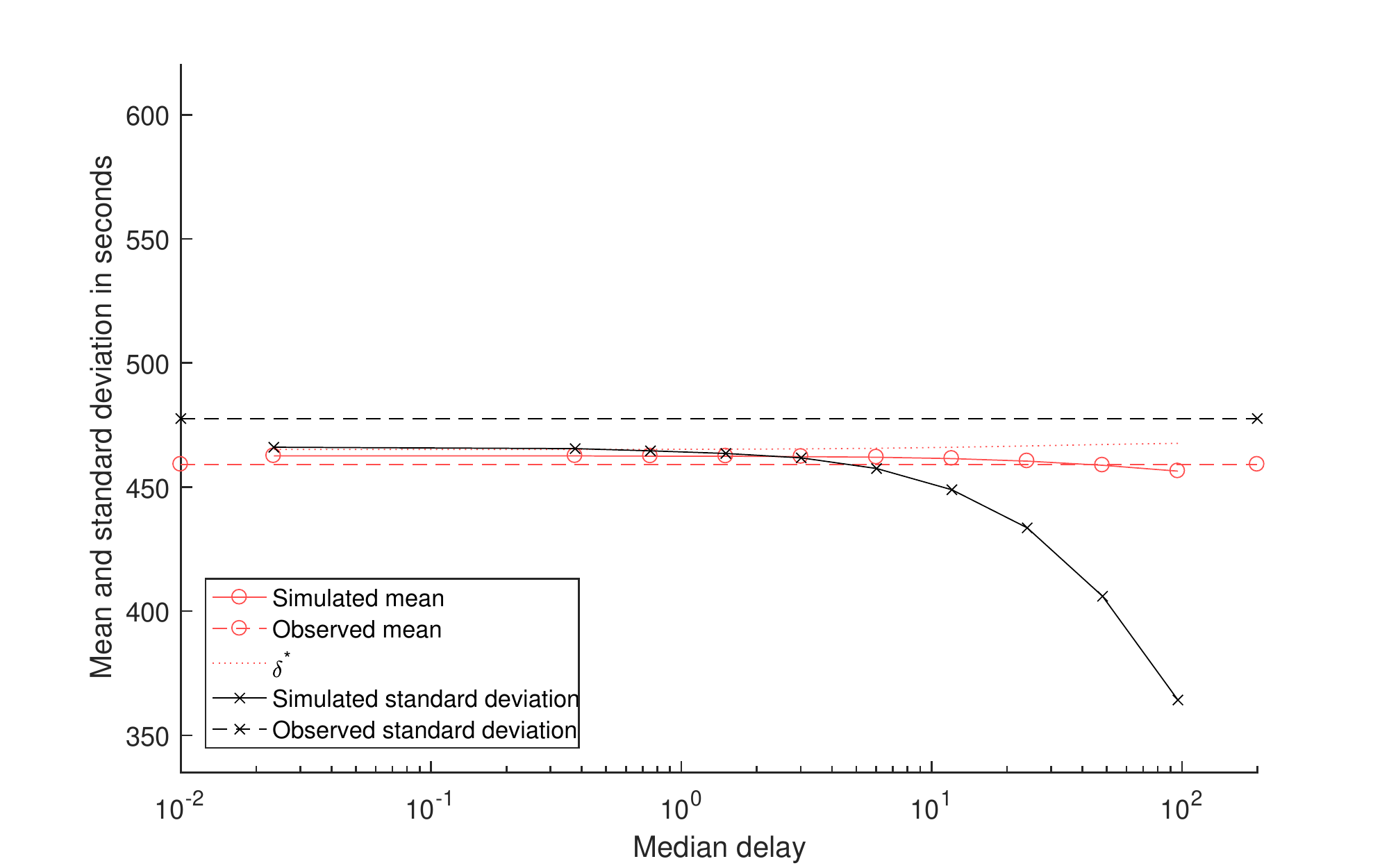}
\caption[]%
{{\small Exponential hash rate function, random difficulty changes}} 
\label{fig:exp3}
\end{subfigure}
\hfill
\begin{subfigure}[b]{0.475\textwidth} 
\centering 
 \includegraphics[width=\textwidth]{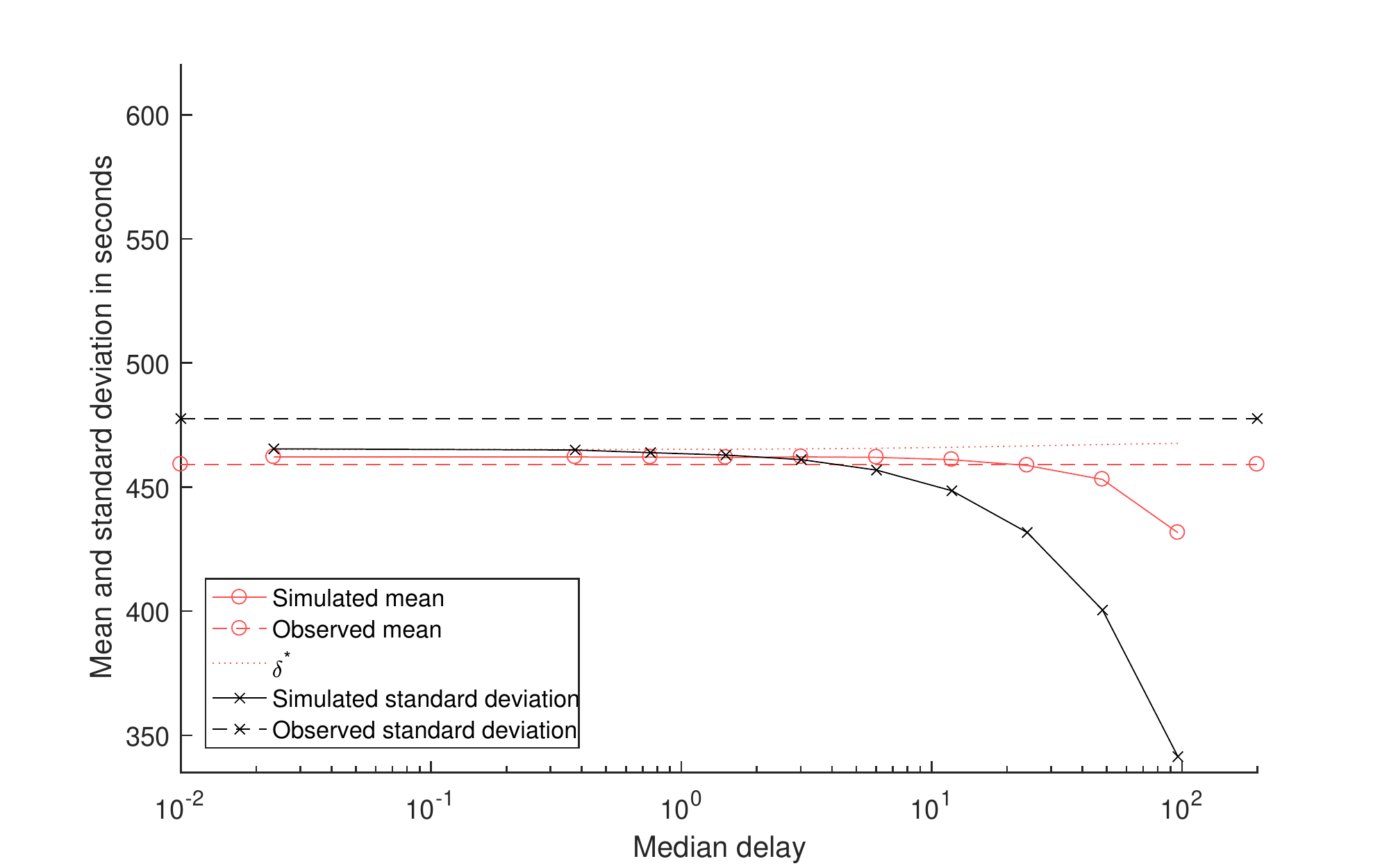}
\caption[]%
{{\small Exponential hash rate function, deterministic difficulty changes}} 
\label{fig:dexp3}
\end{subfigure}
\vskip\baselineskip
\begin{subfigure}[b]{0.475\textwidth} 
\centering 
 \includegraphics[width=\textwidth]{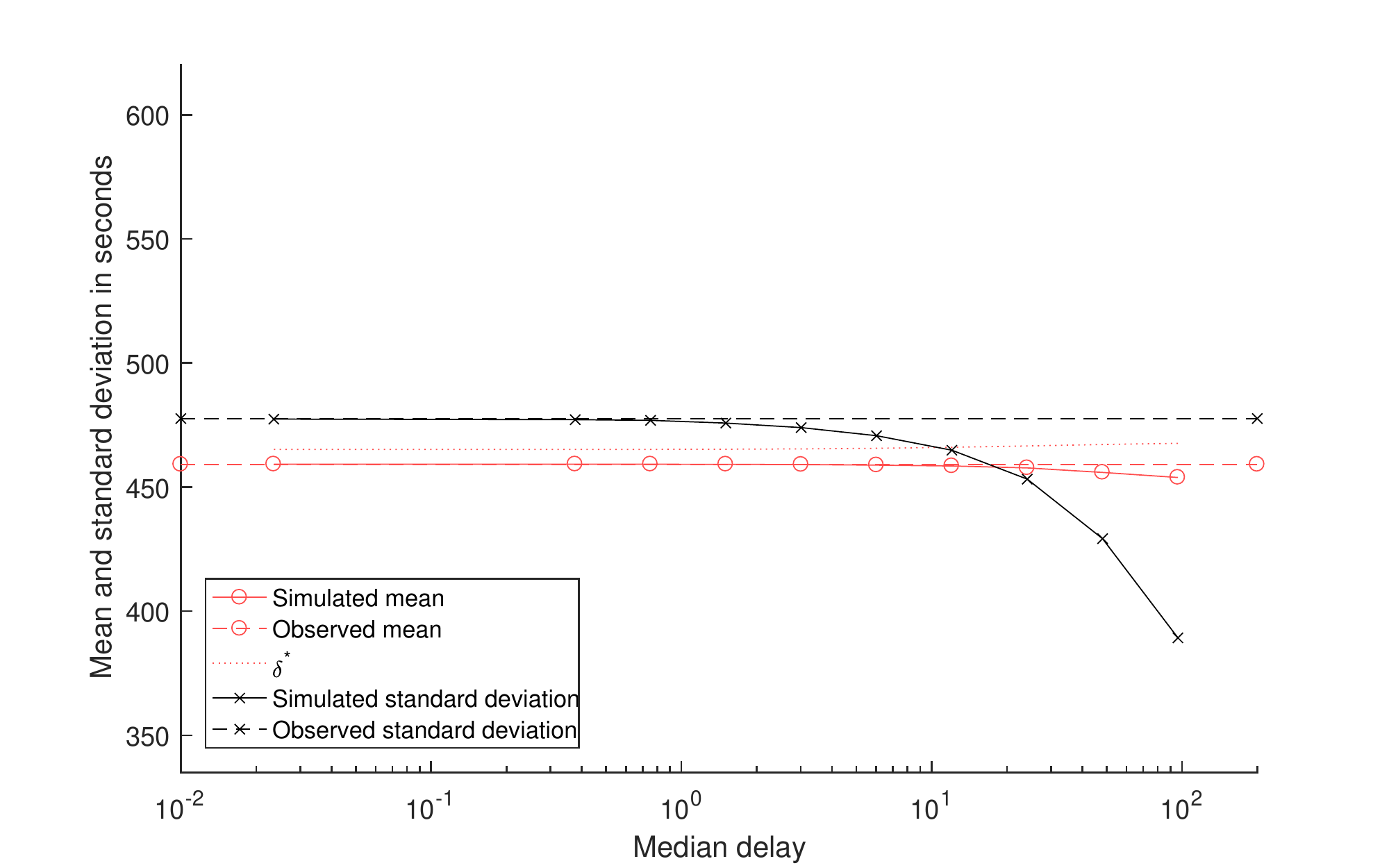}
\caption[]%
{{\small Empirical hash rate function, random difficulty changes}} 
\label{fig:np3}
\end{subfigure}
\quad
\begin{subfigure}[b]{0.475\textwidth} 
\centering 
 \includegraphics[width=\textwidth]{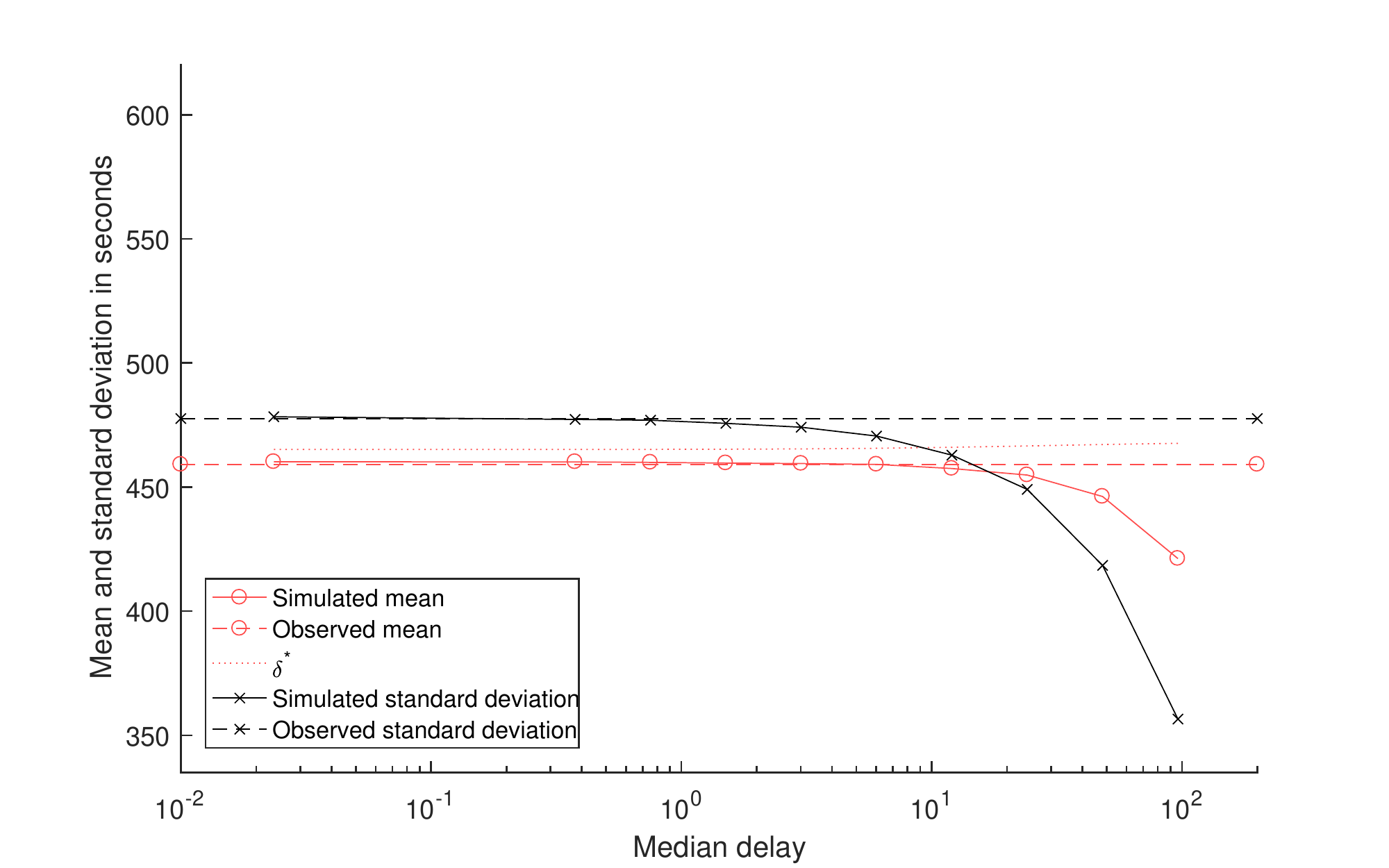}
\caption[]%
{{\small Empirical hash rate function, deterministic difficulty changes}} 
\label{fig:dnp3}
\end{subfigure}
\caption[]%
{\small Interval 3.  Simulating the effect of block propagation delay
(seconds) on the statistics of the inter-arrival time distributions. The
dashed lines indicate the values measured in the blockchain. The dotted
line is the value of the mean predicted in Equation~(\ref{eq:lambert})
by substituting in the value of $a$ estimated from the blockchain
timestamp data.
} 

\label{fig:int3}
\end{figure*}

\begin{figure*}
\centering
\begin{subfigure}[b]{0.475\textwidth}
\centering
 \includegraphics[width=\textwidth]{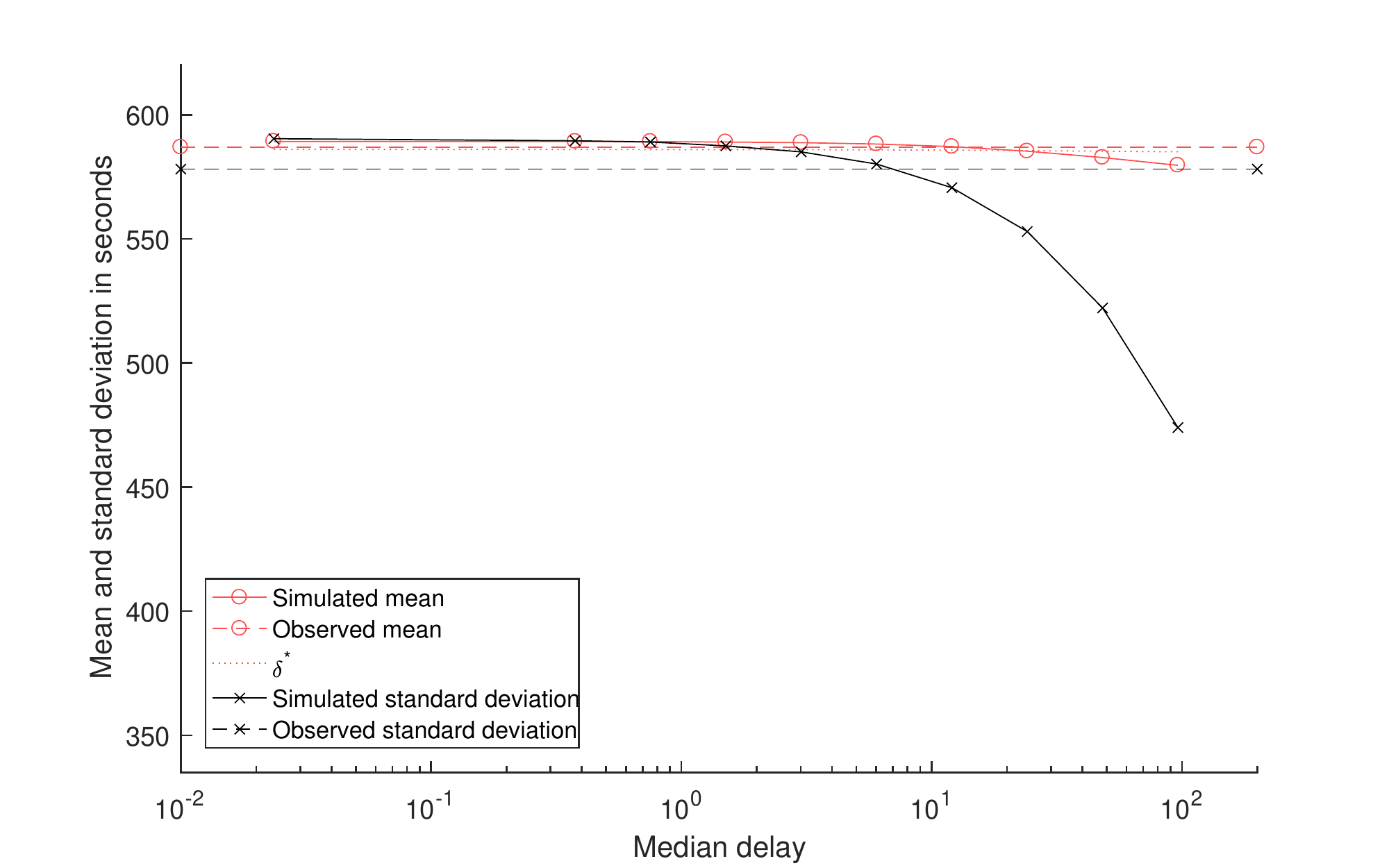}
\caption[]%
{{\small Exponential hash rate function, random difficulty changes}} 
\label{fig:exp4}
\end{subfigure}
\hfill
\begin{subfigure}[b]{0.475\textwidth} 
\centering 
 \includegraphics[width=\textwidth]{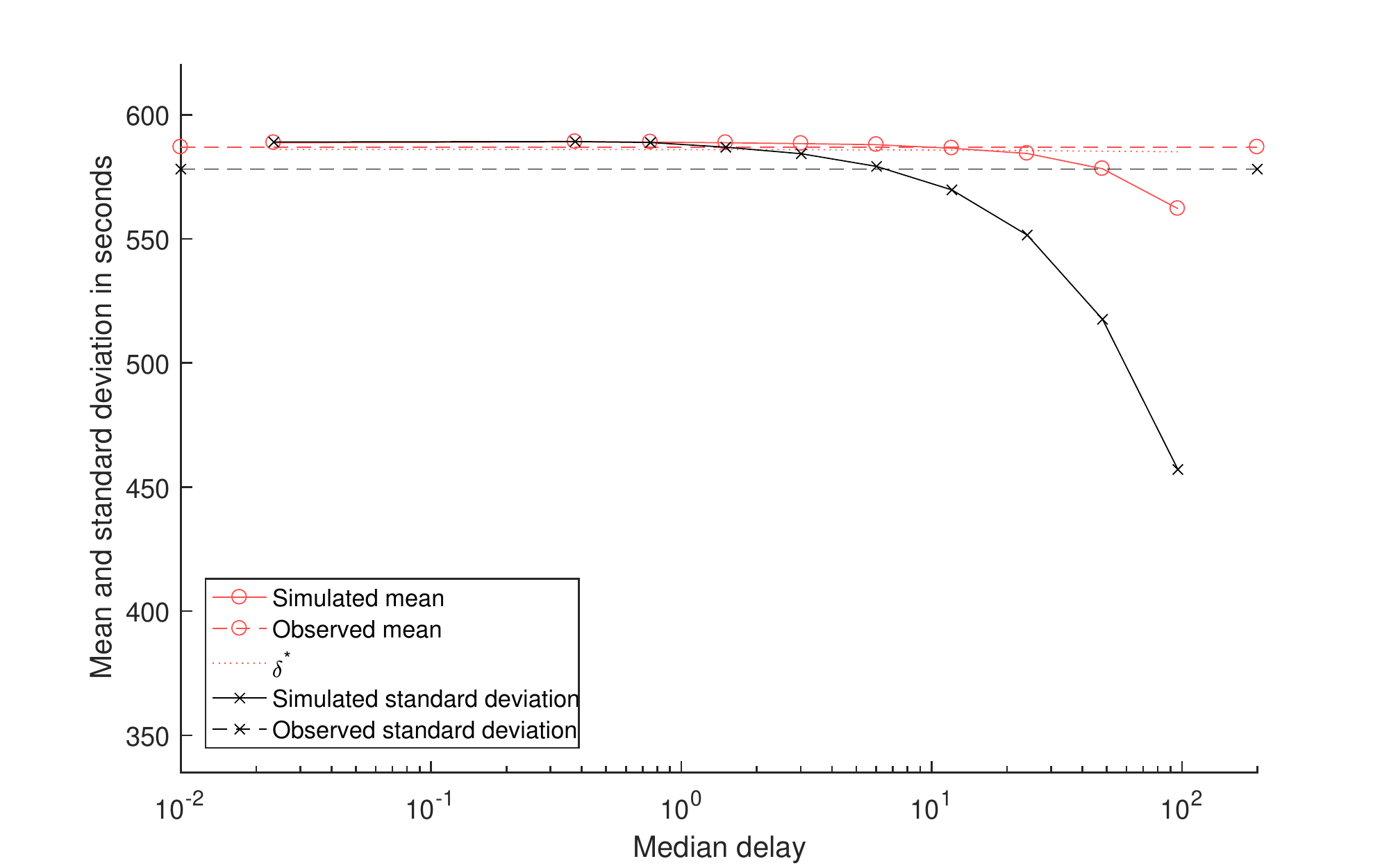}
\caption[]%
{{\small Exponential hash rate function, deterministic difficulty changes}} 
\label{fig:dexp4}
\end{subfigure}
\vskip\baselineskip
\begin{subfigure}[b]{0.475\textwidth} 
\centering 
 \includegraphics[width=\textwidth]{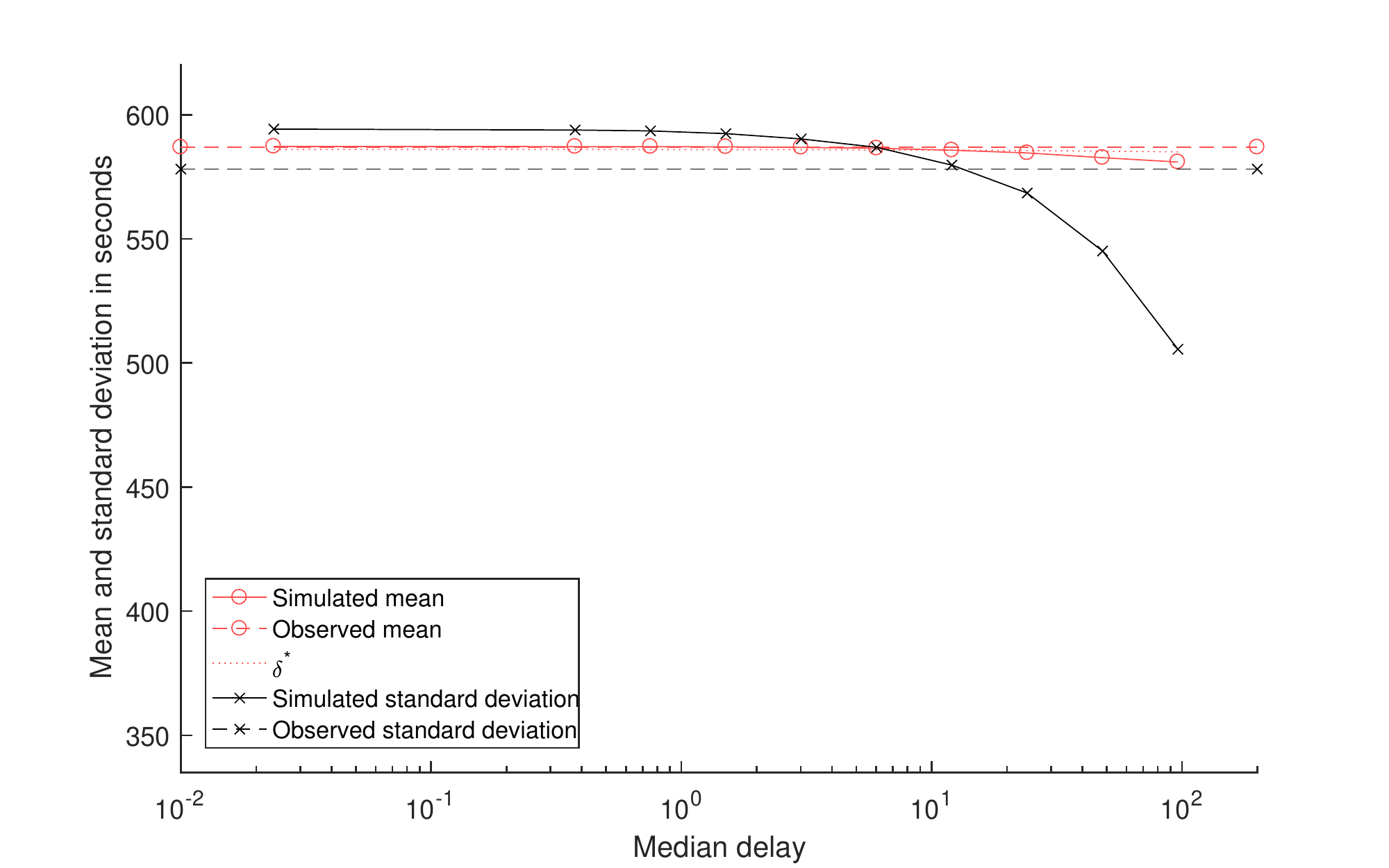}
\caption[]%
{{\small Empirical hash rate function, random difficulty changes}} 
\label{fig:np4}
\end{subfigure}
\quad
\begin{subfigure}[b]{0.475\textwidth} 
\centering 
 \includegraphics[width=\textwidth]{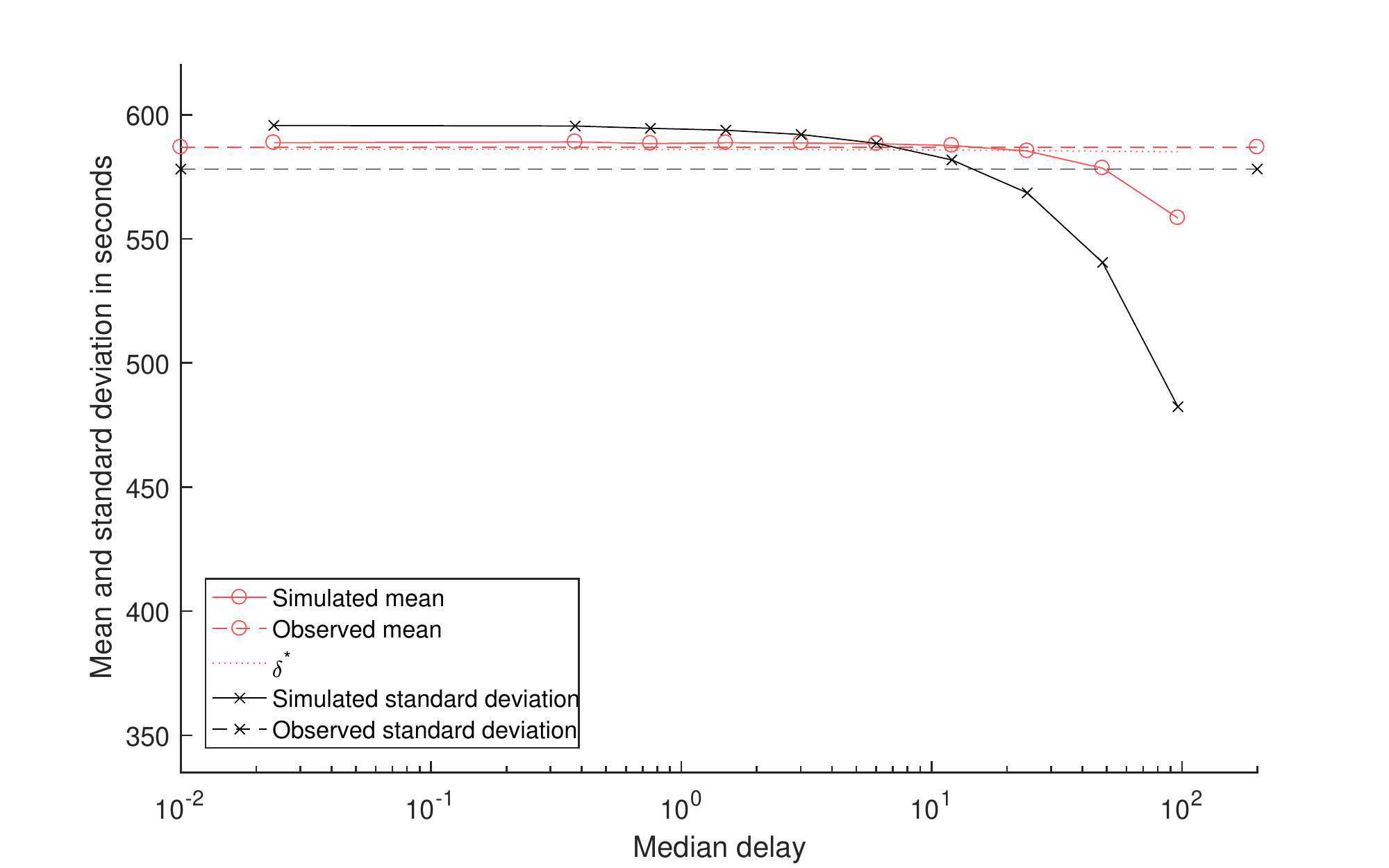}
\caption[]%
{{\small Empirical hash rate function, deterministic difficulty changes}} 
\label{fig:dnp4}
\end{subfigure}
\caption[]%
{\small Interval 4.  Simulating the effect of block propagation delay
(seconds) on the statistics of the inter-arrival time distributions. The
dashed lines indicate the values measured in the blockchain. The dotted
line is the value of the mean predicted in Equation~(\ref{eq:lambert})
by substituting in the value of $a$ estimated from the blockchain
timestamp data.
} 

\label{fig:int4}
\end{figure*}

\begin{figure*}
\centering
\begin{subfigure}[b]{0.475\textwidth}
\centering
 \includegraphics[width=\textwidth]{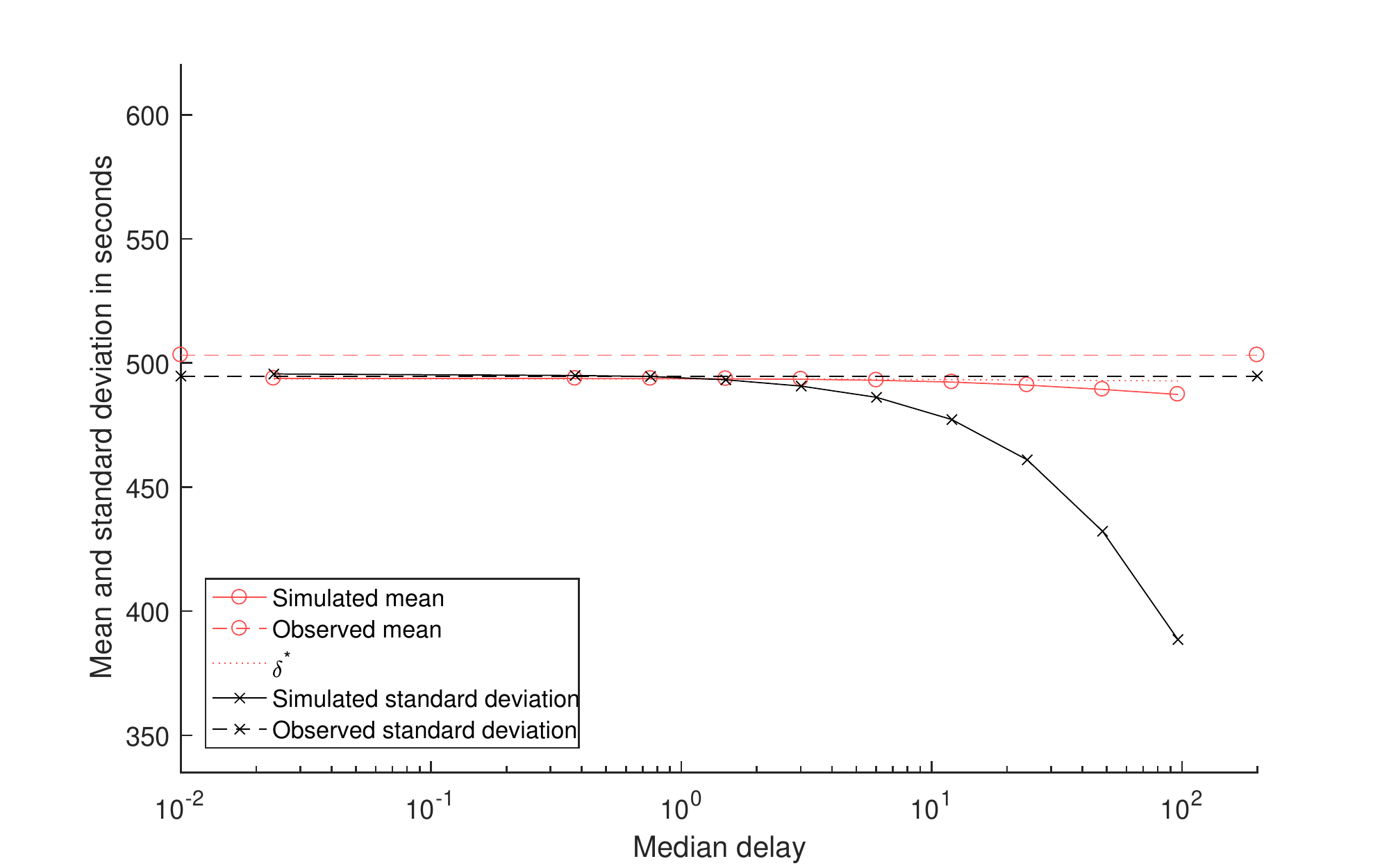}
\caption[]%
{{\small Exponential hash rate function, random difficulty changes}} 
\label{fig:exp5}
\end{subfigure}
\hfill
\begin{subfigure}[b]{0.475\textwidth} 
\centering 
 \includegraphics[width=\textwidth]{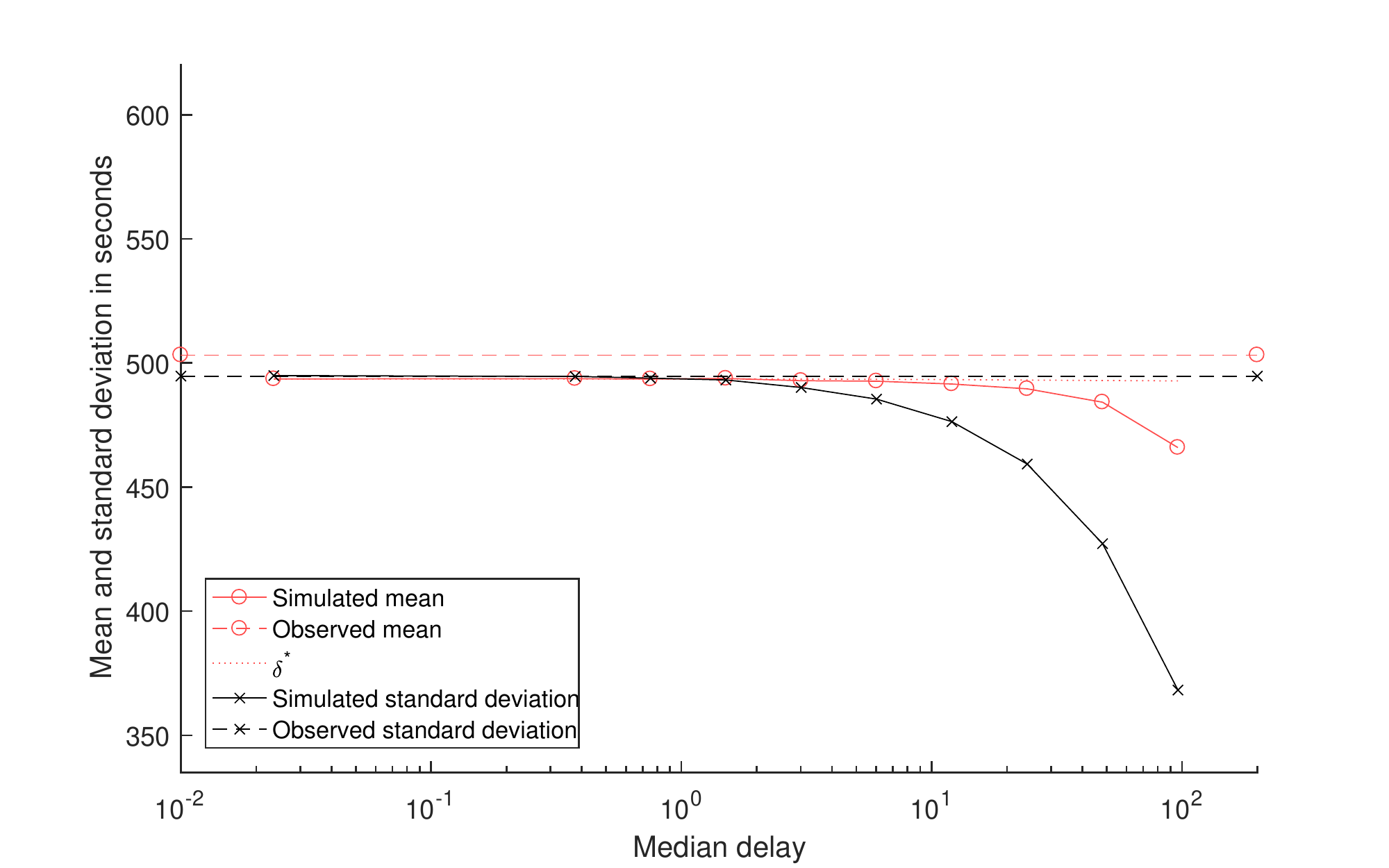}
\caption[]%
{{\small Exponential hash rate function, deterministic difficulty changes}} 
\label{fig:dexp5}
\end{subfigure}
\vskip\baselineskip
\begin{subfigure}[b]{0.475\textwidth} 
\centering 
 \includegraphics[width=\textwidth]{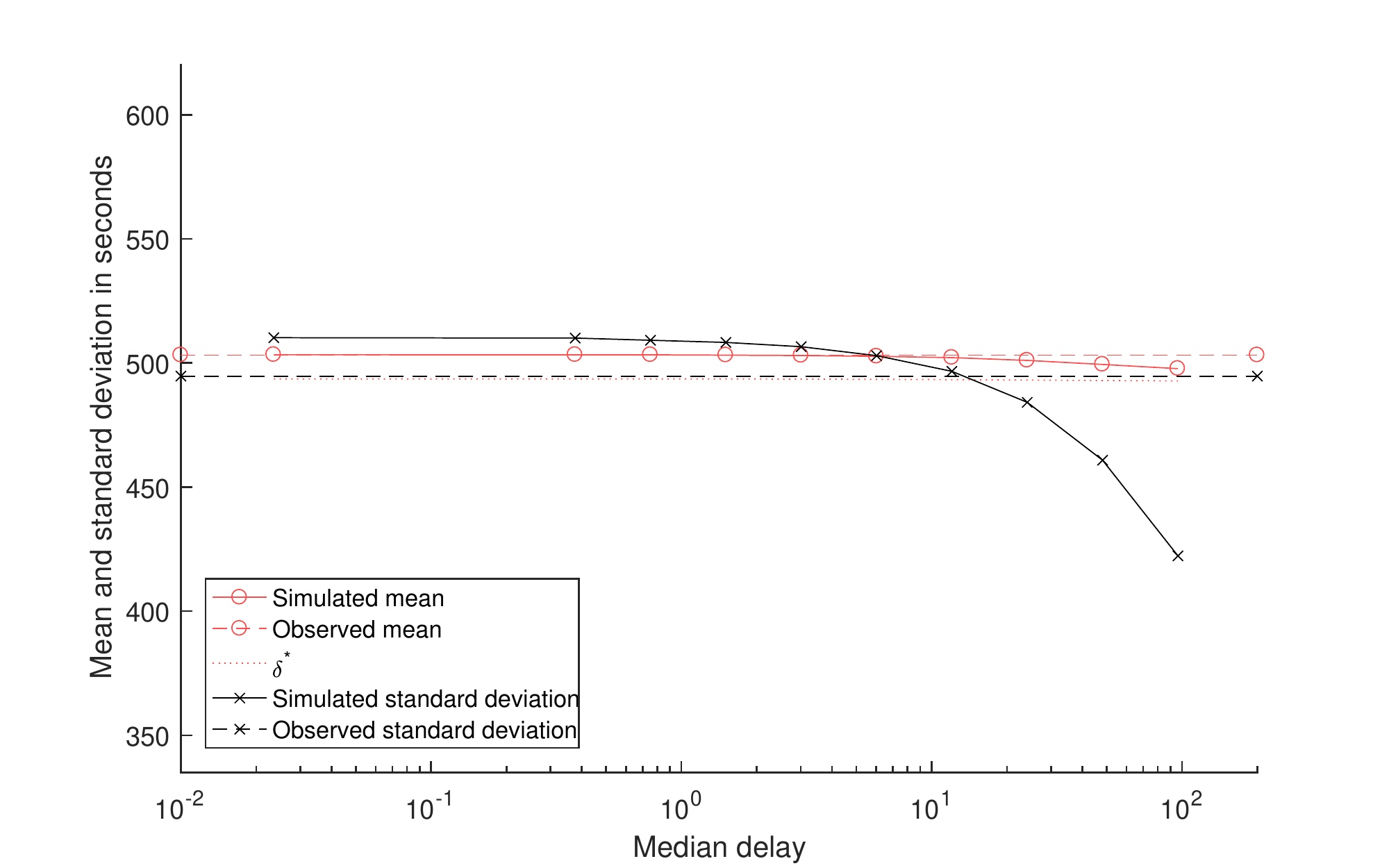}
\caption[]%
{{\small Empirical hash rate function, random difficulty changes}} 
\label{fig:np5}
\end{subfigure}
\quad
\begin{subfigure}[b]{0.475\textwidth} 
\centering 
 \includegraphics[width=\textwidth]{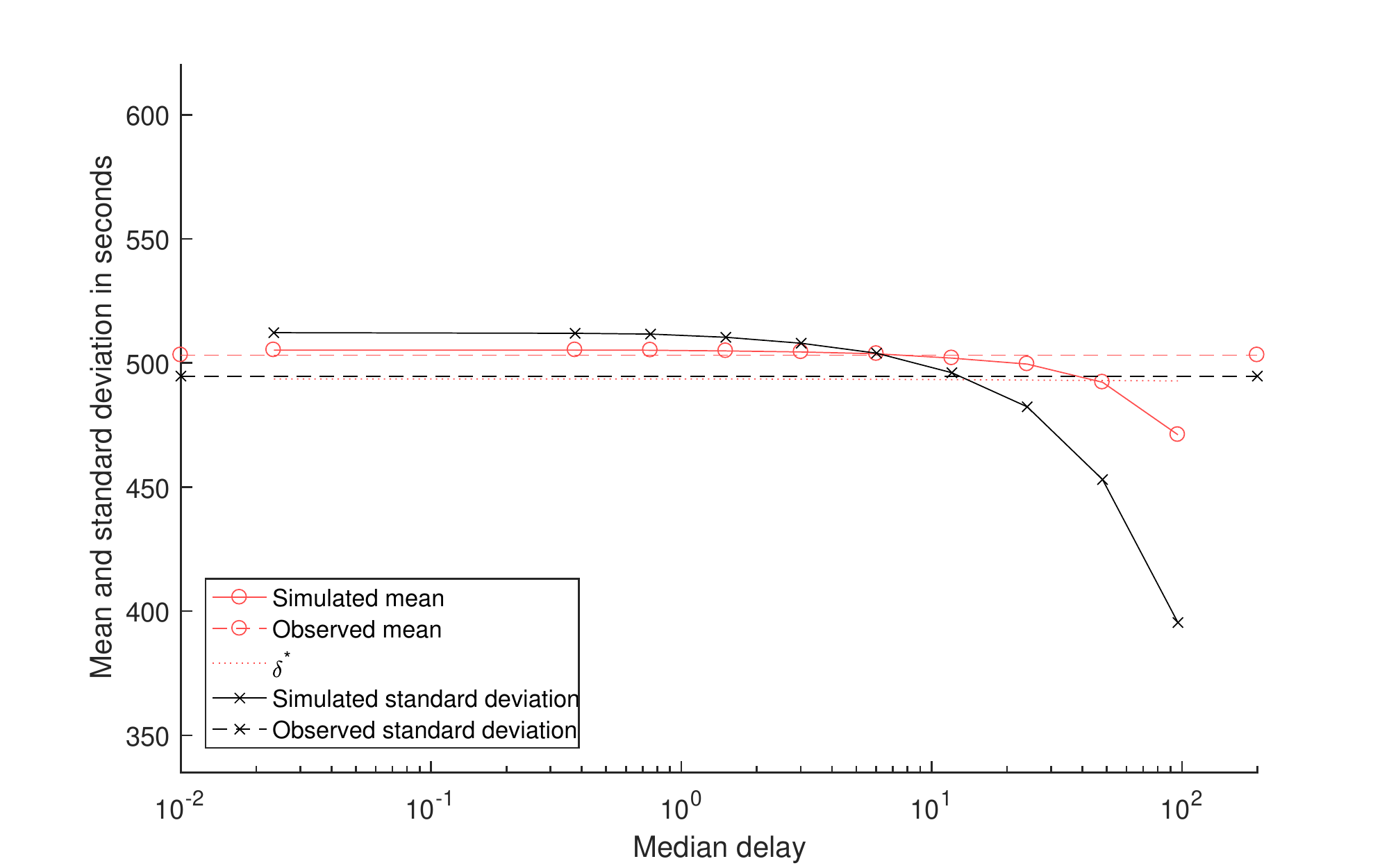}
\caption[]%
{{\small Empirical hash rate function, deterministic difficulty changes}} 
\label{fig:dnp5}
\end{subfigure}
\caption[]%
{\small Interval 5.  Simulating the effect of block propagation delay
(seconds) on the statistics of the inter-arrival time distributions. The
dashed lines indicate the values measured in the blockchain. The dotted
line is the value of the mean predicted in Equation~(\ref{eq:lambert})
by substituting in the value of $a$ estimated from the blockchain
timestamp data.
} 

\label{fig:int5}
\end{figure*}

\begin{figure*}
\centering
\begin{subfigure}[b]{0.475\textwidth}
\centering
 \includegraphics[width=\textwidth]{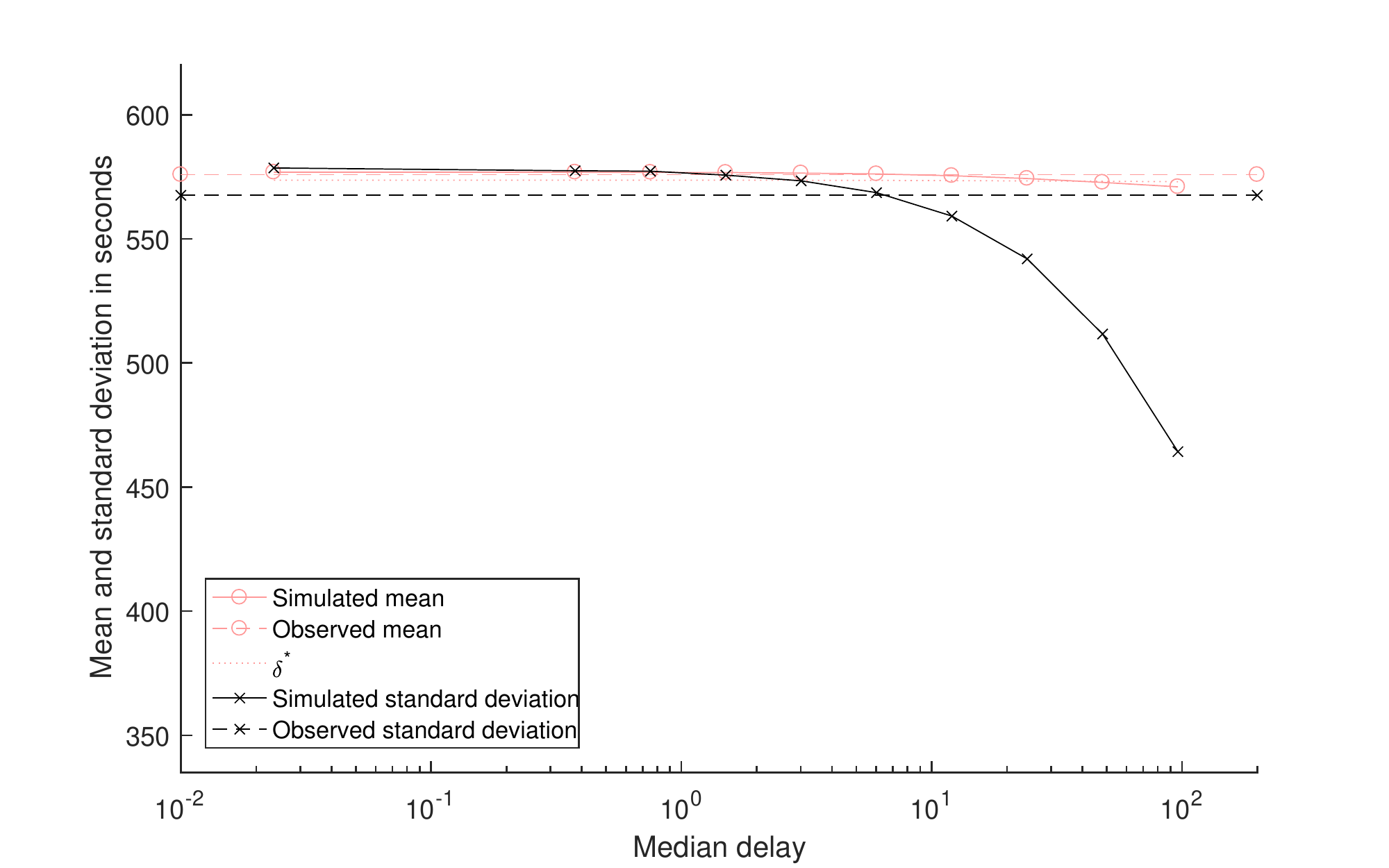}
\caption[]%
{{\small Exponential hash rate function, random difficulty changes}} 
\label{fig:exp6}
\end{subfigure}
\hfill
\begin{subfigure}[b]{0.475\textwidth} 
\centering 
 \includegraphics[width=\textwidth]{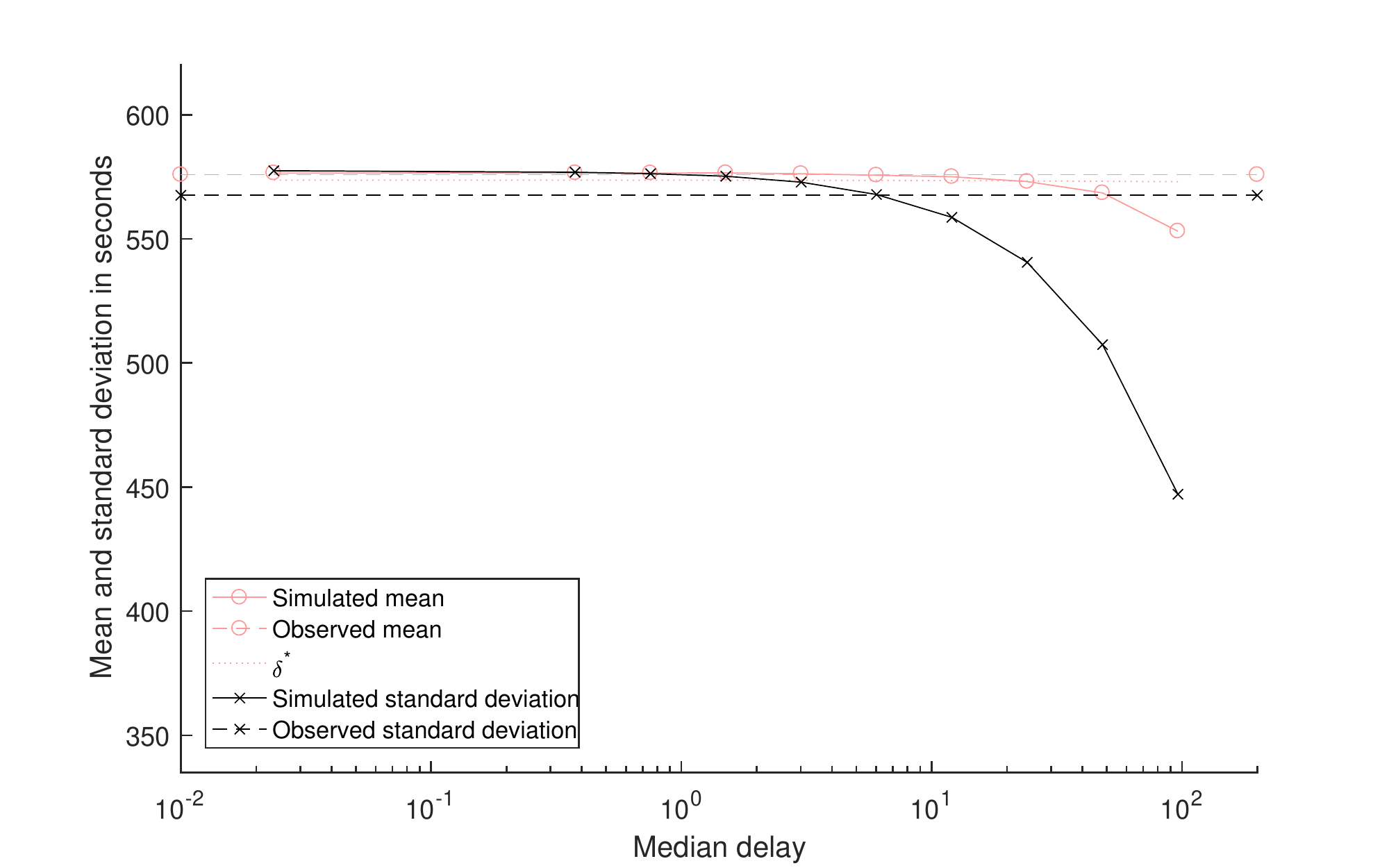}
\caption[]%
{{\small Exponential hash rate function, deterministic difficulty changes}} 
\label{fig:dexp6}
\end{subfigure}
\vskip\baselineskip
\begin{subfigure}[b]{0.475\textwidth} 
\centering 
 \includegraphics[width=\textwidth]{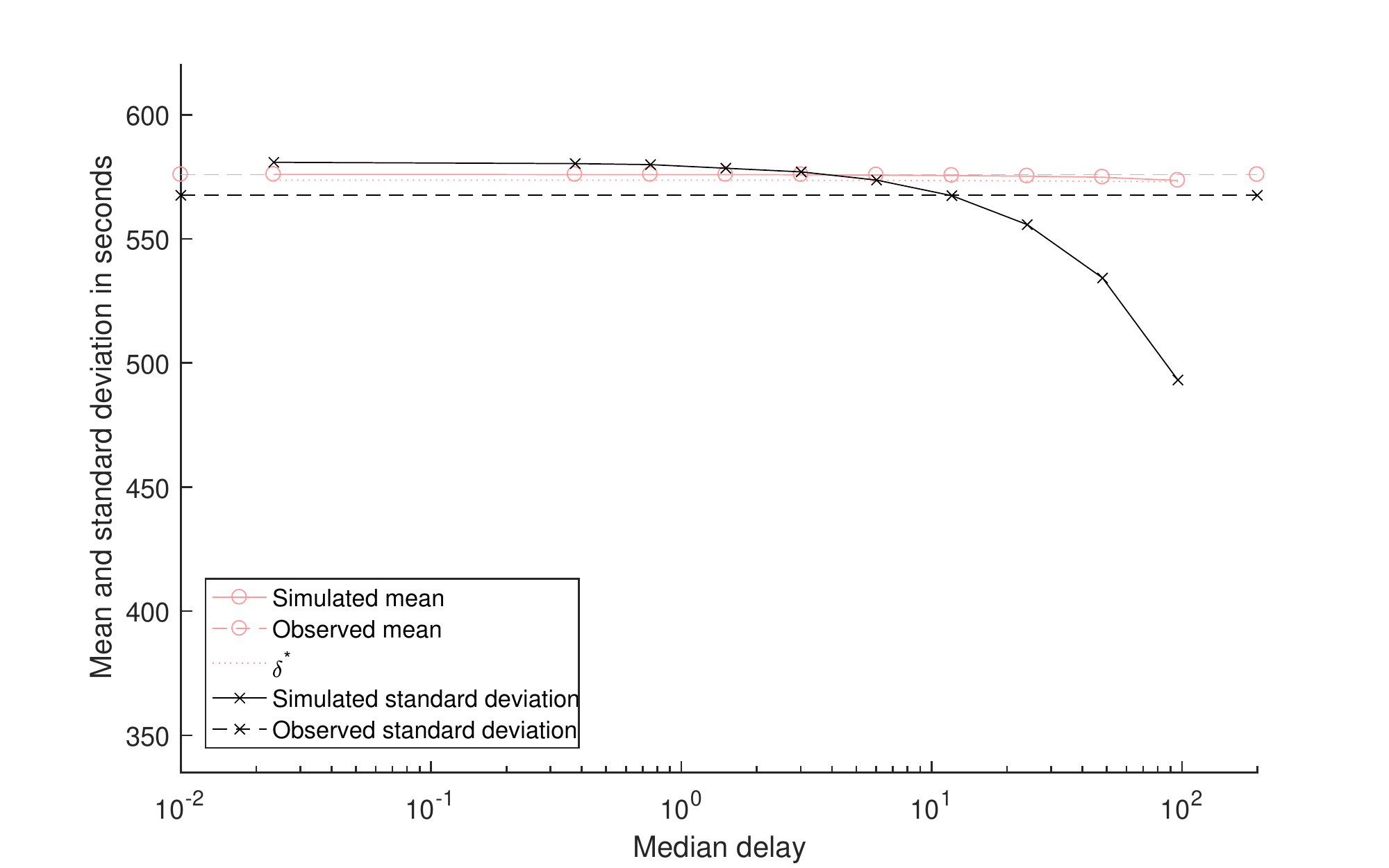}
\caption[]%
{{\small Empirical hash rate function, random difficulty changes}} 
\label{fig:np6}
\end{subfigure}
\quad
\begin{subfigure}[b]{0.475\textwidth} 
\centering 
 \includegraphics[width=\textwidth]{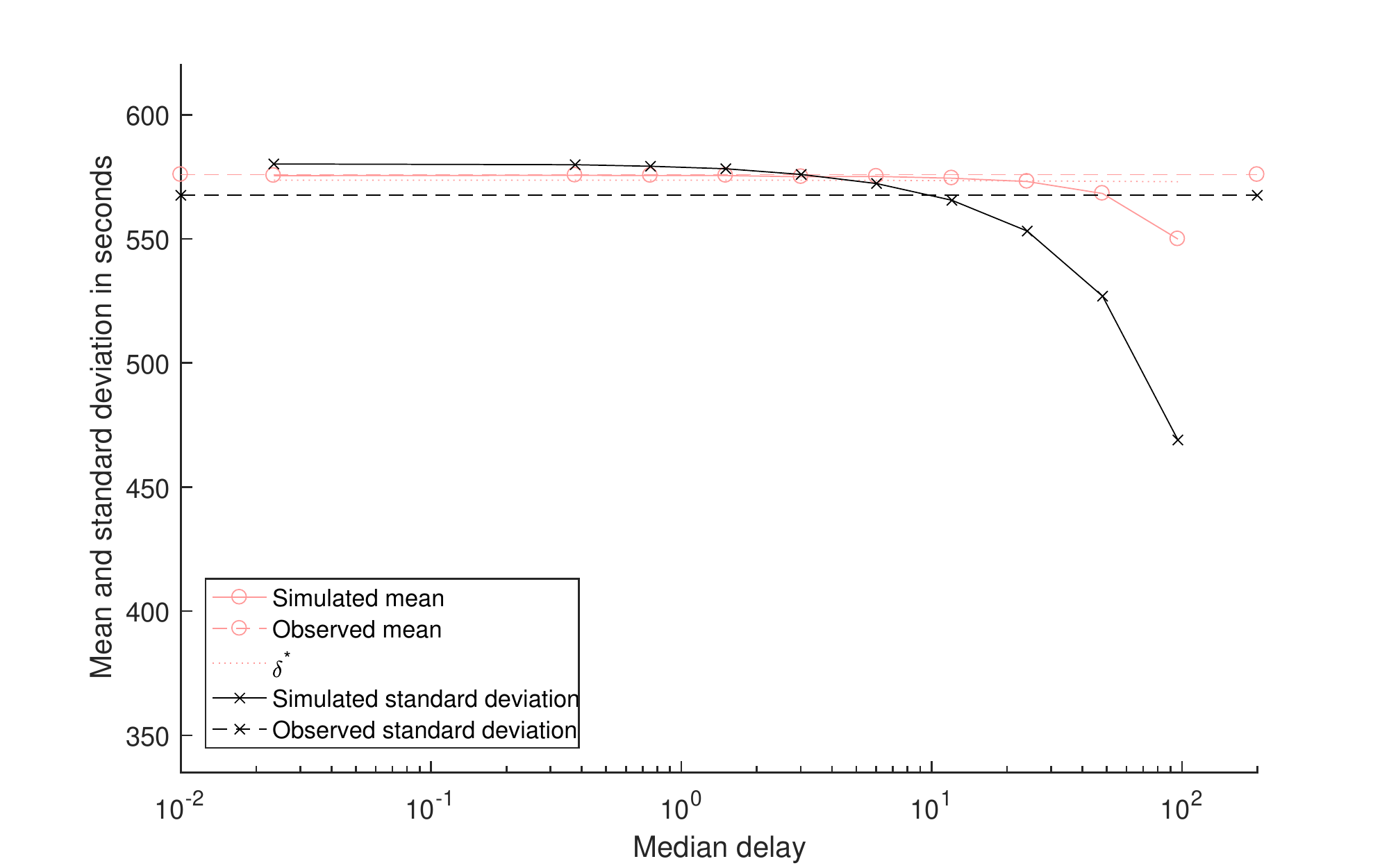}
\caption[]%
{{\small Empirical hash rate function, deterministic difficulty changes}} 
\label{fig:dnp6}
\end{subfigure}
\caption[]%
{\small Interval 6.  Simulating the effect of block propagation delay
(seconds) on the statistics of the inter-arrival time distributions. The
dashed lines indicate the values measured in the blockchain. The dotted
line is the value of the mean predicted in Equation~(\ref{eq:lambert})
by substituting in the value of $a$ estimated from the blockchain
timestamp data.
} 

\label{fig:int6}
\end{figure*}

Figures~\ref{fig:int2} through \ref{fig:int6}, subfigures (b) and (d)
show the the results for simulations of models with deterministic
difficulty changes. Subfigure (b) corresponds to the model with an
exponential hash rate, and subfigure (d) corresponds to the model with
an empirical hash rate. In each case we can see the simulated mean and
standard deviation values are close to those of the corresponding models
with random difficulty changes for low values of propagation delay, but
as the propagation delay increases the deterministic approximation loses
accuracy. For high values of the propagation delay, the deterministic
models uniformly have lower means and standard deviations of
inter-arrival times.

\subsection{A summary of model performance and recommendations}

In this subsection we consider which models and in what circumstances
they should be used. We use the notation outlined in
Section~\ref{sec:summary}.
\begin{enumerate}

\item The most accurate and successful model is \ref{label:1b}
\ref{label:2b} \ref{label:3b}
-- empirical hash rate, random difficulty changes and propagation delay
included. We would recommend this model whenever it is possible to
closely estimate the hash rate and accurate results are required. This
is also the most complicated model.\label{mod1}

\item The next most accurate model is \ref{label:1a} \ref{label:2b}
\ref{label:3b}
-- this is the same as above, but with an exponential hash rate. This
model is accurate whenever the growth in the hash rate is expected to be
to be approximately exponential. This would include the present day and
could include a short period into the future, so this model might be
reasonable for forecasting.\label{mod2}

\item The simplest model that we recommend is \ref{label:1a}
\ref{label:2a} \ref{label:3a}
-- exponential hash rate, deterministic difficulty changes and no
propagation delay. This model is a nonhomogeneous Poisson process where
the hash rate function is periodic. The rate increases exponentially
over each segment  then instantaneously returns to its initial value at
the end of each segment.  This model provides a reasonable approximation
to the block arrival process while still remaining analytically
tractable.\label{mod3}

\end{enumerate}
For the latter two options, if the long term block arrival rate is all
that is required it can be calculated with Equation (\ref{eq:time_vs_a})
or (\ref{eq:lambert}).

\section{Conclusion}\label{sec:conc}

In this paper we have presented a suite of model\cremoveR{}{s} for the
point process of block mining epochs in the Bitcoin blockchain, and
tested it with both simulation and data available from the blockchain
itself. The main challenges in doing this are: 
\begin{enumerate}

\item the unknown global hash rate that drives the rate of block
discovery; and 

\item the historically known, but random, mining difficulty value. 

\end{enumerate}
Furthermore, the data on the times at which blocks arrive is
comprehensive, but not completely reliable. We postulate a point process
model in which the block arrival process behaves as a nonhomogeneous
Poisson process on times between difficulty changes with rate
proportional to the ratio of the hash rate to the difficulty, but which
is dependent on itself when difficulty changes are taken into account.
For a given global hash rate, the times at which the difficulty changes
(and hence the block arrival rate changes) are determined by the
sample-path of the process. 
self-correcting.  Nevertheless, the global hash rate is consistently
rapidly increasing and the difficulty feedback mechanism is sufficiently
delayed so that the rate of block arrivals has been around 11.5\%
greater than the base rate of 6 per hour\footnote{For blocks arriving
after December 30, 2009.}. This means that overall, the transaction
throughput and the total miner income from bounties are higher than in
the base case.  Furthermore, the times when the block mining bounty is
halved and the time when all bitcoins have been created will occur
earlier than they would have if blocks were mined at a rate of six per
hour.

In addition to giving a model for the block arrival process, we have
derived the relationship between block arrival rates and the rate of
exponential increase of the hash rate. We have also provided a practical
approximation that exhibits limiting behaviour independent of initial
conditions and perturbations of the block arrival process, and verify
this approximation with simulation and measurements from the blockchain.

\raggedright

\section*{Acknowledgments} \label{sec:acknowledgement}

The work of Anthony Krzesinski is supported by Telkom SA Limited.

\Np
The work of Peter Taylor and Rhys Bowden is supported by the Australian Research Council
Laureate Fellowship FL130100039 and the ARC Centre of Excellence for
Mathematical and Statistical Frontiers (ACEMS).


\bibliographystyle{elsarticle-num}




\clearpage

\begin{appendices}

\onecolumn

\section{Proof of Equation~(\ref{eq:expon_expect})}\label{app:a}

For $1 \leq n \leq 2016$ and $a > 0$
\begin{align}
\E\lbrack X_n|X_0=0\rbrack
&= \int_0^{\infty} x f_{X_n|X_0 = 0}(x)\,dx\nonumber \\
&= \int_0^{\infty} x e^{ax} e^{-\frac{1}{a}(e^{ax}-e^{0a})} (\frac{1}{a}(e^{ax}-e^{0a}))^{n-1} \frac{1}{(n-1)!}\,dx\notag\nonumber \\
&= \int_0^{\infty} x e^{ax} e^{-\frac{1}{a}(e^{ax}-1)} (\frac{1}{a}(e^{ax}-1))^{n-1} \frac{1}{(n-1)!}\,dx\notag.
\end{align}
Setting $g(t) = e^{at}$ and $h(t)
=\mathrm{exp}(-(e^{at}-1)/a)$ and for $n=1,2,\ldots$ let
\begin{align}
I_n(x) = \int_\infty^x g(t)(g(t)-1)^{n-1}h(t)\,dt .
\end{align}
Integration by parts gives
\begin{align}
\E\lbrack X_n|X_0=0\rbrack &= \frac{1}{(n-1)!a^{n-1}} \int_0^\infty xg(x)(g(x)-1)^{n-1}h(x)\,dx \notag\\
	&=\frac{1}{(n-1)!a^{n-1}} \left( \Big\lbrack x I_n(x)\Big\rbrack_0^\infty - \int_0^\infty I_n(x) \,dx\right). \label{eq:EXn1}
\end{align}
We proceed by finding an expression for $I_n(x)$. Note first that
$\frac{d}{dx} h(x) = -g(x)h(x)$. Once again applying integration by
parts gives (for $n \geq 2$):
\begin{align}
I_n(x) &= \int_\infty^x gh(g-1)^{n-1}\,dt \notag\\
	&= -h(g-1)^{n-1} - \int_\infty^x ag(g-1)^{n-2}(n-1)(-h)\,dt\notag\\
	&= -h(g-1)^{n-1}+a(n-1)I_{n-1}(x).\notag
\end{align}
Combining this with $I_1(x) = \int_\infty^x gh\,dt = -h(x)$ and setting
the constant of integration to~0 we get (for $n \geq 1$):
\begin{align}
I_n(x) = -h \sum_{i=0}^{n-1}   \frac{(n-1)!}{i!}a^{n-1-i}(g-1)^{i}. \label{eq:In}
\end{align}
Similarly
\begin{align}
\int_\infty^x g^n h\,dt = -h(x)\sum_{i=0}^{n-1} \frac{(n-1)!}{i!} a^{n-1-i} g(x)^{i}.
\end{align}
Thus
\begin{align}
\int_0^\infty  I_n(x)\,dx
&= \int_0^\infty -h \sum_{i=0}^{n-1}   \frac{(n-1)!}{i!}a^{n-1-i}(g-1)^{i} \,dx \notag\\
	&= \int_0^\infty -h \sum_{i=0}^{n-1} \frac{(n-1)!}{i!}a^{n-1-i} \sum_{j=0}^i \binom{i}{j} g^j(-1)^{i-j} \,dx \notag\\
	&=  \sum_{i=0}^{n-1}  (n-1)!a^{n-1-i} \sum_{j=0}^i \frac{ (-1)^{i+j+1}}{j!(i-j)!}\int_0^\infty g^jh \,dx \notag\\
	&=  \sum_{i=0}^{n-1}  (n-1)!a^{n-1-i} \left(\frac{(-1)^{i+1}}{i!} \int_0^{\infty}h\,dx +\sum_{j=1}^i \frac{(-1)^{i+j}}{j!(i-j)!} \bigg\lbrack h\sum_{k=0}^{j-1} \frac{(j-1)!}{k!} a^{j-1-k} g^{k} \bigg\rbrack_0^\infty\right)\notag\\
	&=  \sum_{i=0}^{n-1}  (n-1)!a^{n-1-i} \left(\frac{(-1)^{i}}{i!a}e^{1/a}\mathrm{Ei}(-\frac{1}{a}) +\sum_{j=1}^i \frac{(-1)^{i+j+1}}{j(i-j)!}\sum_{k=0}^{j-1} \frac{a^{j-1-k}}{k!} \right)\label{eq:int_In} \, ,
\end{align}
where the last line 
uses
\begin{eqnarray}
\int_0^\infty h\,dx
&=& \int_0^\infty \mathrm{exp}(-(e^{ax}-1)/a)\,dx\notag \\
&=& \frac{e^{1/a}}{a}\int_{1/a}^\infty \frac{e^{-u}}{u} \,dx\notag
 = -\frac{e^{1/a}}{a}\mathrm{Ei}(-\frac{1}{a}).\notag
\end{eqnarray}


Now we can evaluate (\ref{eq:EXn1})
\begin{align}
\E\lbrack  X_n|X_0=0\rbrack =\frac{1}{(n-1)!a^{n-1}} \left( \lbrack x I_n(x)\rbrack_0^\infty - \int_0^\infty I_n(x) \,dx\right) \,,
\end{align}
which, via equations (\ref{eq:In}) and (\ref{eq:int_In}), becomes
\begin{align}
\E\lbrack & X_n|X_0=0\rbrack \\
=& \frac{1}{(n-1)! a^{n-1}}  \notag
	 \times\left(0 - \sum_{i=0}^{n-1}  (n-1)!a^{n-1-i} \left(\frac{(-1)^{i}}{i!a}e^{1/a}\mathrm{Ei}(-\frac{1}{a}) +\sum_{j=1}^i \frac{(-1)^{i-j+1}}{j(i-j)!}\sum_{k=0}^{j-1} \frac{a^{j-1-k}}{k!} \right) \right)  \nonumber \\
=& \sum_{i=0}^{n-1} a^{-i} \left(\frac{(-1)^{i+1}}{i!a}e^{1/a}\mathrm{Ei}(-\frac{1}{a}) +\sum_{j=1}^i \frac{(-1)^{i+j}}{j(i-j)!}\sum_{k=0}^{j-1} \frac{a^{j-1-k}}{k!} \right) \nonumber \\
=& \sum_{i=0}^{n-1} (-a)^{-i} \left(\frac{-e^{1/a}}{i!a}\mathrm{Ei}(-\frac{1}{a}) +\sum_{j=1}^i \frac{(-a)^{j}}{j(i-j)!}\sum_{k=0}^{j-1} \frac{a^{-1-k}}{k!} \right) \nonumber \\
=& \sum_{i=0}^{n-1} (-a)^{-i-1} \left(\frac{e^{1/a}}{i!}\mathrm{Ei}(-\frac{1}{a}) -\sum_{j=1}^i \frac{(-a)^{j}}{j(i-j)!}\sum_{k=0}^{j-1} \frac{a^{-k}}{k!} \right). \label{eq:intrasegment}
\end{align}

\end{appendices}

\raggedright

\twocolumn

\end{document}